\documentclass[fleqn,usenatbib]{mnras}
\usepackage{comment}
\usepackage{setspace}
\usepackage{braket}
\usepackage{url}
\usepackage{longtable}
\usepackage{pdflscape}

\usepackage{newtxtext,newtxmath}

\usepackage[T1]{fontenc}
\usepackage{ae,aecompl}

\usepackage{graphicx}	
\usepackage{amsmath}	
\usepackage{amssymb}	

\title[GMC properties in the strongly barred galaxy  NGC~1300]{Properties of giant molecular clouds in the strongly barred galaxy NGC~1300}

\author[F. Maeda et al.]{
Fumiya Maeda,$^{1}$\thanks{E-mail: fmaeda@kusastro.kyoto-u.ac.jp}
Kouji Ohta,$^{1}$
Yusuke Fujimoto,$^{2,3}$
and Asao Habe$^{4}$
\\
$^{1}$Department of Astronomy, Kyoto University, Kitashirakawa-Oiwake-Cho, Sakyo-ku, Kyoto, Kyoto 606-8502, Japan\\
$^{2}$Research School of Astronomy \& Astrophysics, Australian National University, Canberra, Australian Capital Territory 2611, Australia\\
$^{3}$Department of Terrestrial Magnetism, Carnegie Institution for Science, 5241 Broad Branch Road, NW, Washington, DC 20015, USA\\
$^{4}$Graduate School of Science, Hokkaido University, Kita 10 Nishi 8, Kita-ku, Sapporo, Hokkaido 060-0810, Japan
}

\date{Accepted XXX. Received YYY; in original form ZZZ}

\pubyear{2019}

\begin{document}
\label{firstpage}
\pagerange{\pageref{firstpage}--\pageref{lastpage}}
\maketitle

\begin{abstract}
Star formation activity depends on galactic-scale environments. To understand the variations in star formation activity, comparing the properties of giant molecular clouds (GMCs) among environments with different star formation efficiency (SFE) is necessary. We thus focus on a strongly barred galaxy to investigate the impact of the galactic environment on the GMCs properties, because the SFE is clearly lower in bar regions than in arm regions. In this paper, we present the $^{12}$CO($1-0$) observations toward the western bar, arm and bar-end regions of the strongly barred galaxy NGC~1300 with ALMA 12-m array at a high angular resolution of $\sim$40 pc. We detected GMCs associated with the dark lanes not only in the arm and bar-end regions but also in the bar region, where massive star formation is not seen. Using the CPROPS algorithm, we identified and characterized 233 GMCs across the observed regions. Based on the Kolmogorov-Smirnov test, we find that there is virtually no significant variations in GMC properties (e.g., radius, velocity dispersion, molecular gas mass, and virial parameter) among the bar, arm and bar-end region. These results suggest that systematic differences in the  physical properties of the GMCs are not the cause for SFE differences with environments, and that there should be other mechanisms which control the SFE of the GMCs such as fast cloud-cloud collisions in NGC~1300.
\end{abstract}



\defcitealias{Colombo:2014ei}{C14}

\section{Introduction}\label{sec:intro}
It is  important to investigate the relation between star formation rate (SFR) and gas density in the context of galaxy evolution. This relation describes how efficiently galaxies convert their gases into stars. Previous studies on the Milky Way and nearby disc galaxies show that there is a tight correlation between the SFR and the surface density of molecular gas; $\Sigma_{\rm SFR} \propto \Sigma_{\rm H_2}^N$, where the index $N$  is between 1 and 2 \citep[e.g.,][]{Kennicutt1998ARA&A,Bigiel2008AJ,Schruba2011AJ}. 
However, variations in the relation have been found; star formation activity changes among galactic-scale environments, and the gas surface density is not the only factor controlling the star formation.
\citet{Momose2010ApJ} measured the star formation efficiency (SFE $= \Sigma_{\rm SFR}/\Sigma_{\rm H_2}$) in a barred galaxy of NGC 4303 at a spatial resolution of 500 pc and reported that the SFEs in the arm regions are about two times higher than those in the bar regions \citep[see also][]{Yajima:2019do}. Such variations of the SFE with environments are reported for other nearby galaxies and the Milky Way  \citep[e.g.,][]{Watanabe2011,Leroy2013AJ.146,Longmore2013MNRAS.429,Hirota:2014bt,Usero2015AJ,Gallagher:2018ep}, but physical mechanism which controls
the SFE within galaxies is still unclear.

To understand the variation of the SFE, it is important to unveil the properties of giant molecular clouds (GMCs) where star formation occurs. In particular, comparing the GMC properties among environments with different SFEs is necessary. In the Milky Way, GMCs' mass, size, and velocity dispersion are typically $\sim 10^{4-6}~M_\odot$, $\sim 20-100$ pc, and $2 - 10~\rm km~s^{-1}$, respectively \citep[e.g.,][]{Solomon87,Heyer2009ApJ}. In recent years, thanks to high sensitive interferometers, extragalactic GMCs can be observed at a high angular resolution of 20$\sim$50 pc; 
e.g.,
M51 \citep{Colombo:2014ei},
M83 \citep{Hirota:2018jp},
PHANGS survey
(A. K. Leroy et al. 2019, in preparation).
For example, \citet{Hirota:2018jp} investigated GMC properties in an intermediate-type barred galaxy of M83 where the SFE in the bar regions is about two times smaller than that in the arm regions. They found the virial parameter, which is a measure for gravitational binding of GMCs, in the bar ($\sim$1.6) is larger than that in the arm ($\sim$1.0). This result is consistent with the idea that the many clouds in the bar are gravitationally unbound which makes the SFE low \citep[see also][]{Egusa:2018hq}. However, only a few such studies on the relation between GMC properties and the variation of star formation activity with environments are made so far.

\begin{figure}
	\includegraphics[width=\hsize]{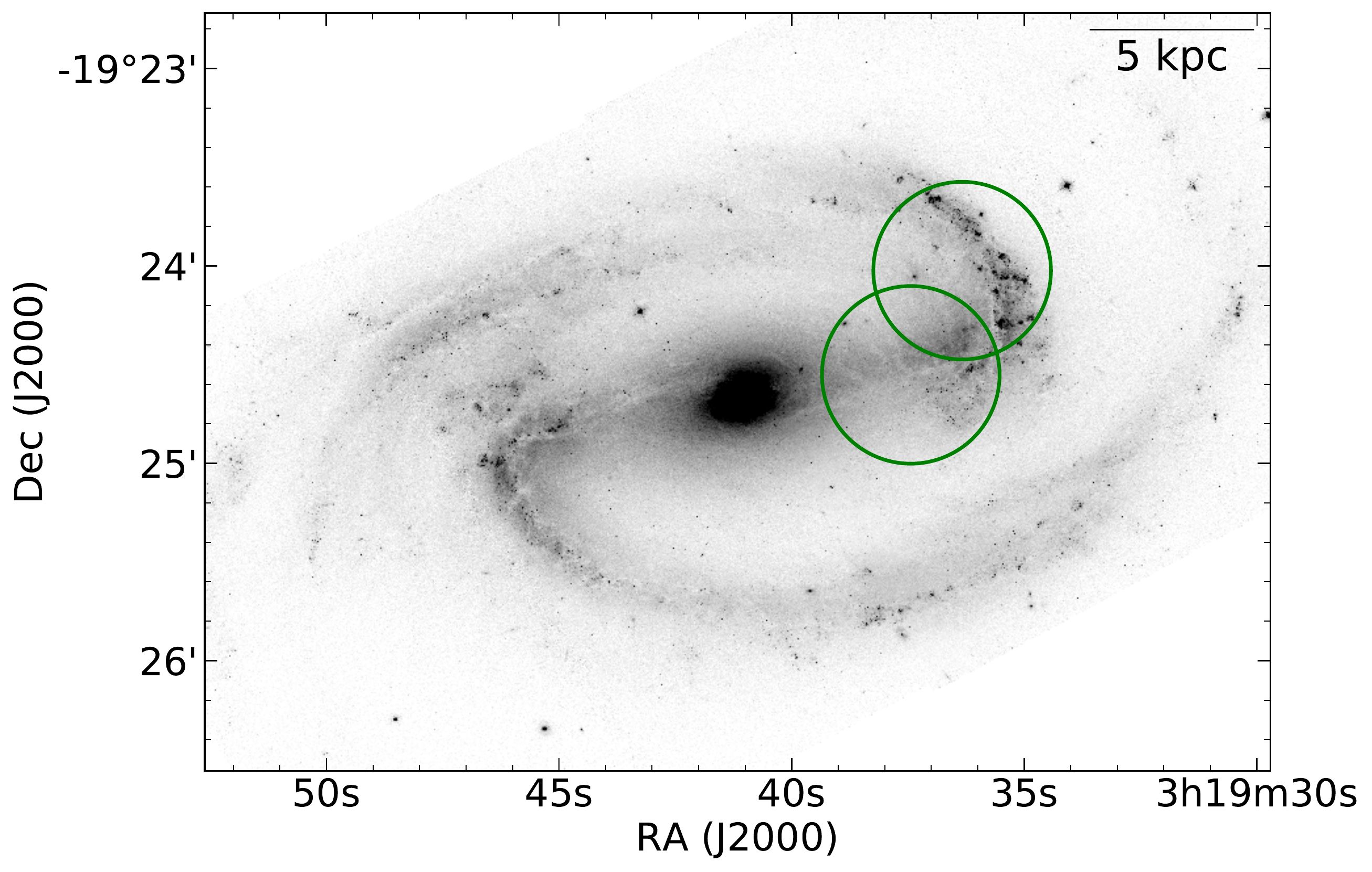}
    \caption{
    V-band image of NGC~1300 taken with F555W filter on Advanced Camera for Surveys (ACS) of the Hubble Space Telescope (HST).   We obtained this image from the Hubble Legacy Archive (\url{https://hla.stsci.edu/}). Green circles  represent field of views (FoVs; HPBW $= 54^{\prime\prime}\phi$) observed with ALMA (see Section~\ref{sec:obs}).}
 \label{fig:NGC1300_Ha_continuum}
\end{figure}

A strongly barred galaxy is a suitable laboratory for studying the impact of the galactic environments on the GMCs properties, because the absence of star formation is clearly seen in bar regions; remarkable dust lane is seen in the bar regions without prominent H\textsc{ii} regions, while in the arm regions H\textsc{ii} regions are associated with dust lanes. In this study, therefore, we focus on a nearby prototype strongly barred galaxy of NGC~1300 (Fig. ~\ref{fig:NGC1300_Ha_continuum}). \citet{Maeda:2018bg} carried out $^{12}$CO($1-0$) observations toward the bar and arm region of NGC~1300 with a single dish telescope of Nobeyama 45-m telescope. They find that the $\Sigma_{\rm H_2}$ in the bar region is comparable to that in the arm region ($\Sigma_{\rm H_2} \sim 10~ M_\odot~{\rm pc^{-2}}$ ), indicating the star formation activity in the bar region is clearly suppressed.

In this paper, we report $^{12}$CO($1-0$) observations toward the western bar, arm, and bar-end regions in NGC~1300 at a high angular resolution of  $\sim$40 pc with Atacama Large Millimeter/submillimeter Array (ALMA). Our main goals are to measure the properties of GMCs and investigate whether there are differences in the properties with environments. In addition, we compare the properties of GMCs in NGC~1300 with those in a normal spiral galaxy. As pointed out by \citet{Hughes2013ApJ}, for comparison with different data cubes, it is essential to match the spatial and spectral resolution and  line sensitivity to minimize observational bias. In this study, we compare the properties of GMCs in spiral arms of proto-type spiral galaxy of M51 by \citet[\citetalias{Colombo:2014ei}]{Colombo:2014ei} because the spatial resolution and line sensitivity are mostly comparable to those of our observations.

This paper is structured as follows: In Section~\ref{sec:obs}, we describe the CO($1-0$) observations and data reduction. The resultant CO($1-0$) distribution is presented in Section~\ref{sec:GMCdis}. In Section~\ref{sec:GMC identification and characterization}, we summarize the method used to identify GMCs. The basic properties of the GMCs and examination of the variations of GMC physical properties with environments are described in Section~\ref{sec:basicproperties}. In Sections~\ref{sec: Scaling relations} and~\ref{sec: GMC mass spectra}, we examine scaling relations and mass spectra of the GMCs. In Section~\ref{sec:discussion}, we discuss what physical mechanism controls the SFE of the GMC. Our conclusions are presented in Section~\ref{sec:summary}. We discuss the reliability of the measurements of the GMC properties in Appendix~\ref{apx:CPROPS}. The catalog of GMCs we identified is presented in Appendix~\ref{apx:catalog}. Table~\ref{tab:NGC1300} summarizes parameters of NGC~1300 adopted throughout this paper.

\begin{table}
 \caption{Adopted parameters of NGC~1300}
 \label{tab:NGC1300}
 \begin{tabular}{lc}
  \hline
  Parameter & Value \\
  \hline
  Morphology$^*$ & SB(s)bc \\
  Center position (J2000.0)$^\dagger$ & $\rm 03^h19^m41^s.036$  \\
  & $\rm -19^\circ24^\prime40^{\prime\prime}.00$\\
  Inclination$^\S$ & $50^\circ.2$ \\
  Distance$^\ddagger$ & 20.7 Mpc \\
  Linear scale & 100 $\rm pc~arcsec^{-1}$ \\
\hline
\multicolumn{2}{l}{{\small$^*$  \citet{Sandae_Tammann}}} \\
\multicolumn{2}{l}{{\small $^\dagger$  The peak of Fig. ~\ref{fig:NGC1300_Ha_continuum}.}}\\
\multicolumn{2}{l}{{\small$^\S$  \citet{England1989a}}}\\
\multicolumn{2}{l}{{\small$^\ddagger$ We adopted the systemic velocity with corrections for }}\\
\multicolumn{2}{l}{{\small  the Virgo cluster, the Great Attractor, and the Shapley}}\\
\multicolumn{2}{l}{{\small  concentration of $1511~{\rm km~s^{-1}}$ \citep{MouldEtAl00} and }}\\
\multicolumn{2}{l}{{\small  the Hubble constant of $73~{\rm km~s^{-1}~Mpc^{-1}}$. }}
 \end{tabular}
\end{table}

\begin{table*}
 \caption{Observation settings}
 \label{tab:obscondition}
 \begin{tabular}{lcccc}
  \hline
Execution Block &  Start time (UTC) & On source  & Number of & Baseline\\
&    & time (min) & antennas &  range \\
\hline
uid\_\_\_A002\_Xc889b6\_X21a3.ms & 2017/12/30 02:27:10 & 45.55& 46& 15.1m$-$2.5km \\
uid\_\_\_A002\_Xc889b6\_X2691.ms & 2017/12/30 03:36:38 & 45.53& 46& 15.1m$-$2.5km \\
uid\_\_\_A002\_Xc8d560\_X4fa.ms  & 2018/01/06 23:17:13 & 45.55& 44& 15.1m$-$2.5km \\
uid\_\_\_A002\_Xc8d560\_X9e4.ms  & 2018/01/07 00:26:35 & 45.48& 44& 15.1m$-$2.5km \\
uid\_\_\_A002\_Xc8d560\_X9737.ms & 2018/01/07 22:10:49 & 45.50& 43& 15.1m$-$2.5km \\
uid\_\_\_A002\_Xc91189\_X727.ms  & 2018/01/13 01:59:20 & 45.52& 44& 15.1m$-$2.5km \\
uid\_\_\_A002\_Xc92fe3\_X8e21.ms & 2018/01/16 22:21:23 & 45.50& 44& 15.1m$-$1.4km \\
\hline
 \end{tabular}
\end{table*}

\section{Observations and Data Reduction}\label{sec:obs}
We carried out $^{12}{\rm CO}(1-0)$ line observations of NGC~1300 on 2017 December 30 and 2018 January 6, 7, 13, and 16 with ALMA (program ID: 2017.1.00248.S, PI = F. Maeda). To cover the regions observed with $^{12}$CO($1-0$) by \citet{Maeda:2018bg}, two pointings were centered at $({\rm RA, Dec})$ $=$ $(3^{\rm h}19^{\rm m}37^{\rm s}.435, -19^\circ24^\prime33^{\prime\prime}.24)$ and $(3^{\rm h}19^{\rm m}36^{\rm s}.326, -19^\circ24^\prime01^{\prime\prime}.42)$ in J2000 as shown with the green circles in Fig. ~\ref{fig:NGC1300_Ha_continuum}. The ALMA primary beam for the 12-m array is 54$^{\prime\prime}$ at HPBW.

The ALMA observations were taken during five separated periods with seven execution blocks in total. Table~\ref{tab:obscondition} summarizes the details of the ALMA observations. The total on-source time was 5.31 hours (2.65 hours for each position). For all observations, we used about 44 antennas with C43-5 configuration. The projected baseline length ranged from  15.1 m to 2.5 km, which corresponds to a maximum recoverable scale (MRS) of $\sim 21.4^{\prime\prime}$ at 115 GHz. We used the Band 3 receiver with the central frequency of 114.664 GHz, channel width of 244.1 kHz ($\sim 0.64~\rm km~s^{-1}$), and bandwidth of 468.8 MHz ($\sim 1225~\rm km~s^{-1}$). Bandpass and phase were calibrated with J0423-0120 and J0340-2119, respectively. J0423-0120 was also used as a flux calibrator. The $T_{\rm sys}$ and PWV were typically 100$\sim$150 K and  1$\sim$5 mm during the observations, respectively. 

We calibrated raw visibility data using the Common Astronomy Software Applications (CASA) ver. 5.1.1. and the observatory-provided calibration script. We reconstructed the two-field mosaic image using CASA ver. 5.4.0. First, we combined the calibrated visibility data of the 2 pointing positions using \verb|concat| task. Then, we imaged the concatenated visibility using the \verb|multiscale| CLEAN algorithm \citep{Cornwell2008}, which models an image by a collection of multiple Gaussians with different FWHM values. As shown by \citet{Rich2008AJ}, this method works well in reducing the depth of negative emission features (so-called `negative bowl') and recovering extended emission in comparison to the standard CLEAN algorithm, which models an image by a collection of point sources. We used the task \verb|tclean| for imaging by adopting the factors of 1, 2, and 5 times the synthesized beam as multiscale parameter and Briggs weighting with robust = 0.5. 
To ensure consistency with \citetalias{Colombo:2014ei} and to achieve a reasonable S/N, we choose a velocity resolution of $5~\rm km~s^{-1}$ and a pixel size of $0.^{\prime\prime}12$.
The resultant rms noise is 0.51 $\rm mJy~beam^{-1}$, corresponding to 0.36~K. Finally, we applied the primary beam correction on the output restored image. In this study, we used the region within the primary beam correction factor smaller than 2.0. The final data cube has an angular resolution of $0.44^{\prime\prime} \times 0.30^{\prime\prime}$.

\section{GMC scale molecular gas distribution in NGC~1300}\label{sec:GMCdis}
\begin{figure*}
\begin{center}
 \includegraphics[width=180mm]{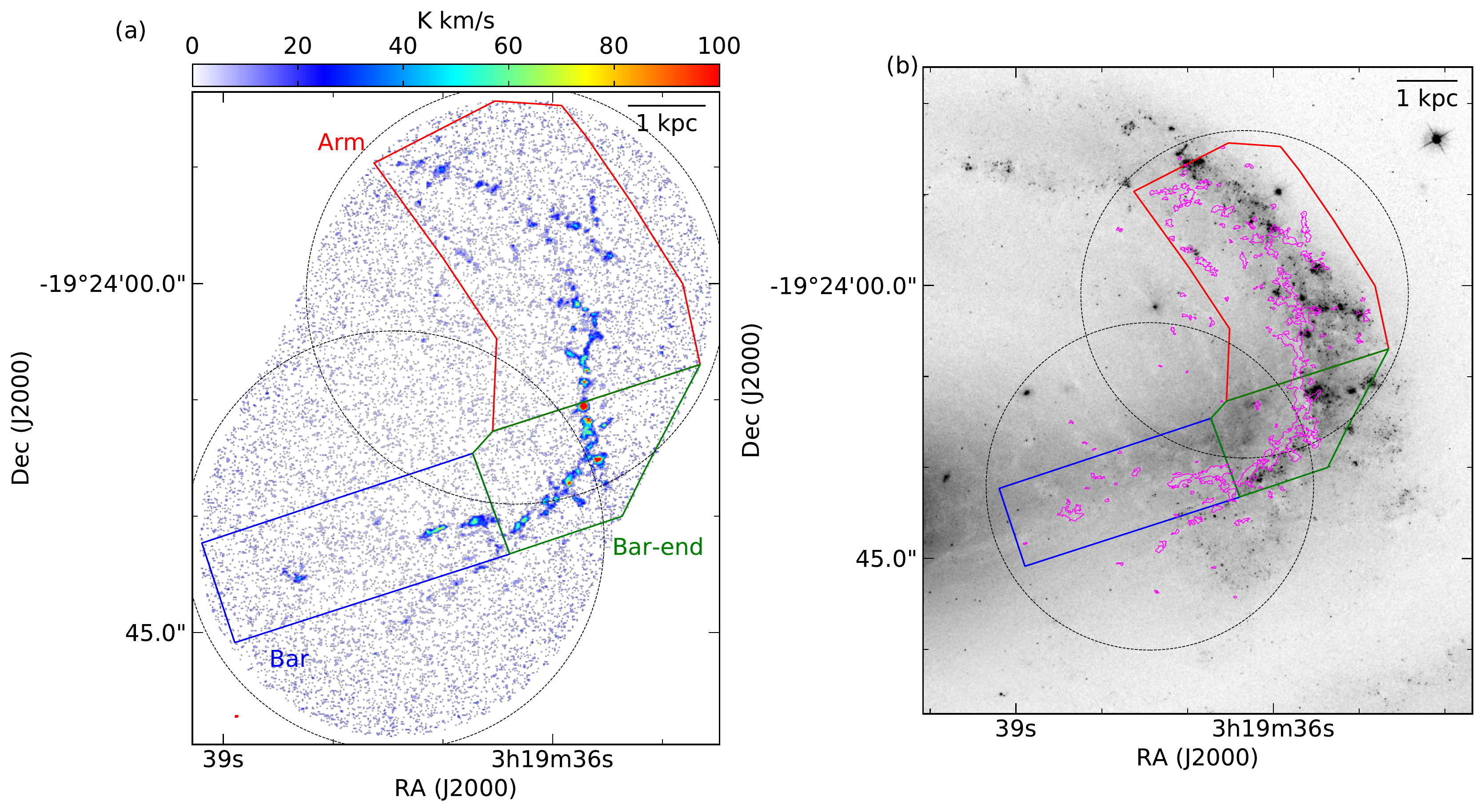}
 \caption{(a) Velocity integrated intensity $^{12}$CO($1-0$) image of NGC~1300 generated from the ALMA data cube. 
We used channel maps within the velocity range $1540-1690~\rm km~s^{-1}$ clipped at 3$\sigma$ (1.53 $\rm mJy~beam^{-1}$).
Black dotted circles ($54^{\prime\prime}\phi$) represent FoV observed with ALMA.
Color solid lines indicate the definition of the environmental mask.
{\it Bar}, {\it Arm}, and {\it Bar-end} are indicated with blue, red, and green lines, respectively.
(b) CO contours in magenta are superimposed on the F555W image obtained from HST archive data. 
CO contours are created from the significant emission identified by CPROPS.
The contour level of the CO map is 8.0 $\rm K~km~s^{-1}$.
}
 \label{fig:NGC1300_CO}
\end{center}
\end{figure*}

Fig. ~\ref{fig:NGC1300_CO}(a) shows a map of the velocity-integrated intensity ($I_{\rm CO}$). This map is created from channel maps within the LSR velocity range $1540-1690~\rm km~s^{-1}$ clipped at 3$\sigma$ ($=~1.53 \rm~mJy~beam^{-1}$). The clipping is only for visualization purpose.
We detected $^{12}$CO($1-0$) emission from the western arm to the bar region. Thanks to the high angular resolution of $\sim 40~\rm pc$, the CO emission is resolved into GMC scale.
Fig. ~\ref{fig:NGC1300_CO}(b) compares the CO distribution with  the map of H\textsc{ii} regions and dust lanes.  The gray scale image is the V-band image (same as Fig. ~\ref{fig:NGC1300_Ha_continuum}). The magenta contours in Fig. ~\ref{fig:NGC1300_CO}(b) show the CO emission with 8.0 $\rm K~km~s^{-1}$. These contours are created from the significant emission identified by CPROPS (so-called `island' regions; see section~\ref{sec:GMC identification}).

In this study, to compare physical properties of GMCs among galactic environments, we separated NGC~1300 into {\it Bar}, {\it Arm} and {\it Bar-end} regions, according to Fig. ~\ref{fig:NGC1300_CO}. Blue rectangle is defined as the {\it Bar} region, which covers the dark lane and associated spurs that are connected almost perpendicularly to the dark lane. The {\it Arm} region is defined as a red polygon. Green polygon that covers the intersection region of the bar and the arm is defined as the {\it Bar-end} region. These color codes will be kept throughout this paper. The deprojected area is 5.1, 9.9, and 3.6 $\rm kpc^2$ in {\it Bar}, {\it Arm} and {\it Bar-end}, respectively. 

Here we see the feature of the CO emissions in each region. In {\it Arm} and {\it Bar-end}, the mean $I_{\rm CO}$ within the region we detected significant emission is 20 and 32~$\rm K~km~s^{-1}$ in  {\it Arm} and {\it Bar-end}, respectively. These values correspond to 55 and 90~$M_\odot~\rm pc^{-2}$ assuming the standard CO-to-H$_2$ conversion factor ($\alpha_{\rm CO}$) of $4.4~ M_\odot~\rm (K~km~s^{-1}~pc^2)^{-1}$. In these regions, most of CO emissions are associated with H$\alpha$ emissions and coexist with dust lanes. This good correspondence of the CO emissions to the dust lanes and displacement of the H$\alpha$ and CO emissions are often seen in spiral galaxies at a high angular resolution of $\sim 50$ pc  \citep[e.g., M51;][]{Schinnerer:2013jy}. This is naturally interpreted with respect to the spiral density wave; the gases go into the spiral density wave, then dense shocked gas regions form in the spiral arms which result in the formation of GMCs. In the GMCs, stars form, and massive stars born produce H\textsc{ii} regions. The stellar feedback (e.g., stellar wind, supernova) would disperse the molecular gas, thus contributing to the observed offset.

In {\it Bar}, we detected GMC-like CO emissions on the dust lanes. The mean $I_{\rm CO}$ is 22 $\rm K~km~s^{-1}$, corresponding to 60 $M_\odot~\rm pc^{-2}$. Unlike {\it Arm}, however, no prominent H\textsc{ii} region is associated with the CO emissions. This result indicates that massive star formation in the bar region of the strongly barred galaxy is suppressed despite the presence of GMC-like molecular gases. Note that we did not detect significant CO emission above 4$\sigma$ of $T_{\rm peak} = 1.44~\rm K$, corresponding to an average surface density of 20 $M_\odot~\rm pc^{-2}$ for a single pixel and spectral channel, in part of the dust lane in the bar region.

\section{GMC identification and characterization}\label{sec:GMC identification and characterization}
\subsection{GMCs Identification}\label{sec:GMC identification}
The choice of an algorithm for identifying GMCs can be the largest source of uncertainty in measuring and comparing GMC properties. In this study, we used 3-D clumps finding algorithm  CPROPS \citep{RosolowskyLeroy} which is designed to identify GMCs well even at low sensitivities. The spatial resolution and line sensitivity of our data cube are comparable to those of the data cube of M51 by \citet{Schinnerer:2013jy} which was  observed with the Plateau de Bure Interferometer and IRAM 30-m telescope at a spatial resolution of $\sim 40$ pc and a line sensitivity of $0.4$ K at 5 $\rm km~s^{-1}$ bin. Using CPROPS, \citetalias{Colombo:2014ei} identified GMCs in M51 from the data cube. Therefore, we adopt almost the same CPROPS parameters adopted in \citetalias{Colombo:2014ei} to make a proper comparison and minimize systematics.

Here, we provide a brief summary of CPROPS. CPROPS has been fully described by \citet{RosolowskyLeroy}. This algorithm begins with the identification of regions with significant emissions within the data cube (called the `islands' in CPROPS). CPROPS identifies pixels in which the signal is above $t \sigma_{\rm rms}$ in at least two adjacent velocity channels, where $\sigma_{\rm rms}$ is the rms noise of the data cube. Since the data cube is corrected for the primary beam pattern, the noise in the map is non-uniform. Thus, we calculated a spatially varying of the rms noise in the map by calling \verb|NONUNIFORM| flag in CPROPS. CPROPS then grows these regions to include adjacent pixels
in which the signal is above $e \sigma_{\rm rms}$. The $t$ and $e$ are the \verb|THRESHOLD| and \verb|EDGE| parameters in CPROPS, respectively. We chose $t = 4.0$ and $e = 1.5$ to ensure consistency with  \citetalias{Colombo:2014ei}.

After the identification of the islands, the islands are divided into individual GMCs using a modified watershed algorithm. CPROPS searches for local maxima within a box of three times the beam and channel width, corresponding to  $\rm \sim 120~pc \times 120~pc \times 15~km~s^{-1}$. All local maxima are required to lie at least 2$\sigma_{\rm rms}$ above the merge level with another maximum. Only emission that is uniquely associated with each local maximum is given an assignment (i.e., only that emission which is above all merge levels with other local maxima). 

For every pair of local maxima in an island, CPROPS compares the values of the moments at contour levels just above and just below the merge level. If the moments jump by more than a user-defined fraction, called \verb|SIGDISCONT| parameter, the two local maxima are categorized as distinct, otherwise they are merged into a single cloud. The default parameter of \verb|SIGDISCONT| $= 1.0$ requires 100\% or more fractional change. As described in \citetalias{Colombo:2014ei}, however, setting of 
\verb|SIGDISCONT| $> 0.0$ does not work well for the CO bright regions where a lot of local maxima are connected at a very low contour level; CPROPS misses objects that visual inspection suggests are GMCs (see Appendix B in \citetalias{Colombo:2014ei}).
In our data cube, some GMCs are missed with \verb|SIGDISCONT| $> 0.0$ in Bar-End where CO emission is the brightest. To avoid this failure, we set \verb|SIGDISCONT| 
to 0.0 which makes no attempt to merge the two local maxima into a single cloud. In other words, each local maximum is assigned to an individual independent cloud.
The remainder of the emission is considered to be in the watershed, and CPROPS does not assign it to any cloud.

A large number of objects were found whose peak temperature is above 4$\sigma_{\rm rms}$ but temperature of the adjacent pixels in the velocity axis is lower than 4$\sigma_{\rm rms}$. These objects seem to be GMCs because their size, velocity dispersion, and luminosity are similar to those in M51 spiral arms. However, they are missed by CPROPS because CPROPS does not identify objects which do not contain two consecutive pixels above 4$\sigma_{\rm rms}$ in  the velocity axis. Therefore, to identify such GMCs for which signal above 4$\sigma_{\rm rms}$ is detected in only a single channel,  we allowed CPROPS to identify an island which has pixels above $4~\sigma_{\rm rms}$ in only one channel. As we will discuss in Section~\ref{sec:basicproperties},  the possibility that these GMCs are spurious is very low.

\subsection{Derivation of basic properties}
\subsubsection{Size, velocity width and CO luminosity}\label{sec:derivation of R}

CPROPS determines the size, velocity width and flux of GMCs using moment methods. CPROPS corrects for the sensitivity by extrapolating GMC properties to those we would expect to measure with perfect sensitivity (i.e., 0 K) and the resolution by deconvolution for the beam and channel width. Details of the measurement method of CPROPS are described in \citet{RosolowskyLeroy}.

The effective radius of the GMC is defined as $R = (3.4/\sqrt{\pi})\sigma_{r}$, where $\sigma_r$ is geometric mean of the second intensity-weighted moments along the major and minor axis (i.e., rms size) and $3.4/\sqrt{\pi}$ is an empirical factor defined by \citet{Solomon87}. If the extrapolated rms size is smaller than the beam size of $15.4$ pc, we set this value as an upper limit of the rms size. The velocity width, $\sigma_v$, is computed as the intensity-weighted second moment of the velocity after the deconvolution to correct the velocity resolution ($\Delta V_{\rm chan}$). CPROPS approximates the shape of the channel as a Gaussian with an integral area equal to that of the channel width (i.e., $\Delta V_{\rm chan}/\sqrt{2 \pi}$).
The CO luminosity, $L_{\rm CO}$, is derived from
\begin{eqnarray}
 \left( \frac{L_{\rm CO}}{\rm K~km~s^{-1}~pc^2} \right) = 
\left( \frac{F_{\rm CO}}{\rm K~km~s^{-1}~arcsec^{2}}\right) \left( \frac{D}{\rm pc} \right)^2
    \left( \frac{\pi}{180 \times 3600} \right)^2,
\end{eqnarray}
where $F_{\rm CO}$ is the zeroth moment of the intensity and the $D$ is the distance to NGC~1300 of 20.7~Mpc.

The median ratio of the extrapolated (i.e., without deconvolution) radius and velocity dispersion to the directly measured (i.e., without deconvolution and extrapolation) are 1.5 and 1.4, respectively. For the CO luminosity, the median ratio of the extrapolated  value to the directly measured value is 2.1. These correction factors are comparable to those of the GMCs in M51 \citepalias{Colombo:2014ei}.

\subsubsection{Derived Properties}
The molecular gas mass of a GMC, $M_{\rm mol}$,
is  converted from the CO luminosity by adapting an $\alpha_{\rm CO}$:
\begin{equation}
\label{eq:molecular gas mass}
    \left( \frac{M_{\rm mol}}{M_\odot} \right) = \alpha_{\rm CO} 
    \left( \frac{L_{\rm CO}}{\rm K~km~s^{-1}~pc^2} \right).
\end{equation}
We adopt the standard $\alpha_{\rm CO}$ of $4.4~M_\odot~(\rm K~km~s^{-1}~pc^2)^{-1}$ to ensure consistency with \citetalias{Colombo:2014ei}. Note that this $\alpha_{\rm CO}$ includes a factor of 1.36 to account for the presence of helium.

Assuming a spherical GMC with the power-law density profile of $\rho \propto r^{-n}$, we calculate the virial mass of a GMC, $M_{\rm vir}$, from the size and velocity dispersion as 
\begin{eqnarray}
    M_{\rm vir} &=& \frac{3(5-2n)}{3-n}\frac{R \sigma_v^2}{G} \nonumber \\
    &=& 1040 \left( \frac{R}{\rm pc} \right) \left( \frac{\sigma_v}{\rm km~s^{-1}} \right)^2 [M_\odot].
\end{eqnarray}
We adopt the profile of $n = 1$  to ensure consistency with previous studies. We assume that magnetic fields and external pressure are negligible.

The average molecular gas surface density, $\Sigma_{\rm mol}$, is defined as 
\begin{equation}
    \Sigma_{\rm mol} = \frac{M_{\rm mol}}{\pi R^2}.
\end{equation}

The virial parameter, $\alpha_{\rm vir}$, is a useful measure of the gravitational binding  and is defined as 
\begin{equation}
    \alpha_{\rm vir} = \frac{5 \sigma_{v}^2 R}{G M_{\rm mol}}
\end{equation}
 by \citet{BertoldiandMckee}. This definition assumes $n = 0$, but the impact of the difference between $n = 0$ and $1$ is only about 10\%.
GMCs with $\alpha_{\rm vir} \lesssim 2$ are typically considered to be bound. A value of $\alpha_{\rm vir} > 2$ indicates that the GMC is gravitationally unbound.

The scaling coefficient, $c$,  characterizes the scaling between the size and the velocity dispersion of a GMC as
\begin{equation}
    c \equiv \frac{\sigma_v}{R^{0.5}}.
\end{equation}
For a GMC with $a_{\rm vir} = 1$, $c$ can be related to the molecular gas surface density as $c = (\pi G \Sigma_{\rm mol}/5)^{0.5}$.

CPROPS estimates the uncertainty of these GMC properties using a bootstrapping method, and the final uncertainty is the standard deviation of the bootstrapped values scaled up by the square root of the oversampling rate. The oversampling rate is equal to the number of pixels per beam, which accounts for the nonindependence of the pixels. \citetalias{Colombo:2014ei} found that 50 bootstrapping measurements provided a reliable estimation of the uncertainty, thus we adopted this number.

\begin{figure}
\begin{center}
 \includegraphics[width=85mm]{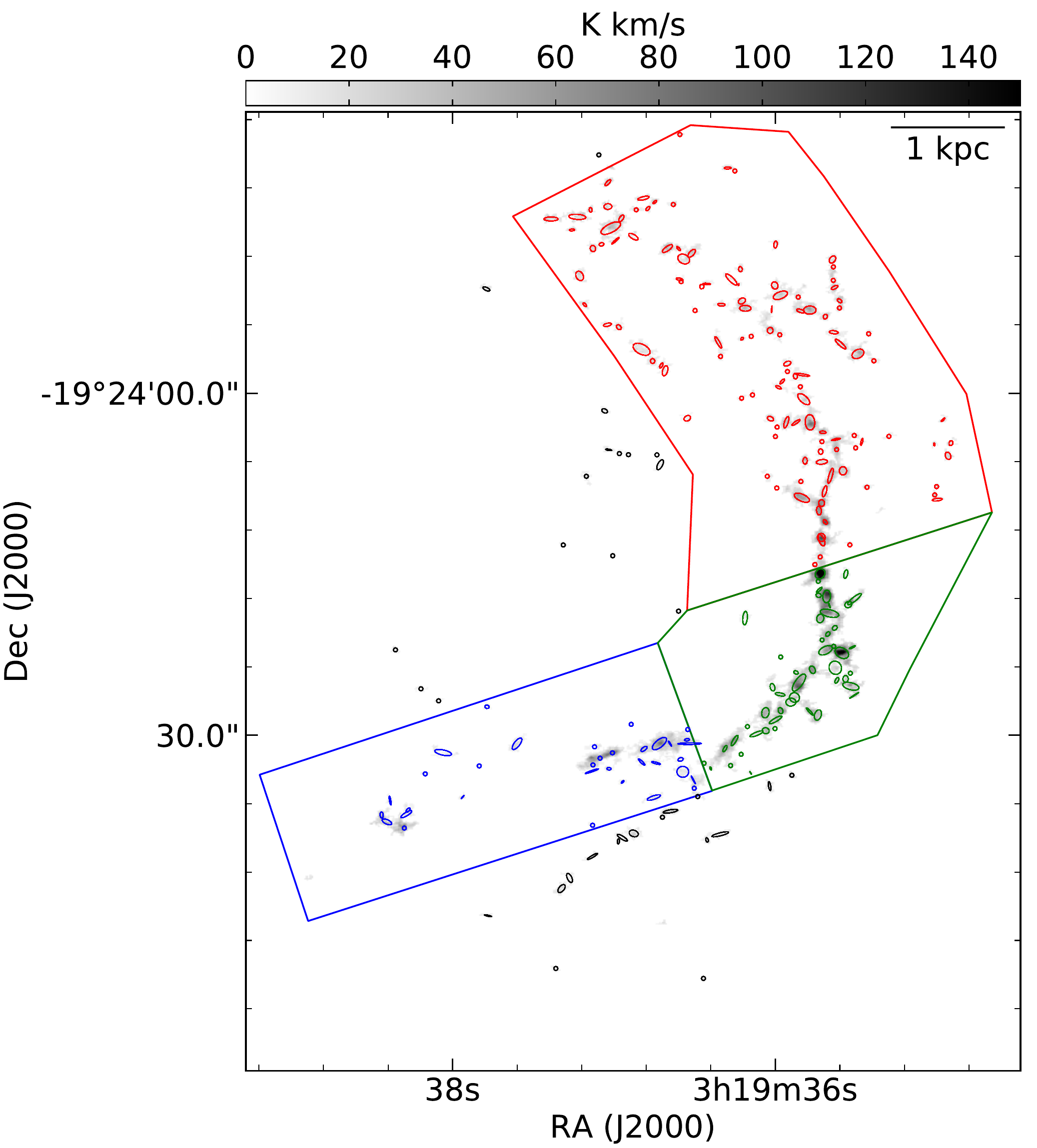}
 \caption{GMC distribution in NGC~1300 superimposed on the velocity integrated CO map (gray scale). The GMCs are represented as ellipses with the extrapolated and deconvolved major and minor axes, oriented according to the measured position angle. Colors indicate the environment defined in Fig. ~\ref{fig:NGC1300_CO}. }
 \label{fig:NGC1300_GMC_coord}
\end{center}
\end{figure}

\begin{figure}
\begin{center}
 \includegraphics[width=70mm]{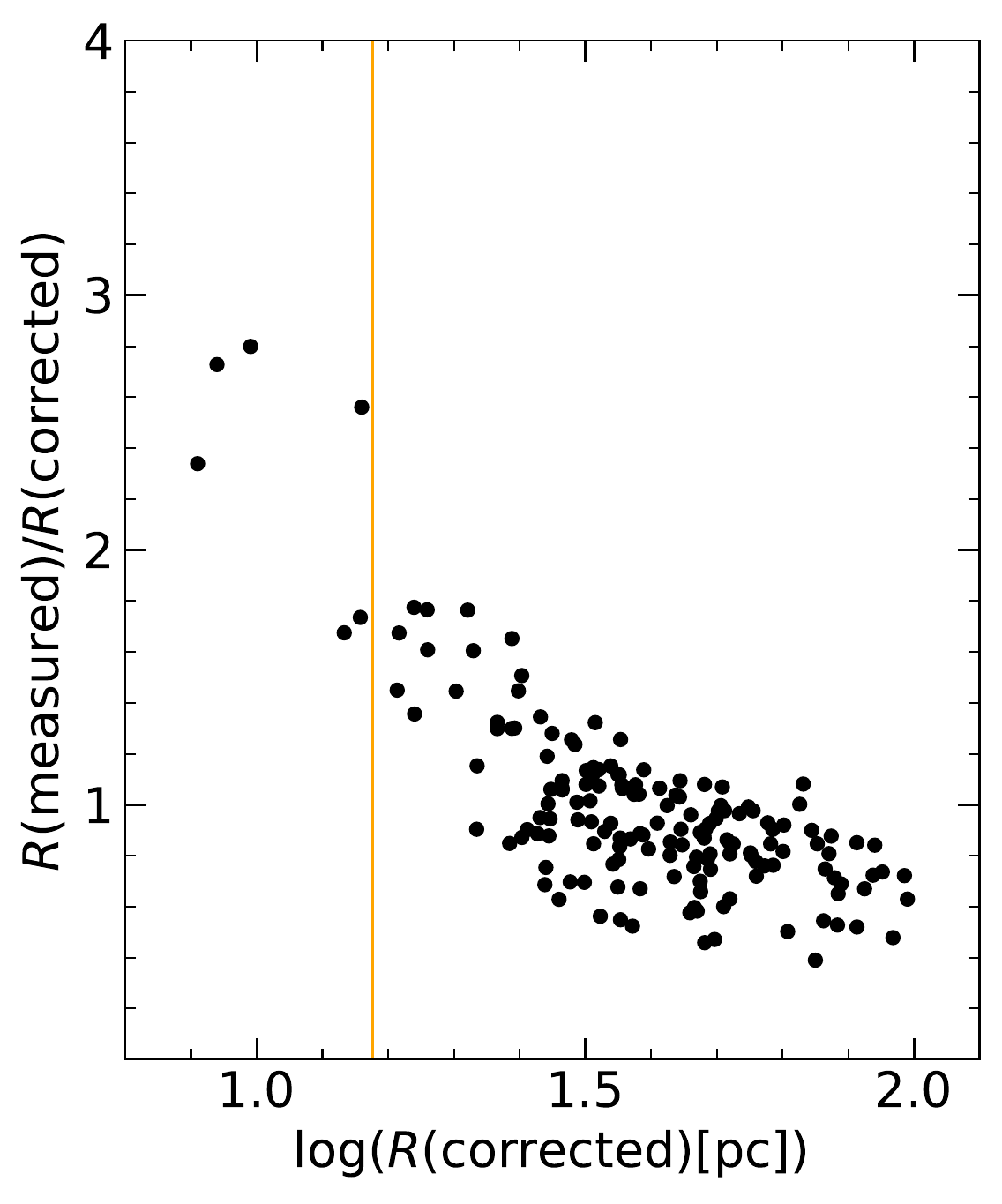}
 \caption{Ratio of measured radius to corrected (i.e., with deconvolution and extrapolation) radius for all resolved GMCs against corrected radius. Vertical orange line indicates 15 pc.}
 \label{fig:correction_factor}
\end{center}
\end{figure}

\section{Basic properties of GMCs}\label{sec:basicproperties}
CPROPS identified 233 GMCs in the data cube. 166 of the GMCs have at least two consecutive pixels above 4$\sigma_{\rm rms}$ in the velocity axis direction. The remaining 67 GMCs have pixels above 4$\sigma_{\rm rms}$ in only one channel. Fig. ~\ref{fig:NGC1300_GMC_coord} shows GMC distribution in NGC~1300. We detected 34, 119, and 49 GMCs in {\it Bar}, {\it Arm}, and {\it Bar-end}, respectively.
31 GMCs are outside the three regions (i.e., in interarm regions).
The GMC catalog in NGC~1300 is presented in Appendix~\ref{apx:catalog}.

We inspected the CO line profiles from each GMC CPROPS identified visually, and an obvious spurious signal was not found. To check contamination by spurious sources, we counted the number of local maxima in the data cube scaled by $-1$ with the same settings described in Section~\ref{sec:GMC identification}. We found no false positives which have two consecutive pixels  above 4$\sigma_{\rm rms}$ in the velocity axis direction. For the 166 GMCs, therefore, we believe there is no contamination by spurious sources. On the other hand, CPROPS identified three false positives which have pixels above $4~\sigma_{\rm rms}$ in only one channel. Thus, $\sim$ 3 of the 67 ($\sim 4~\%$) GMCs for which signal above 4$\sigma_{\rm rms}$ is detected in only a single channel are expected to be spurious.

With the adopted parameters for CPROPS, GMCs with  $T_{\rm peak} > 4 \sigma_{\rm rms}$, which corresponds to $\sim 1.44~\rm K$, were detected. The mass completeness limit is estimated to be $2.0 \times 10^5 ~M_\odot$ by assuming that a minimum detectable GMC has a size and a velocity width comparable to the beam size and the channel width, respectively, and $\alpha_{\rm CO}$ is $4.4~M_\odot~\rm (K~km~s^{-1}~pc^2)^{-1}$. We find that 154 of these 233 GMCs are resolved by the synthesized beam. When the observed size is similar to the beam size,  uncertainties of $R$ are greatly magnified by the deconvolution process \citep[e.g.,][]{Faesi:2018gg}. Fig. ~\ref{fig:correction_factor} shows the ratio of directly measured radius to the corrected radius (i.e., with extrapolation and deconvolution) as a function of the corrected radius. Below 15 pc ($\log R = 1.17$), the ratio is often larger than a factor of 2, or even higher.  Further, we find only 47~\% of the GMCs with $M_{\rm mol} < 5.0 \times 10^5~M_\odot$  are resolved, while  95~\% of the  GMCs with $M_{\rm mol} \geq 5.0 \times 10^5~M_\odot$  are resolved. Therefore,  we defined the GMCs with $M_{\rm mol} \geq 5.0 \times 10^5~M_\odot$ and $R > 15$~pc as a resolved sample  for the investigation of $R$, $M_{\rm vir}$, $\Sigma_{\rm mol}$, $\alpha_{\rm vir}$, and $c$, and defined the GMCs with  $M_{\rm mol} > 2.0 \times 10^5~ M_\odot$  as a mass completed sample for the investigation of $T_{\rm peak}$, $\sigma_v$, and $M_{\rm mol}$.

\begin{figure*}
	\includegraphics[width=\hsize]{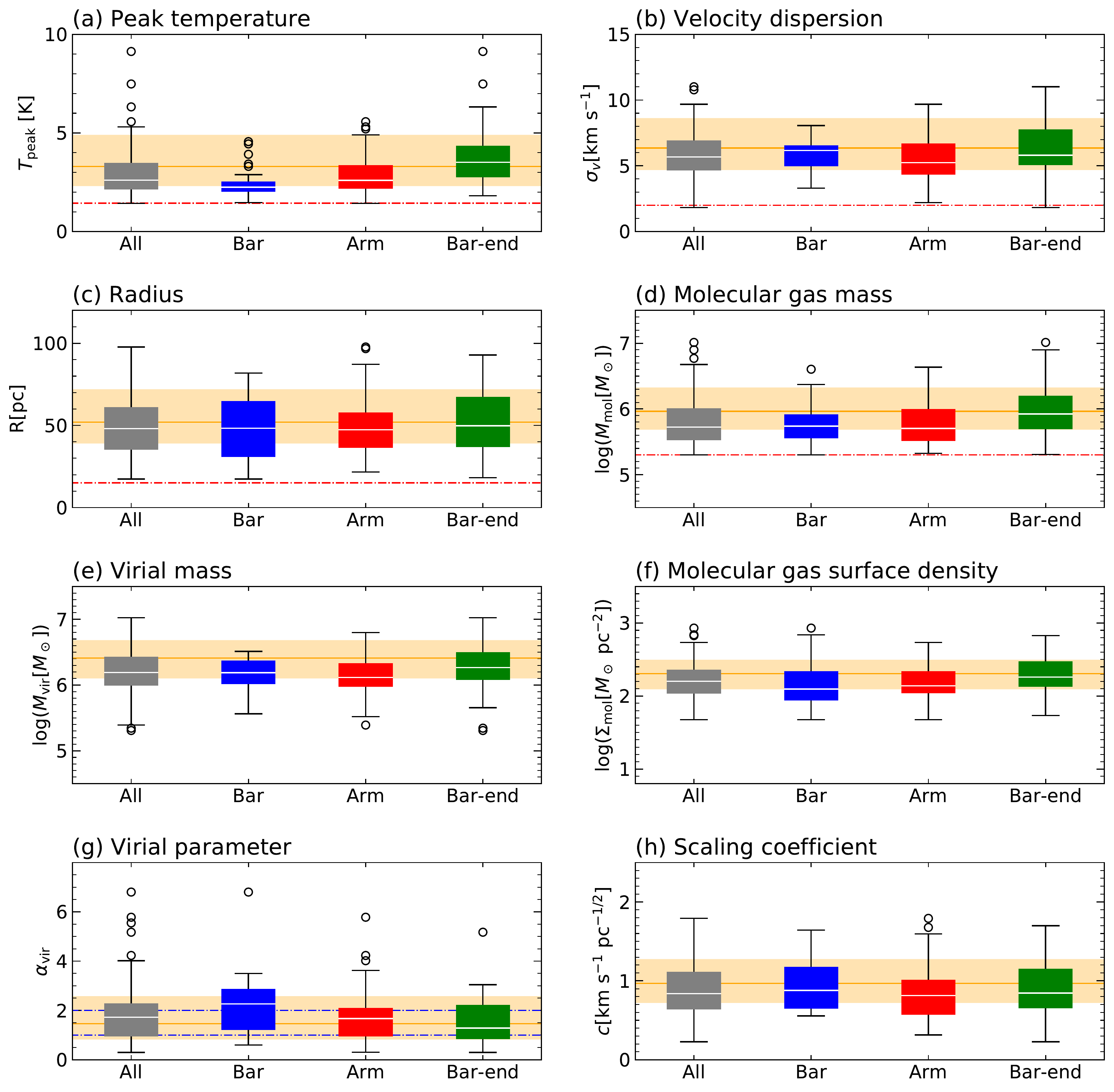}
    \caption{Box-and-whisker plots for GMC's 
 (a) peak temperature ($T_{\rm peak}$), (b) velocity dispersion ($\sigma_v$), (c) radius ($R$), (d) molecular gas mass ($M_{\rm mol}$),  (e) virial mass ($M_{\rm vir}$), (f) molecular gas surface density ($\Sigma_{\rm mol}$), (g) virial parameter ($\alpha_{\rm vir}$), and (h) scaling coefficient ($c$) in the whole region (gray) and each environment ({\it Bar}: blue, {\it Arm}: red, and {\it Bar-end}: green) of NGC~1300. We used the GMCs with $M_{\rm mol} \geq 2.0 \times 10^5~M_\odot$ for $T_{\rm peak}$, $\sigma_{\rm v}$, and $M_{\rm mol}$  and the GMCs with $M_{\rm mol} \geq 5.0 \times 10^5~M_\odot$ and $R \geq 15$ pc for other properties.  The median value is represented as a white horizontal line  within each box. Upper and lower edges of each box indicate the  upper and lower quartile, respectively ($Q_1$ and $Q_3$, respectively).  The upper whisker expands to the largest data point below  $Q_3+1.5\Delta Q$ and the lower whisker expands to the smallest data point above $Q_1-1.5 \Delta Q$. The data points outside the whiskers are considered outliers and shown as open circles. The orange solid line and bands show the median value and the range from $Q_1$ to $Q_3$ of each property of GMCs in M51 spiral arms \citepalias{Colombo:2014ei}. Horizontal red dash-dotted lines indicate 1.44 K (= 4 $\sigma$), 2.0 $\rm km~s^{-1}$ ($= \Delta V_{\rm chan}/\sqrt{2 \pi}$), 15 pc, and  $2.0 \times 10^5~M_\odot$.  Horizontal blue dash-dotted lines in panel (g) indicate $\alpha_{\rm vir} = 1.0$ and $2.0$.}
 \label{fig:boxplot}
\end{figure*}

\begin{table*}
 \caption{GMC properties in the different environments of NGC~1300}
 \label{tab:GMCproperties}
 \begin{tabular}{lcccccccccc}
  \hline
 Envir. &Area & \# $^*$ & $T_{\rm peak}$ $^\dagger$  & $\sigma_{\rm v}$ $^\dagger$ & $R$ $^\ddagger$ & $M_{\rm mol}$ $^\dagger$ & $M_{\rm vir}$ $^\ddagger$  & $\Sigma_{\rm mol}$ $^\ddagger$  & $\alpha_{\rm vir}$ $^\ddagger$  & $c$ $^\ddagger$  \\
   & ($\rm kpc^2$) &   & (K) & ($\rm km~s^{-1}$) & (pc) & ($10^5~M_\odot$) &  ($10^5~M_\odot$) &  ($M_\odot \rm ~pc^{-2}$) &  & ($\rm km~s^{-1}~pc^{-0.5}$) \\
\hline
All              & 37.9 & 203(104) & $2.6^{+0.9}_{-0.4}$ & $5.2^{+1.3}_{-0.8}$ & $48.0^{+12.6}_{-12.4}$ & $5.3^{+4.7}_{-1.9}$ & $15.5^{+10.7}_{-5.5}$ & $159.2^{+65.9}_{-49.7}$  & $1.7^{+0.5}_{-0.8}$  & $0.8^{+0.3}_{-0.2}$  \\
 \hline
Bar              & 5.1  & 28(12)   & $2.2^{+0.2}_{-0.2}$ & $5.7^{+0.8}_{-0.9}$ & $48.3^{+16.2}_{-17.0}$ & $5.5^{+2.7}_{-1.8}$ & $15.4^{+7.8}_{-4.7}$  & $125.5^{+87.9}_{-36.8}$  & $2.3^{+0.6}_{-1.0}$  & $0.9^{+0.3}_{-0.2}$ \\
Arm              & 9.9  & 108(55)  & $2.6^{+0.7}_{-0.4}$ & $5.0^{+1.5}_{-0.8}$ & $47.4^{+10.1}_{-10.7}$ & $5.1^{+4.6}_{-1.7}$ & $13.1^{+8.0}_{-3.5}$  & $137.5^{+77.3}_{-25.5}$  & $1.7^{+0.4}_{-0.7}$  & $0.8^{+0.2}_{-0.2}$ \\
Bar-end          & 3.6  & 43(31)   & $3.5^{+0.8}_{-0.7}$ & $5.6^{+2.0}_{-0.9}$ & $49.7^{+17.3}_{-12.6}$ & $8.4^{+7.1}_{-3.4}$ & $18.4^{+12.6}_{-6.2}$ & $181.4^{+111.3}_{-45.0}$ & $1.3^{+0.9}_{-0.4}$  & $0.8^{+0.3}_{-0.2}$ \\
\hline
\hline
M51 SA$^{\star}$ & 14.6 & 584(365) & $3.3^{+1.6}_{-1.0}$ & $6.3^{+2.3}_{-1.7}$ & $52.0^{+20.0}_{-13.0}$ & $9.2^{+12.0}_{-4.4}$& $25.7^{+22.6}_{-13.1}$ &$202.0^{+111.7}_{-78.5}$ &$1.5^{+1.1}_{-0.6}$  &$1.0^{+0.3}_{-0.2}$ \\
\hline
\multicolumn{11}{l}{$^*${\small The number of the GMCs with $M_{\rm mol} \geq 2.0 \times 10^5~M_\odot$. In parentheses, we show the number of the  GMCs}} \\
\multicolumn{11}{l}{{\small with $M_{\rm mol} \geq 5.0 \times 10^5~M_\odot$ and $R \geq 15$ pc.}} \\
\multicolumn{11}{l}{$^\dagger$ {\small For the GMCs with $M_{\rm mol} \geq 2.0 \times 10^5~M_\odot$.}}\\
\multicolumn{11}{l}{$^\ddagger$ {\small For the GMCs with $M_{\rm mol} \geq 5.0 \times 10^5~M_\odot$ and $R \geq 15$ pc.}}\\
\multicolumn{11}{l}{$^\star$ {\small For M51 spiral arms (SA) \citepalias{Colombo:2014ei}. We extract GMCs with the same selection criteria as described in Section~\ref{sec:basicproperties}.}}
 \end{tabular}
\end{table*}

\begin{table*}
 \caption{$p$-values of the Kolmogorov-Smirnov Test for GMC properties}
 \label{tab:KStest}
 \begin{tabular}{lccc}
 \hline
  Prop. & Bar vs. Arm & Bar vs. Bar-end & Arm vs. Bar-end 
 \\
\hline
 $T_{\rm peak}$ $^*$    &$0.019 \pm 0.013 (0.008)$ &$\ll 0.01 (\ll 0.01)$ &$\ll 0.01 (\ll 0.01)$  \\
 $\sigma_{\rm v}$ $^*$ &$0.350 \pm 0.186 (0.281)$ &$0.756 \pm 0.165 (0.667)$ &$0.181 \pm 0.083 (0.142)$  \\
 $R$ $^\dagger$                &$0.672 \pm 0.167 (0.746)$ &$0.726 \pm 0.167 (0.896)$ &$0.479 \pm 0.153 (0.552)$  \\
 $M_{\rm mol}$ $^*$           &$0.876 \pm 0.074 (0.954)$ &$0.027 \pm 0.017 (0.037)$ &$0.007 \pm 0.004 (0.004)$  \\
 $M_{\rm vir}$ $^\dagger$      &$0.794 \pm 0.148 (0.613)$ &$0.545 \pm 0.167 (0.698)$ &$0.133 \pm 0.067 (0.117)$  \\
 $\Sigma_{\rm mol}$ $^\dagger$ &$0.588 \pm 0.183 (0.355)$ &$0.206 \pm 0.102 (0.221)$ &$0.123 \pm 0.068 (0.128)$  \\
 $\alpha_{\rm vir}$ $^\dagger$ &$0.183 \pm 0.114 (0.050) $ &$0.206 \pm 0.106 (0.172)$ &$0.813 \pm 0.135 (0.938)$  \\
 $c$ $^\dagger$                &$0.655 \pm 0.172 (0.778)$ &$0.886 \pm 0.092 (0.978)$ &$0.430 \pm 0.130 (0.592)$  \\
\hline
\multicolumn{4}{l}{{\small In parentheses, we show the $p$-value for observed values.}} \\
\multicolumn{4}{l}{$^*${\small For the GMCs with $M_{\rm mol} \geq 2.0 \times 10^5~M_\odot$.}} \\
\multicolumn{4}{l}{$^\dagger${\small For the GMCs with $M_{\rm mol} \geq 5.0 \times 10^5~M_\odot$ and $R \geq 15$ pc.}} 
 \end{tabular}
\end{table*}

\subsection{Variation of GMC physical properties with environment} \label{sec:Basic properties}
Fig. ~\ref{fig:boxplot} shows the distribution of GMC properties. We use a box and whiskers plot  as \citetalias{Colombo:2014ei} used for M51. In this plot, the ends of the box represent the upper and lower quartiles ($Q_1$ and $Q_3$, respectively). The median value or $Q_2$ is marked by a vertical line in the box. The upper whisker expands to the largest data point below $Q_3 + 1.5 \Delta Q$ and the lower whisker expands to the smallest data point above $Q_1 - 1.5 \Delta Q$, where $\Delta Q$ is the box length or $Q_3 - Q_1$. For a Gaussian distribution, the region within the whiskers contains $99.3~\%$ ($\pm 2.7 \sigma$) of the population. The data points outside the whiskers are considered to be outliers. Except for $T_{\rm peak}$, $\sigma_v$, and $M_{\rm mol}$, we show a box and whiskers plot for the resolved sample. The orange lines and bands in Fig. ~\ref{fig:boxplot} show the median value and the range from $Q_1$ to $Q_3$ of those in M51 spiral arms from the catalog of \citetalias{Colombo:2014ei}. The median value and the range from $Q_1$ to $Q_3$ of each physical property are listed in Table~\ref{tab:GMCproperties}.

We investigated the environmental variation of GMC properties in NGC~1300 and compared to the GMC properties in M51 spiral arms. We used the two-sided Kolmogorov-Smirnov (K-S) test to statistically evaluate differences in GMC property distributions for different environments in NGC~1300. We used \verb|stats.ks_2samp| function of python's Scipy package, which calculates $p$-value based on the approximately formula given by \citet{Stephens:1970ic}.  To estimate the uncertainties of the $p$-values, we made resampling with 100 realizations. In one realization, random values of a given property were generated within  the bootstrap uncertainties CPROPS calculated. The most likely $p$-value and its uncertainty are the median and the median absolute deviation (MAD) of 100 realizations, respectively. Table~\ref{tab:KStest} shows the results for each GMC physical parameter. Since CPROPS does not provide uncertainties of the $T_{\rm peak}$, we applied rms noise of the data cube, 0.36 K as the uncertainty. We also show the $p$-value for the observed value in parentheses. Following the conventional criteria, the two cumulative distribution functions are considered to be significantly different  if the $p$-value is less than 0.01. A $p$-value within 0.01 to 0.05 indicates  that the difference is marginally significant. Note that the approximation of $p$-value is quite good for $N_1N_2/(N_1+N_2) \geq 4$, where $N_1$ and $N_2$  are the number of data points in the first and second distribution, respectively \citep{Stephens:1970ic}. Our sample sufficiently meets this criterion, $N_1N_2/(N_1+N_2) \geq 8$.

{\it Peak temperature }(Fig. ~\ref{fig:boxplot} (a)): The $T_{\rm peak}$ in {\it Bar-end} is the highest ($2.8-4.3$ K), followed by {\it Arm} ($2.2-3.4$ K) and {\it Bar} ($2.0-2.4$ K). The K-S test indicates that the differences between {\it Bar-end}  and {\it Arm} or {\it Bar} are significant, and that between {\it Bar} and {\it Arm} is marginally significant. Thus we conclude that the $T_{\rm peak}$ in {\it Bar-end} is significantly higher than those in {\it Arm} and {\it Bar}. This result does not change for the resolved sample. The $\Delta Q$ of $T_{\rm peak}$ distribution in {\it Arm} and {\it Bar-end} is within that in M51 spiral arm, but that in {\it Bar} is lower.

{\it Velocity dispersion }(Fig. ~\ref{fig:boxplot} (b)): We find no environmental variations based on the K-S test: $4.8 - 6.5~\rm km~s^{-1}$ in {\it Bar}, $4.2 - 6.5~\rm km~s^{-1}$ in {\it Arm}, and $4.7 - 7.6~\rm km~s^{-1}$ in {\it Bar-end}. In {\it Bar-end}, some GMCs have a $\sigma_v$ larger than $10~\rm km~s^{-1}$. The $\Delta Q$ of $\sigma_v$ distribution in each region is  roughly within those in M51 spiral arm. One caveat of measurement of $\sigma_v$ is that we may overestimate $\sigma_v$ of GMCs. We discuss this bias in section~\ref{reliability} and appendix~\ref{apx:CPROPS}.

{\it Radius }(Fig. ~\ref{fig:boxplot} (c)): There are no environmental variations of $R$ ($31.3 - 64.5~\rm pc$ in {\it Bar}, $36.7 - 57.5~\rm pc$ in {\it Arm}, and $37.1 - 67.0~\rm pc$ in {\it Bar-end}) and, the $\Delta Q$ of $R$ distribution in each region is roughly within those in M51 spiral arms. Note that we evaluate differences in radius without deconvolution for all sample using the K-S test, and we find no environmental variations.

{\it Molecular gas mass }(Fig. ~\ref{fig:boxplot} (d)): Molecular gas mass is  proportional to a combination of  brightness temperature, velocity dispersion, and size: $M_{\rm mol} \propto L_{\rm CO} \propto \braket{T}R^2\sigma_v$, where $\braket{T}$ is an average brightness temperature. Since the $R$ and $\sigma_v$ are roughly equal in each region, $M_{\rm mol}$ is mainly proportional to $\braket{T}$ in NGC~1300. Therefore, $M_{\rm mol}$ in {\it Bar-end} is expected to be larger than those in {\it Arm} and {\it Bar} like the $T_{\rm peak}$. In fact,  the $M_{\rm mol}$ in {\it Bar-end} is higher ($ (5.0-15.5) \times 10^5~M_\odot$) than those in {\it Bar} ($ (3.7-8.2) \times 10^5~M_\odot$) and {\it Arm} ($ (3.4-9.7) \times 10^5~M_\odot$). The median value in {\it Bar-end} is by a factor of 1.6 larger than those in {\it Bar} and {\it Arm}. The K-S test indicates that  the difference between {\it Bar-end} and {\it Arm} ({\it Bar}) is significant (marginally significant), and there is no significant difference between {\it Bar} and {\it Arm}. The median $M_{\rm mol}$ in {\it Bar-end} is comparable to that  in M51 spiral arms, but those in {\it Arm} and {\it Bar} are about 2 times lower.

{\it Virial mass }(Fig. ~\ref{fig:boxplot} (e)): Since the $R$ and $\sigma_v$ are roughly equal in each region, $M_{\rm vir} \propto R \sigma_v^2$ is expected to be equal. In fact, we find no significant environmental variation: $(1.1 \times - 2.3) \times 10^6~M_\odot$ in {\it Bar}, $(1.0 \times - 2.1) \times 10^6~M_\odot$ in {\it Arm}, and $(1.2 \times - 3.1) \times 10^6~M_\odot$ in {\it Bar-end}. 

{\it Molecular gas surface density }(Fig. ~\ref{fig:boxplot} (f)): Because  $S_{\rm mol} $ is proportional to  $\braket{T} \sigma_v$, $S_{\rm mol}$ is expected to show the same tendency as $T_{\rm peak}$. In fact, $S_{\rm mol}$ in {\it Bar-end}  is the highest ($138.5 - 317.6~M_\odot~\rm pc^{-2}$), followed by {\it Arm} ($112.0-214.8~M_\odot~\rm pc^{-2}$) and {\it Bar} ($88.8-213.4~M_\odot~\rm pc^{-2}$). Based on K-S test, however, we do not find the significant difference.

{\it Virial parameter }(Fig. ~\ref{fig:boxplot} (g)): Because  $\alpha_{\rm vir} $ is proportional to $ \sigma_v/(\braket{T} R)$, $\alpha_{\rm vir}$ is expected to show a reverse tendency of $T_{\rm peak}$. In fact, $\alpha_{\rm vir}$ in {\it Bar}  is the highest ($1.3-2.9$), followed by {\it Arm} ($1.0-2.1$) and {\it Bar-end} ($0.9-2.2$). Based on the resampling method, $58 \pm 8$~\% of the GMCs in {\it Bar} is gravitationally unbound ($\alpha_{\rm vir} > 2$). This percentage is  larger than that in {\it Arm} and {\it Bar-end} ($33 \pm 4$~\% and $29 \pm 3$~\%), and in M51 spiral arm (35~\%). However, the K-S test gives high $p$-value  which indicates there is no significant environmental variation in $\alpha_{\rm vir}$. Note that the K-S test for the observed value  shows the marginally significant difference between {\it Arm} and {\it Bar}. But based on the resampling method, it is rare that K-S test gives a $p$-value less than or equal to 0.05. In conclusion,  there is  no significant environmental variations in $\alpha_{\rm vir}$ in NGC~1300.

{\it Scaling coefficient }(Fig. ~\ref{fig:boxplot} (h)): As expected from the fact that $c = \sigma_v/R^{0.5}$ is a combination of $\sigma_v$ and $R$, we find no environmental variation in $c$. The $c$ in NGC~1300 is comparable to that  in M51 spiral arms.

In summary, there is a significant environmental variation in the $T_{\rm peak}$; the highest value in {\it Bar-end} followed by {\it Arm} and {\it Bar}. Since  $\sigma_v$ and $R$ do not exhibit environmental variation, the variation of $T_{\rm peak}$ is  mainly responsible for the variation of median value of $M_{\rm mol}$, $\Sigma_{\rm mol}$, and $\alpha_{\rm vir}$. But based on the two-sided K-S tests, the (marginally) significant differences are only seen in the distribution of $M_{\rm mol}$ between {\it Bar} and {\it Bar-end}, and {\it Arm} and {\it Bar-end}; there is virtually no significant difference in GMC physical properties among the {\it Bar}, {\it Arm} and {\it Bar-end}. In addition, we find the properties of GMCs in NGC~1300 are roughly comparable to those in M51 spiral arms. In particular, the properties  in {\it Bar-end} are very similar.

\subsection{The reliability of the measurements}\label{reliability}
\subsubsection{CPROPS bias}\label{sec:CPROPS bias summary}
As described in section~\ref{sec:derivation of R}, the extrapolated (but non-deconvolved) $R$, $\sigma_v$ and $L_{\rm CO}$ is typically $1.5 \sim 2.0$ times higher than  the directly measured values. The accuracy of the CPROPS correction (i.e., extrapolation and deconvolution) depends on the sensitivity, spatial resolution and velocity resolution \citep{RosolowskyLeroy}. In order to  assess the reliability of the measurements of GMC properties, we simulated ALMA observation of mock GMCs, which  are three-dimensional Gaussian clouds in position-position-velocity space with a given masses, sizes, and velocity dispersions, in CASA under the same condition of our observations (Section~\ref{sec:obs}). After reconstructing the image from the simulated visibility, we identify the GMCs using CPROPS with the same parameters described in Section~\ref{sec:GMC identification and characterization} and compare the input and output values. The details of the simulation method and results are described in Appendix~\ref{apx:CPROPS},  and we briefly describe the results and discussion here.

We find that the corrected $\sigma_v$ is overestimated by $\sim$ 50~\% if the directly measured velocity width, $2.35\sigma_v$, is less than half the channel width of $5.0~\rm km~s^{-1}$. Such GMC accounts for  56~\%, 64~\%, and 64~\% in  {\it Bar}, {\it Arm}, and {\it Bar-end}, respectively. Therefore,  a large number of cataloged GMCs may be overestimated in the $\sigma_v$ by a factor of $\sim$ 1.5. The CPROPS algorithm performs relatively well for the measurements of  the $M_{\rm mol}$ and $R$ in comparison to $\sigma_v$. We find that the corrected $M_{\rm mol}$ is mostly equal to the input $M_{\rm mol}$ for the GMC with higher S/N ($7 \leq {\rm S/N} \leq 10$), but the corrected $M_{\rm mol}$ is typically  underestimated by  $\sim$ 10~\% for the  GMC with lower S/N ($4 \leq {\rm S/N} < 7$). The corrected $R$ is typically underestimated by $\sim$ 20~\% and  $\sim$ 10~\% for the  GMC with lower S/N  and higher S/N, respectively. We retested the environmental variation of the GMC properties using the catalog corrected for the over/underestimation factor. As a result of the K-S test, $p$-values do not change much (see Table~\ref{tab:re-KStest}). Thus we concluded that the over/underestimation of the measurements  does not influence on the discussion about the  environmental variations described in Section~\ref{sec:Basic properties}.

\subsubsection{Impact of the missing flux} \label{sec:Missing flux}

Measurements of GMC properties may be affected by the missing flux \citep[e.g.,][]{RosolowskyLeroy,Pan:2017gy}. Our ALMA observations did not recover the total CO($1-0$) flux due to the lack of Atacama Compact Array (ACA) measurements. To estimate the recovery fraction of the total CO($1-0$) flux, we compared the CO($1-0$) flux obtained from ALMA with those from Nobeyama 45-m single-dish telescope by \citet{Maeda:2018bg}, which presented single-pointing observations of several regions in NGC~1300. First, we convolved the CO($1-0$) data cube to $13^{\prime\prime}.5$, the beam size of the Nobeyama 45-m telescope at 115 GHz. We then obtained spectra at the pointing positions of the 45-m telescope and measured the recovery fraction (i.e., the ratio of the two flux).
We find that the average recovery fraction is 54 \% with typical uncertainty of 5 \%; about half of the total flux was missed, which could affect the measurements of GMC properties.

\begin{figure}
\begin{center}
	\includegraphics[width=\hsize]{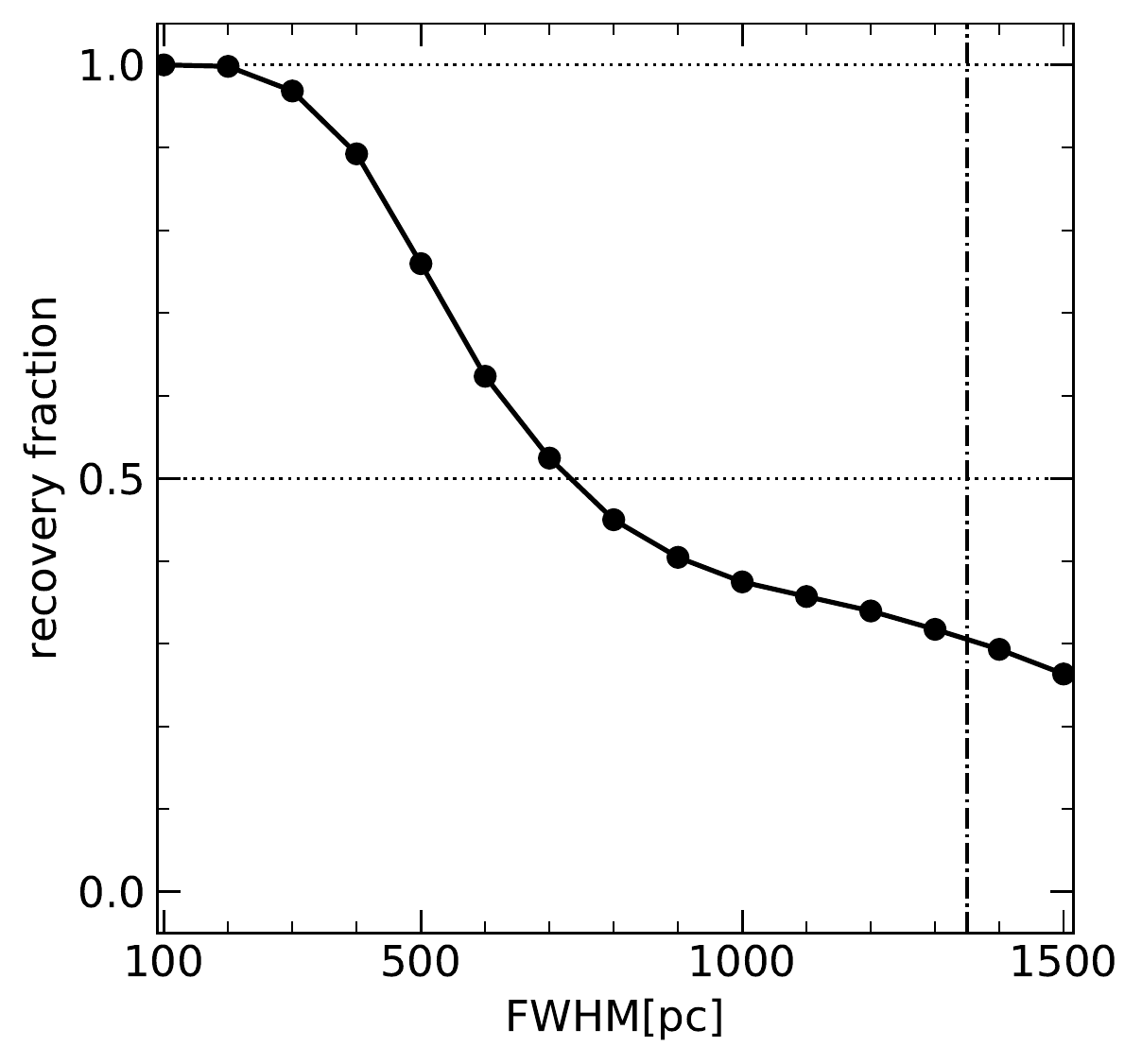}
    \caption{Recovery fraction of the Gaussian component as a function of the FWHM. We simulated ALMA observation of the mock Gaussian component with a given FWHM under the same configuration and noise-free condition. Horizontal dashed lines indicate the recovery fraction of 1.0 and 0.5.
    Vertical dash-dotted line indicates 1350 pc, corresponding to the beam size of Nobeyama 45-m telescope.}
 \label{fig:recovery_fraction}
\end{center}
\end{figure}

To investigate the relationship between the spatial scale of the gas distribution and missing flux, we simulated ALMA observation of a mock Gaussian component. We created a fits image of a circular Gaussian component with total flux of 1.0 Jy and given FWHM. The FWHM was set from 100 pc to 1500 pc every 100 pc. 
In order to extract the effect originated from the uv-distribution,
we simulated the observation under the same configuration and noise-free condition by using the task of \verb|simobserve| in CASA. 
After reconstructing the image with \verb|tclean| task, we measured the flux of the component and recovery fraction. Fig. \ref{fig:recovery_fraction} shows the recovery fraction of the Gaussian component as a function of the FWHM.
We find that the recovery fraction of the Gaussian component with FWHM $\leq 300$ pc is $\sim 1.0$ and that with FWHM $> 300$ pc is under 1.0. For the large Gaussian component with the FWHM $\geq 700$ pc, more than half of the flux is missed.

According to the simulation results, total flux of a GMC smaller than $\sim 300$ pc can be recovered in our observations. Since the radius of the GMCs we identified is smaller than $300$ pc (the maximum is $\sim 100$ pc), their properties would not be affected by the missing flux and the over/underestimates described in \ref{sec:CPROPS bias summary} would be mainly due to the sensitivity. In our observations, a molecular gas structure larger than $\sim 700$ pc is mostly resolved out.
This result would mainly be responsible for the missing flux in our data and imply the presence of the large scale gas structures such as diffuse molecular gas components, as have been seen in other galaxies \citep[e.g., M51;][]{Pety:2013fw}. 
Note that the missing flux would also depend on shape of the GMCs and distribution of them.  Thus it is difficult to figure out the real impact of the missing flux on the GMC properties by simulations. This issue will be expected to be clear by the new ACA observations of NGC 1300 in Cycle 7 (PI: F. Maeda).

\section{Scaling relations}\label{sec: Scaling relations}
In this section, we present the analysis of the scaling relations for the  GMCs, which are commonly referred to as Larson's laws \citep{Larson1981}. Here, we investigate the size-velocity dispersion relation, the virial-luminous mass relation,  the size-mass relation, and the relationship between $\sigma_v^2/R$ and $\Sigma_{\rm mol}$ for the resolved GMCs ($M_{\rm mol} \geq 5.0 \times 10^5~M_\odot$ and $R > 15$ pc) in NGC~1300. In each analysis, we calculate the Spearman's rank correlation coefficient ($r_s$)  to evaluate the strength of a link between two sets of data. Considering the number of GMCs, we regard $r_s \geq 0.8$ as a sign of strong correlation and $0.5 < r_s < 0.8$ as that of moderate correlation.

\subsection{Size-velocity dispersion relation}
\begin{figure*}
 \includegraphics[width=175mm]{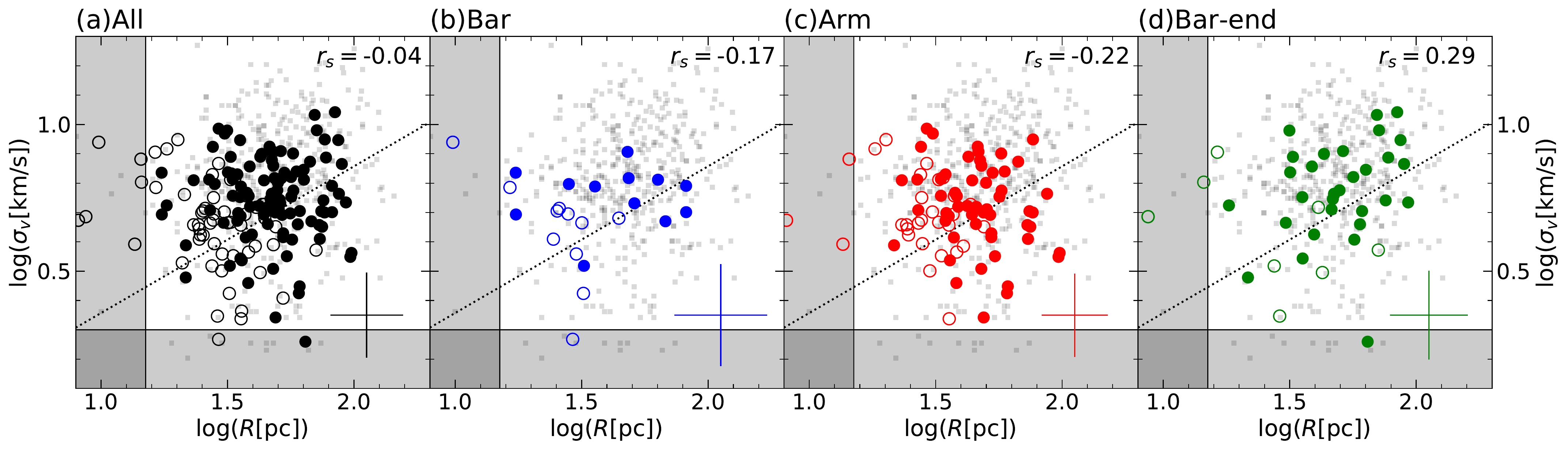}
 \caption{(a) Size-velocity dispersion relation for the GMCs in the whole region in NGC~1300. (b) - (d) Same as panel (a) but for the different environments in NGC~1300. We show the GMCs with $M_{\rm mol} \geq 5.0 \times 10^5~M_\odot$ and $R \geq 15$ pc in filled circles and the rest in open circles. The average error bars are indicated as a cross in the bottom right corner of each panel. The range below the spatial (15 pc) and spectral (2.0 $\rm km~s^{-1}$) sensitivity limit is indicated as shaded regions. The Spearman's correlation rank, $r_s$, for the filled circle is given in the top right corner of each panel. Gray squared show the relationship for GMCs in M51 spiral arms \citepalias{Colombo:2014ei}. Black dotted line indicates the Galactic fit, $\sigma_v = 0.72 R^{0.5}$ \citep{Solomon87}. }
 \label{fig:R_sigv}
\end{figure*}

Fig. ~\ref{fig:R_sigv} shows  the size-velocity dispersion relation in NGC~1300. The relation in M51 spiral arms from \citetalias{Colombo:2014ei} is shown as gray squares. In panel (a), although the majority of the data points lies around the Galactic fit \citep{Solomon87}, the correlation is not seen ($r_s = -0.04$) which is also observed in M51 spiral arms ($r_s = 0.16$). Even with division into subregions, the  correlations are not seen ($|r_s| < 0.3$).

The existence of the size-velocity dispersion relation has been discussed. In M51, there was no correlation in not only spiral arms but also center and inter-arm regions \citepalias{Colombo:2014ei}.  \citet{Hirota:2018jp} reported the apparent lack of correlation in the  various environments including arm and bar regions of M83. On the contrary, \citet{Bolatto2008ApJ} found a correlation of $\sigma_v \propto R^{0.60 \pm 0.10}$ for the multigalaxy GMC population including spiral galaxies and dwarf galaxy.  \citet{Faesi:2018gg} found also a correlation of $\sigma_v \propto R^{0.48 \pm 0.05}$ in NGC 300. In section~\ref{sec:sigma2R_Sigmamol}, we will discuss the cause for the weak correlation in NGC~1300.

\subsection{Virial Equilibrium}
\begin{figure*}
 \includegraphics[width=175mm]{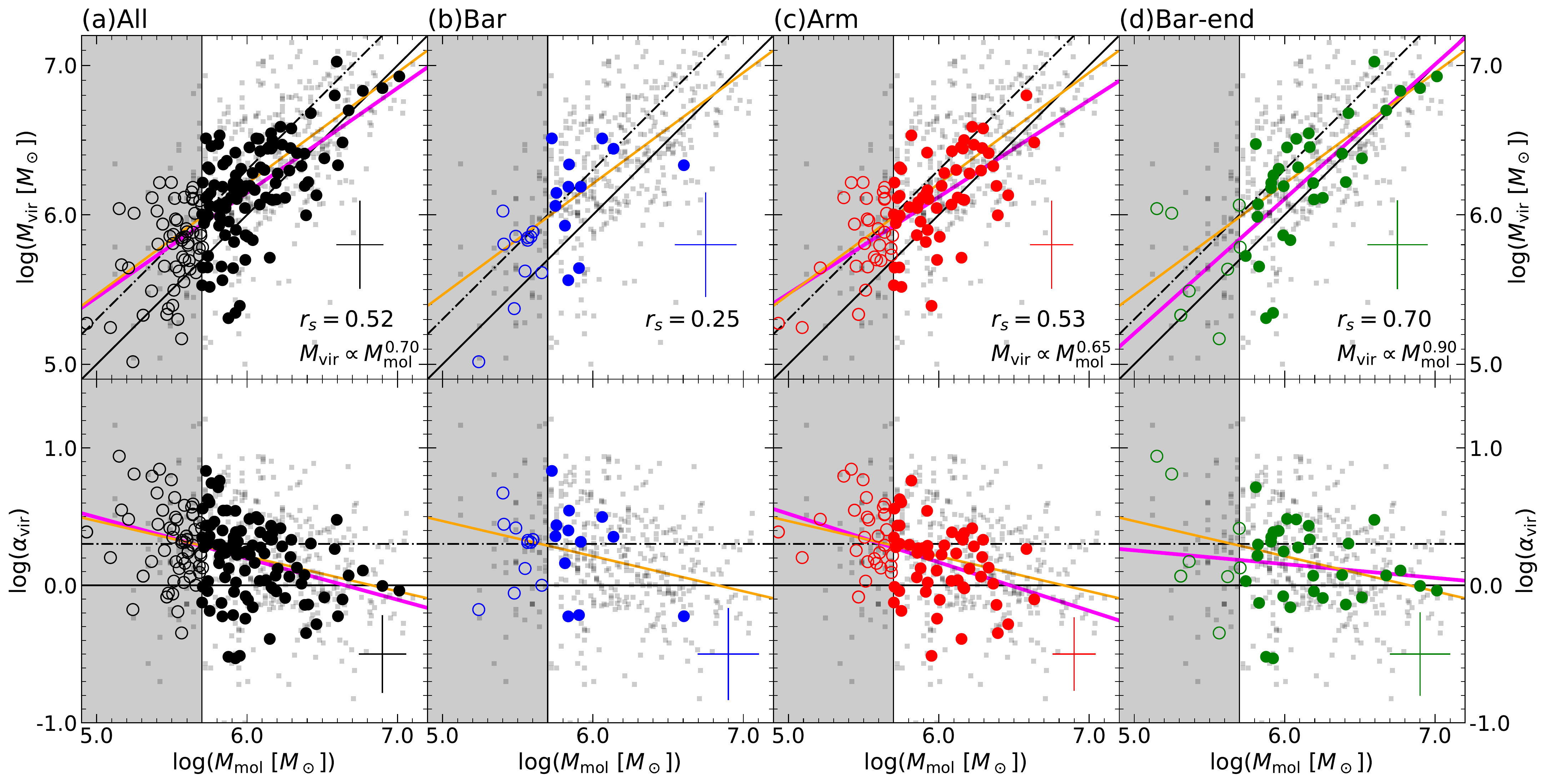}
 \caption{(a) Virial mass-luminosity based mass relation (upper) and virial parameter-luminosity based mass relation (lower) for the GMCs in the whole region of NGC~1300. (b) - (d) Same as panel (a) but for the different environments in NGC~1300. We show the GMCs with $M_{\rm mol} \geq 5.0 \times 10^5~M_\odot$ and $R \geq 15$ pc in filled circles and the rest in open circles. The average error bars are indicated as a cross in each panel. The range below $M_{\rm mol} = 5.0 \times 10^5 ~M_\odot$ is indicated as a shaded region. Black solid and dotted line indicate the line for $\alpha_{\rm vir} = 1.0$ and $\alpha_{\rm vir} = 2.0$, respectively. For upper panels, the Spearman's correlation rank, $r_s$, for the filled circle is given in the bottom right corner (BRC) of each panel. If $r_s$ is larger than 0.5, we show the best fitted line as a magenta solid line and its slope is given in the BRC. Gray squares and orange solid line show the relationship and the best fitted line  for GMCs in M51 spiral arms \citepalias{Colombo:2014ei}, respectively.
}
 \label{fig:Lco_Mvir}
\end{figure*}

Upper panels in Fig. ~\ref{fig:Lco_Mvir} show the relation between virial mass and  the molecular gas mass derived from CO luminosity in NGC~1300. We find  moderate correlations for all GMCs ($r_s = 0.52$) in {\it Arm} ($0.53$) and in {\it Bar-end} ($0.70$), while there is no apparent correlation in {\it Bar} ($0.25$). In the panels with $r_s \geq 0.5$, a magenta solid line represents the best-fitted line  determined by the ordinary least-squares method: $M_{\rm vir} \propto M_{\rm mol}^{0.70 \pm 0.35}$ in the whole region, $M_{\rm vir} \propto M_{\rm mol}^{0.65 \pm 0.55}$ in {\it Arm}, and $M_{\rm vir} \propto  M_{\rm mol}^{0.90 \pm 0.53}$ in {\it Bar-end}. These relationships agree with that  in M51 spiral arms shown as orange solid line ($M_{\rm vir} \propto M_{\rm mol}^{0.74}$). In lower panels of Fig.  ~\ref{fig:Lco_Mvir}, we plot the virial parameter ($\alpha_{\rm vir}$) as a function of $M_{\rm mol}$. As inferred from the slope lower than 1.0 in upper panels in Fig. ~\ref{fig:Lco_Mvir}, $\alpha_{\rm vir}  \cong M_{\rm vir}/M_{\rm mol}$ decreases with increasing  $M_{\rm mol}$. This suggests that  the high mass GMCs  tend to be more strongly bound than low mass GMCs in {\it arm} and {\it Bar-end} of NGC~1300 as seen in M51.

\subsection{Size-mass relation}
\begin{figure*}
 \includegraphics[width=175mm]{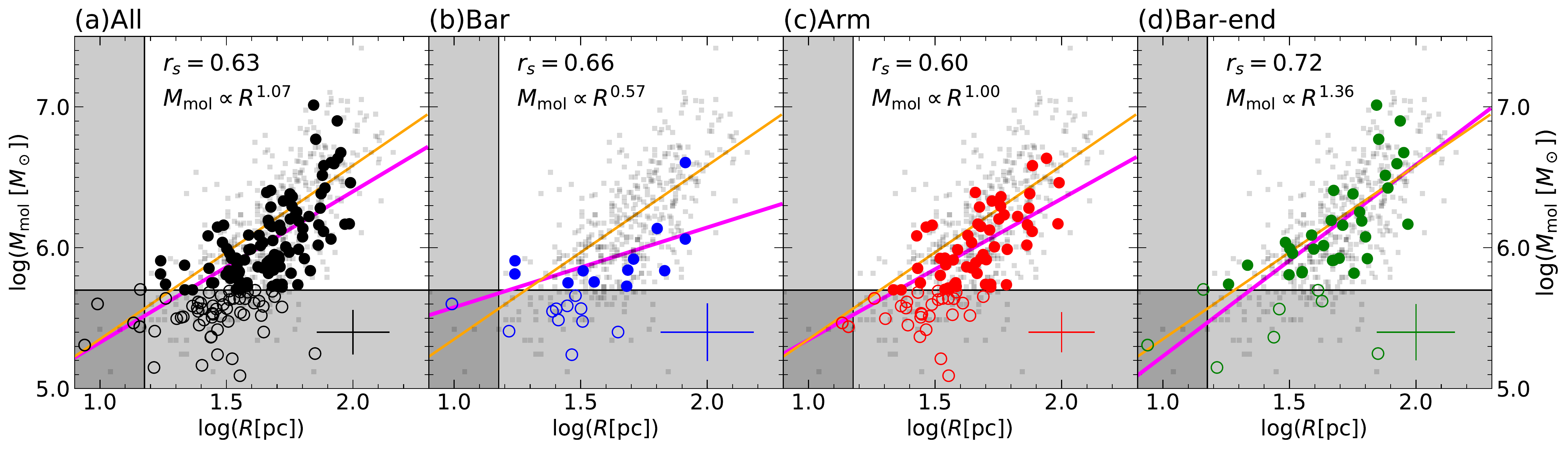}
 \caption{
 (a) Size-molecular gas mass relation for the GMCs in the whole region of NGC~1300. (b) - (d) Same as panel (a) but for the different environments in NGC~1300. We show the GMCs with $M_{\rm mol} \geq 5.0 \times 10^5~M_\odot$ and $R \geq 15$ pc in filled circles and the rest in open circles. The average error bars are indicated as a cross in each panel. The range below $M_{\rm mol} = 5.0 \times 10^5 ~M_\odot$ and $R = 15$ pc is indicated as shaded regions.  The Spearman's correlation rank, $r_s$, for the filled circle is given in the top left corner (TLC) of each panel.  We show the best fitted line as a magenta solid line and its slope is given in the TLC.  Gray squares and orange solid line show the relationship and the best fitted line  for GMCs in M51 spiral arms \citepalias{Colombo:2014ei}, respectively.}
 \label{fig:R_Lco}

 \includegraphics[width=175mm]{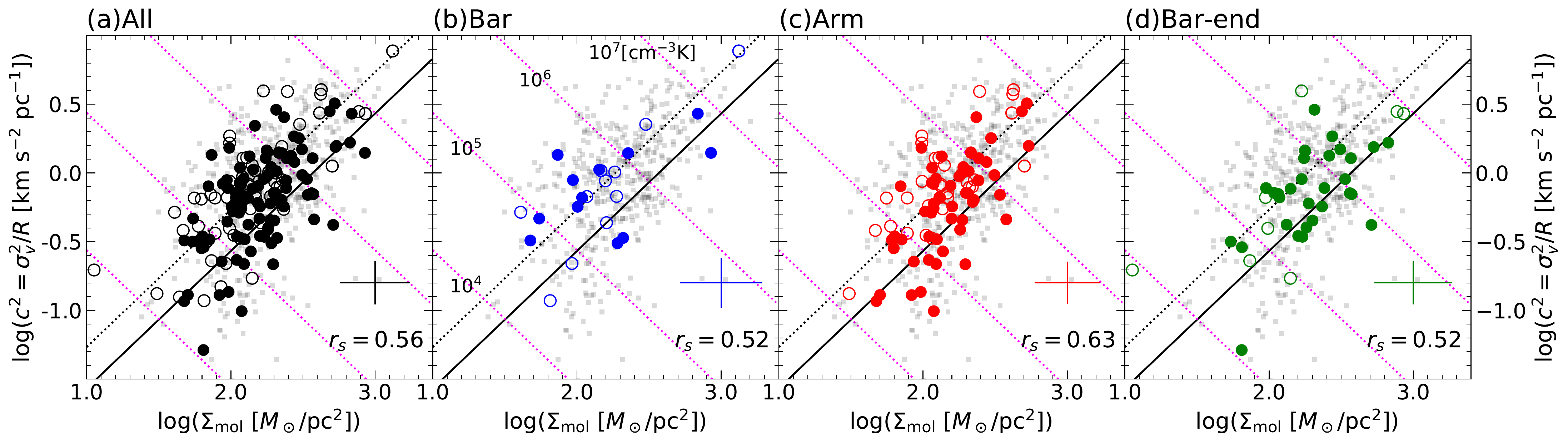}
 \caption{
 (a) Scale coefficient as a function of molecular gas surface density in the whole region of NGC~1300. (b) - (d) Same as panel (a) but for the different environments in NGC~1300. We show the GMCs with $M_{\rm mol} \geq 5.0 \times 10^5~M_\odot$ and $R \geq 15$ pc in filled circles and the rest in open circles. The average error bars are indicated as a cross in each panel. The Spearman's correlation rank, $r_s$, for the filled circle is given in the BRC of each panel. Gray squares show the relationship for the GMCs in M51 spiral arms \citepalias{Colombo:2014ei}.  Black solid and dotted line indicate the line for $\alpha_{\rm vir} = 1.0$ and $ 2.0$, respectively.  Magenta dotted lines indicate the line for $P_{\rm int}/k = 10^4, 10^5, 10^6$, and $10^7~{\rm [cm^{-3}~K]}$.}
 \label{fig:Smol_Scale}
\end{figure*}

Fig. ~\ref{fig:R_Lco} shows the size-mass relation in NGC~1300. There is a moderate correlation in each panel ($0.60 < r_s < 0.75$). Magenta solid line represents the best fit line  determined by the ordinary least-squares method. The slope of the relation in {\it Bar} (0.57) is the shallowest, followed by {\it Arm} (1.00) and {\it Bar-end} (1.36). The fact that the index is smaller than 2.0, the molecular gas surface density ($\Sigma_{\rm mol}$) of GMCs is not roughly constant but decreases with increasing GMC size in NGC~1300. The difference in the index shows that the molecular gas surface density ($\Sigma_{\rm mol}$) in {\it Bar} is lower than GMCs with similar size in {\it Arm} and {\it Bar-end}, which is consistent with the tendency of $\Sigma_{\rm mol}$ (Fig. ~\ref{fig:boxplot} (g)).  Orange solid line represents the best fit line for the GMCs ($M_{\rm mol} \geq 5.0 \times 10^5~M_\odot$ and $R \geq 15$ pc) in M51 spiral arms and slope is 1.23. The $\Sigma_{\rm mol}$ in {\it Bar} and {\it Arm} is lower than GMCs with similar size in M51 spiral arms.

\subsection{Relationship between $\sigma_v^2/R$ and $\Sigma_{\rm mol}$}\label{sec:sigma2R_Sigmamol}
Recent studies on molecular gas at GMC scale argued that the relationship between $\sigma_v^2/R = c^2$ and $\Sigma_{\rm mol}$ is useful to investigate the physical state of the GMCs \citep[e.g.,][]{Heyer2009ApJ,Leroy:2015ds,Sun:2018ib}. This is because the position of the GMC in $\sigma_v^2/R$-$\Sigma_{\rm mol}$ space gives information about not only virial parameter, $\alpha_{\rm vir}$, but also internal turbulent pressure, $P_{\rm int}$ of the GMC. Here, the $P_{\rm int}$ can be expressed as
\begin{eqnarray}
\frac{P_{\rm int}}{k} &=& \frac{\rho \sigma_v^2}{k} \nonumber \\
    &=& 3695  \left( \frac{\Sigma_{\rm mol}}{M_\odot~\rm pc^{-2}}\right)
    \left( \frac{c}{\rm km~s^{-1}~pc^{-0.5}}\right)^2
    ~ [{\rm cm^{-3}~K}],
\end{eqnarray}
where $\rho$ is the molecular gas volume density of the GMC.

Fig. ~\ref{fig:Smol_Scale} plots the $\sigma_v^2/R$-$\Sigma_{\rm mol}$ relation of NGC~1300. The $\alpha_{\rm vir} = 1$ and 2 are represented by a black solid and dotted line, respectively, and magenta dotted lines indicate constant $P_{\rm int}$. Note that the $\sigma_v^2/R$-$\Sigma_{\rm mol}$ relation is  mathematically equivalent to the $M_{\rm vir}$-$M_{\rm mol}$ relation (Fig. ~\ref{fig:Lco_Mvir}). We can see a moderate correlation in the whole region ($r_s = 0.63$) , {\it Bar} (0.66), {\it Arm} (0.60), and {\it Bar-end} (0.72) in NGC~1300, which shows the scaling coefficient, $c$, increases with increasing  $\Sigma_{\rm mol}$. The same tendency is seen in Milky Way; \citet{Heyer2009ApJ} pointed out that  $c$ is proportional to   $\Sigma_{\rm mol}^{1/2}$.

As shown in panel (a), $P_{\rm int}$ of the GMCs in NGC~1300 is spread about 3 dex ($10^4 \leq P_{\rm int}/k/[{\rm cm^{-3}~K}] \leq 10^7$). The median values and the $\Delta Q$ of $P_{\rm int}/k$ are roughly constant across the different environments: $2.2^{+6.2}_{-1.4}$, $2.9^{+4.6}_{-2.5}$, and $3.3^{+10.5}_{-1.6} \times 10^5~\rm cm^{-3}~K$ in {\it Bar}, {\it Arm}, and {\it Bar-end}, respectively.  The K-S tests give high $p$-values of $> 0.1$. These values are slightly lower than that in M51 spiral arms of $6.7^{+8.3}_{-4.3}\times 10^5~\rm cm^{-3}~K$.

Fig.~\ref{fig:Smol_Scale}  can be interpreted to explain the cause for weak correlation between size and velocity dispersion (Fig. ~\ref{fig:R_sigv}). The GMCs in NGC~1300 cover a wider range of $\Sigma_{\rm mol}$ ($\sim$1.5 dex) than  those in Milky Way \citep[$\sim$ 1.0 dex;][]{Solomon87}. Therefore, the range of $c$ in NGC~1300 becomes wider than that in Milky Way, leading to a decorrelation between $\sigma_v$ and $R$. The dependence of the $c$ on the $\Sigma_{\rm mol}$ also may be the part of the cause of weak correlation seen in M51 and M83 \citep[see][]{Hirota:2018jp}. Note that the limited dynamic range of  the size parameter and the errors in the measurements with CPROPS could be responsible for the weak correlation.

\begin{figure}
\begin{center}
 \includegraphics[width=77mm]{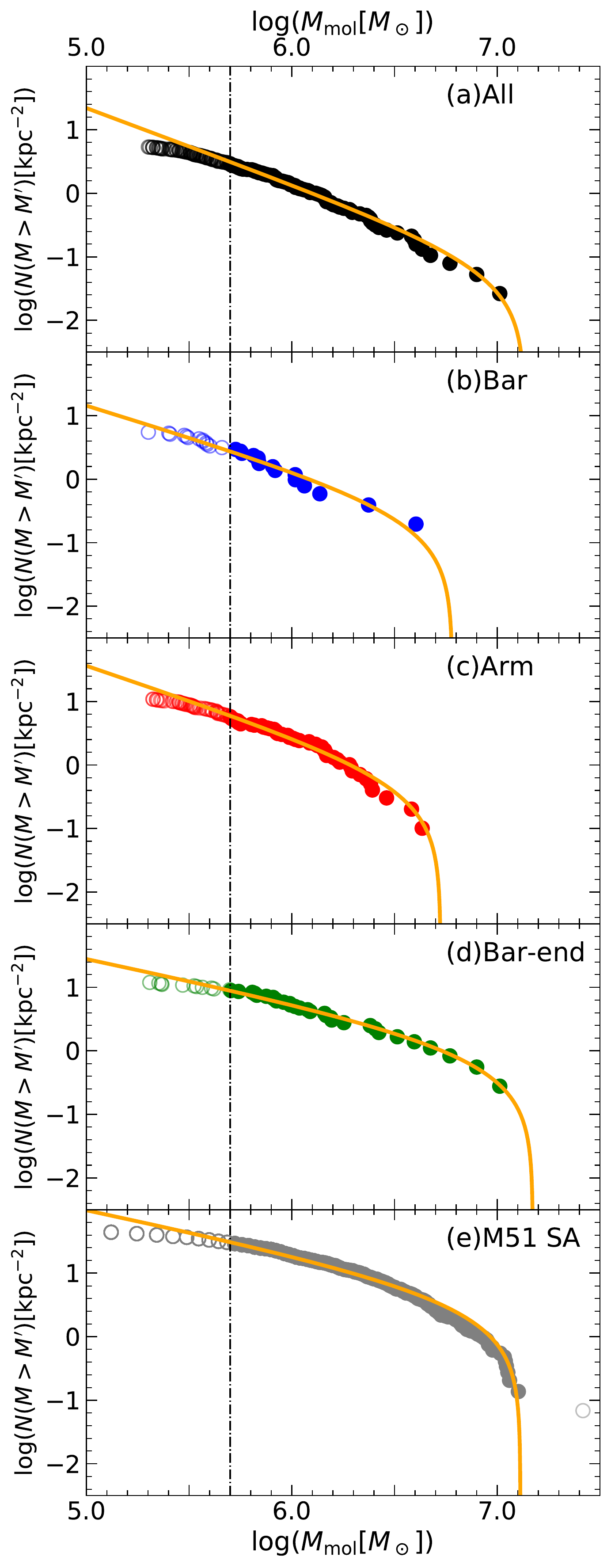}
 \caption{(a) Cumulative mass spectrum for GMCs in the whole region of NGC~1300. (b)-(d) Same as panel (a), but for the different environments in NGC~1300. (e) Same as panel (a), but for GMCs in M51 spiral arms \citepalias{Colombo:2014ei}. In each plot, filled and open circles indicate the GMCs that are and are not used for fitting the spectrum with equation (\ref{eq:CMF}), respectively.  The orange solid line indicates the best-fitted function.  The dash-dotted vertical line indicates the lower mass limit, $5 \times 10^5~M_\odot$, for the fitting of mass spectra. 
 }
 \label{fig:CMF}
 \end{center}
\end{figure}

\section{GMC mass spectra}\label{sec: GMC mass spectra}
In this section, we investigate the mass spectra of GMCs in NGC~1300. The mass spectrum provides information about GMCs formation and destruction processes \citep[e.g.,][]{Kobayashi2017ApJ}, and is usually  expressed in differential form known to follow a power law relation as
\begin{equation}
    \frac{dN}{dM} \propto M^{\gamma},
\end{equation}
where $M$ is the molecular gas mass, $N$ is the number of molecular clouds and $\gamma$ is an index of the power law relation.

Integration of this expression gives a cumulative mass distribution. Several studies reported the GMC mass spectra are underpopulated at higher masses, i.e., it is thought that there is an upper limit to the GMC mass \citep[e.g.,][]{WilliamsMcKee1997ApJ,Fukui2001PASJ,Fukui2008ApJS,Rosolowsky2007ApJ,Gratier2012A&A}. Considering the existence of the upper cutoff mass, a truncated power-law is suitable as a cumulative mass spectrum \citep{WilliamsMcKee1997ApJ}:
\begin{equation}
\label{eq:CMF}
    N(>M) = - \frac{N_u}{\gamma + 1} \left[ \left( \frac{M}{M_u}\right)^{\gamma + 1} - 1\right],
\end{equation}
where $M_u$ is an upper cutoff mass of GMCs and $N_u$ is a measure of the number of GMCs at the upper cutoff mass. The index $\gamma$ shows how the GMC mass is distributed. If $\gamma$ is larger than $-2$, massive GMCs dominate the total cloud mass.

The parameters of the mass spectrum ($\gamma$ and $M_u$) is considered to be determined in the balance between the formation and destructive processes of GMCs. Enhance of the formation process of massive GMCs  \citep[e.g., agglomeration of small clouds and self gravity;][]{Dobbs:2008ez} leads to a steeper mass spectrum and a higher $M_u$. On the contrary, the destruction process (e.g., stellar feedback, cloud-cloud collision, and large-scale shear motion) plays a role to make mass spectrum shallow and  decrease the $M_u$.

Fig. ~\ref{fig:CMF} shows the cumulative GMC mass spectrum in different environments in NGC~1300  (panel (a) - (d)) and in M51 spiral arms \citepalias[panel (e);][]{Colombo:2014ei}. 
The y-axis shows the GMC number density. The number density of high mass GMC ($M_{\rm mol} \geq 1.0 \times 10^6~M_\odot$) in {\it Bar-end} is the highest (5.2 $\rm kpc^{-2}$), followed by in {\it Bar}(1.2 $\rm kpc^{-2}$) and {\it Arm}(2.6 $\rm kpc^{-2}$). Although these values depend on the definition of the environmental mask, there is no doubt that the number densities of high mass GMC with $\geq 1.0 \times 10^6~M_\odot$  in {\it Bar} (5.1 kpc$^2$) and {\it Arm} (9.9 kpc$^2$) are lower than that in {\it Bar-end} (3.6 kpc$^2$). In particular, GMCs more than $ 5.0 \times 10^6~M_\odot$ are only observed in {\it Bar-end}. Note that the number density in M51 spiral arms is 18.2 $\rm kpc^{-2}$, and is the factor of $4 \sim 10$ larger than that in any environments of NGC~1300. 

We fitted the cumulative GMC mass spectrum with equation (\ref{eq:CMF}). Although the mass completeness limit is expected to be $2.0 \times 10^5~M_\odot$ in uncrowded regions (see section~\ref{sec:basicproperties}), the completeness limit might be effectively raising in a crowded region like {\it Bar-end} \citep[e.g.,][]{Colombo:2014ei,Hirota:2018jp}. Thus, we fitted the mass spectrum using the GMCs with $M_\odot \geq 5.0 \times 10^5~M_\odot$ indicated as a vertical black dash-dotted line in Fig. ~\ref{fig:CMF}. To  estimate  the  fitting uncertainties, we  made  resampling  with  100  realizations. In  one realization,  random  values  of  a molecular gas mass  were generated  within  the uncertainties CPROPS calculated. The median and the MAD of 100 realizations for each parameter were adopted as the best-fitted value and the confidence interval (Table~\ref{tab:CMF}). The orange solid line in Fig. ~\ref{fig:CMF} shows the best-fitted cumulative mass function. We list the $p$-values of the K-S tests as an indication of the goodness-of-fit in the last column of Table~\ref{tab:CMF}. Here we fitted the mass spectrum of GMCs in M51 spiral arms in the same way.

No clear difference was found in the shape of mass spectra between in {\it Bar} and {\it Arm}: the slope and upper mass limit is $\gamma \sim -2.0$ and $M_u \sim 5.0 \times 10^6~M_\odot$, respectively. However, the mass spectrum in {\it Bar-end} is obviously different from those in {\it Bar} and {\it Arm}. The slope in {\it Bar-end} is flatter than $-2$, which indicates that massive GMCs  account for a large population of total cloud mass in {\it Bar-end}. The upper mass limit in {\it Bar-end} of $M_u = (1.3 \pm 3.4) \times 10^7~M_\odot$ is twice as large as that in Bar and Arm regions. The presence or absence of differences in the mass spectrum between environments  is consistent with that in the box plot of $M_{\rm mol}$ (see Section~\ref{sec:Basic properties}). The $\gamma$ and $M_u$ in {\it Bar-end} are similar to those in M51 spiral arms.

Since the $\gamma$ in {\it Bar-end} is larger than those in {\it Bar} and {\it Arm}, {\it Bar-end} would be an environment where the massive GMCs forms more than in {\it Bar} and {\it Arm}. The similarity of the feature betwen {\it Bar-end} and M51 spiral arms suggests the mechanism which regulate the formation and destruction of GMCs is similar to that in M51 spiral arms. Although the mass spectra in {\it Bar} and {\it Arm} is similar, the GMC destruction mechanism would be different because the star formation activity is apparently different (Section~\ref{sec:intro}). The stellar feedback process may be predominant in {\it Arm}, but the dynamical effect (e.g., cloud-cloud collision and large-scale shear motion)  may be important process in {\it Bar}.

\begin{table}
 \caption{Truncated Power-law Fits to the GMC Mass Spectra in Different Environments in NGC~1300}
 \label{tab:CMF}
 \begin{tabular}{lcccc}
 \hline
  Envir. & $\gamma$ & $M_u$ & $N_u$ & $p$-value\\
  &  & ($10^6 ~M_\odot$) & &  \\
\hline
 All       & $-2.20 \pm 0.04$ & $13.6 \pm 0.67$ & $2.74  \pm 0.40$ & 0.22\\
 \hline
 Bar       & $-1.97 \pm 0.20$ & $6.26 \pm 3.04$ & $1.36  \pm 0.97$ & 0.59\\
 Arm       & $-2.08 \pm 0.10$ & $5.31 \pm 0.36$ & $5.45  \pm 1.07$ & 0.40\\
 Bar-end   & $-1.66 \pm 0.07$ & $14.8 \pm 0.93$ & $2.56  \pm 0.42$ & 0.85\\
 \hline \hline
 M51 SA    & $-1.70 \pm 0.03$ & $13.0 \pm 0.34$ & $35.6 \pm 2.46$ & $10^{-4}$\\
\hline
\multicolumn{5}{l}{{\small Slopes $\gamma$, upper cutoff mass $M_u$, and a measure of the number }} \\
\multicolumn{5}{l}{{\small of GMCs at the upper cutoff mass $N_u$ of the truncated power-}} \\
\multicolumn{5}{l}{{\small law fits to the GMC mass spectra of the different environments}} \\
\multicolumn{5}{l}{{\small in NGC~1300 The error are obtained through 100 resampling}} \\
\multicolumn{5}{l}{{\small interaction. In the last column, we list the $p$-values of the K-S}} \\
\multicolumn{5}{l}{{\small tests as an indication of the goodness-of-fit.}} \\
 \end{tabular}
\end{table}

\section{Discussion}\label{sec:discussion}
As described in Sections ~\ref{sec:intro} and~\ref{sec:GMCdis}, SFE differences with environments are clearly seen in typical strongly barred galaxies. In the arms, H\textsc{ii} regions are associated with dust lanes, and GMCs coexist with the dust lanes. However,  in bar regions,  prominent H\textsc{ii} regions are often not seen while the GMCs do exist. What physical mechanism controls the SFE of the GMC? Based on the K-S tests, there is no significant variation in the physical properties ($\sigma_v$, $R$, $M_{\rm mol}$, $M_{\rm vir}$, $\Sigma_{\rm mol}$, $\alpha_{\rm vir}$, and $c$) of the GMCs between {\it Bar} and {\it Arm}  (Table~\ref{tab:KStest}). Comparing to the GMCs between {\it Bar} and {\it Bar-end}, and {\it Arm} and {\it Bar-end}, (marginally) significant difference is only seen in the distribution of $M_{\rm mol}$. Therefore, it appears that systematic differences in the GMC properties are not the cause for the SFE differences with environments.

Many previous studies investigated the cause for the low SFE in the bar regions. Some previous studies proposed that GMCs can not form due to a strong shock and/or shear along the bar  \citep[e.g.,][]{Tubbs1982ApJ,Athanassoula1992MNRAS,ReynaudDownes1998A&A}. Recent studies suggest molecular clouds in bar regions may be gravitationally unbound due to the strong internal turbulence of the clouds.  \citet{Sorai2012PASJ} made $^{12}$CO($1-0$) map of Maffei 2 at an angular resolution of 200 pc, and pointed out a possibility that clouds in the bar regions are gravitationally unbound, which causes the low SF activity. \citet{Nimori2013MNRAS} performed a two-dimensional hydrodynamical simulation and also found the unbound clouds in bars. In NGC~1300,  the number fraction of GMC with $\alpha_{\rm vir} > 2$ in {\it Bar} (60~\%) is larger than that in {\it Arm} and {\it Bar-end} (30~\%), which may  partly contribute to decreasing the SFE in the bar region.  However, there is no significant difference in $\alpha_{\rm vir}$  based on the K-S test (Table~\ref{tab:KStest}). This result suggests that the lack of massive star formation in the strong bar of NGC~1300 can not be explained by a systematic difference of $\alpha_{\rm vir}$ only.

\citet{Hirota:2018jp} found the $\alpha_{\rm vir}$ of the GMCs in the bar of M83 (median is $\sim 1.6$) is larger than that in the arm ($\sim 1.2$). This result seems to be inconsistent with our results. However, a direct comparison with our results is not straightforward because the algorithm for identifying GMCs they used was different from ours. It is necessary to compare the GMC properties between M83 and NGC~1300 under the same data quality and methodology. This comparison is important but beyond our scope in this paper, which remains as a future subject.

Although the galactic environments seem not to affect the physical properties of GMCs in NGC~1300, there is a possibility that the  interaction of GMCs is affected.
Cloud-cloud collisions (CCCs), which induces clump formation by shock compressions, has been suggested as the mechanism of massive star formation  \citep[e.g.,][]{Habe1992PASJ,Fukui2014ApJ...780}.
Recent studies suggest that the efficiency of massive star formation strongly depends on the collision speed.  \citet{Takahira:2014cq,Takahira:2018gx} performed hydrodynamical simulations of CCCs. They found that massive clumps ($\sim 100 M_\odot$) finally form in the case of slower CCCs ($\sim 5~\rm km~s^{-1}$). However, in faster CCCs ($ > 10 ~\rm km~s^{-1}$), clouds are highly compressed, but the duration of the collision is not long enough for the clump mass to grow via gas accretion and no massive clump is formed. Based on a high resolution ($\sim$ a few pc) 3D hydrodynamical simulations aiming at modeling a barred galaxy M83, \citet{Fujimoto:2014eb} found that collision speed among clouds in bar regions is faster that in arm regions and proposed this difference makes SFE variations.

CO observations towards NGC~1300 with a single-dish telescope of Nobeyama 45-m show the higher velocity dispersion in bar regions than in arm regions at a kpc-scale resolution \citep{Maeda:2018bg}. 
This result suggests
the relative velocity among the clouds in the bar regions is larger than that in the arm regions and 
is qualitatively consistent with the fast CCC scenario in the bar region.
Other CO observations towards barred galaxies with a single-dish telescope show the same tendency \citep[e.g.,][]{Regan1999,MorokumaMatusi2015PASJ,Muraoka:2016ip,Yajima:2019do}. \citet{Egusa:2018hq} finds that a probability distribution function of velocity dispersion in the bar of M83 is systematically larger than in the arm \citep[see also][]{Sun:2018ib}.
\citet{Querejeta:2019jl} found a significant anti-correlation between the SFE and velocity dispersion of the dense gas at a 100 pc scale in M51. The velocity dispersion of the dense gas is thought to be largely reflecting velocity dispersion among different clumps. According to these results, the difference in SFE between the arm and bar regions would be due to the difference in CCC speed rather than the difference in the physical properties of the GMCs.

Fujimoto et al. (2019, submitted) present  a hydrodynamical simulation of a strongly barred galaxy, using a stellar potential model of NGC~1300. They found that there is no significant environmental dependence of cloud properties including the virial parameter, which is consistent with our result presented in this paper. Further, they show that the collision speed in the bar is significantly faster than the other region due to the elongated global gas motion by the stellar bar.
The fraction of colliding clouds with collision speed
more than $20~\rm km~s^{-1}$ in bar regions ($\sim$ 40\%) is significantly larger than those in other regions ($<$ 10\%).
They conclude that the physical mechanism that causes the lack of active star-forming regions in the bar of strongly barred galaxies is the high-speed CCCs.

The CCC occurs only a few times within 1 Myr in the bar and arm region and the effects of collision do not last for long time; the excited internal gas motion induced by the collision decays quickly within at most a few Myr, which is thought to be shorter than cloud lifetimes \citep[10-40~Myr; e.g.,][]{Kawamura2009ApJS,Meidt2015ApJ,Chevance2019MNRAS,Fujimoto2019MNRAS}. Therefore, in a snapshot of the galaxy, there are a few or less GMCs affected by the collision in the bar and arm regions.
Thus the GMC properties observed would not be affected by this effect.

As a next step of this study, we should directly compare the simulation of Fujimoto et al. (2019, submitted) and our observational data to investigate the relative velocity among GMCs we detected.
Although direct measurement of collision speed from observation data is difficult, the velocity deviation between the GMCs and their surrounding GMCs can be an observable indicator of the collision speed of clouds.
Fujimoto et al. (2019, submitted) found that the velocity deviation in the bar region is larger than that in the arm region, which reflects the fast CCCs in the bar.
Whether the same tendency is seen in NGC 1300 or not will be clarified by further investigations using our GMC catalog (Maeda et al. 2020, in preparation).

\section{Summary}\label{sec:summary}
We made $^{12}$CO($1-0$) observations towards the strongly barred galaxy NGC~1300 at a high angular resolution about 40 pc with ALMA. We detected CO emissions from the western arm to the bar region. Using the CPROPS algorithm, we identified 233 GMCs (34, 119, and 49 in {\it Bar}, {\it Arm}, and {\it Bar-end}, respectively) with S/N $> 4$. We measured $R$, $\sigma_v$, and $L_{\rm CO}$ of these GMCs and then derived  $M_{\rm mol}$, $M_{\rm vir}$, $\Sigma_{\rm mol}$, $\alpha_{\rm vir}$, and $c$ of them. We focus on a mass completed sample with $M_{\rm mol} > 2.0 \times 10^5~M_\odot$ for the investigation of $T_{\rm peak}$, $\sigma_v$, and $M_{\rm mol}$ and a resolved sample with  $M_{\rm mol} > 5.0 \times 10^5~M_\odot$ and $R > 15$ pc for the investigation of $R$, $M_{\rm vir}$, $\Sigma_{\rm mol}$, $\alpha_{\rm vir}$, and $c$. We compare the GMCs properties among the environments. The main results are as follows:

\begin{enumerate}
    \item Based on the two-sided K-S tests, there is a significant environmental variation in the $T_{\rm peak}$; the highest value in {\it Bar-end} followed by {\it Arm} and {\it Bar}. However, there is hardly any significant variations in GMC physical properties ($\sigma_v$, $R$, $M_{\rm mol}$, $M_{\rm vir}$, $\Sigma_{\rm mol}$, $\alpha_{\rm vir}$, and $c$; Fig. ~\ref{fig:boxplot}); (marginally) significant difference is only seen in the distribution of $M_{\rm mol}$ between {\it Bar} and {\it Bar-end}, and {\it Arm} and {\it Bar-end} (Table~\ref{tab:KStest}). The properties of GMCs in NGC~1300 are roughly comparable to those in M51 spiral arms. In particular, the properties  in {\it Bar-end} are very similar.


    \item We find no obvious $R-\sigma_v$ relation although the majority of the data points lies around the Galactic fit (Fig. ~\ref{fig:R_sigv}). For the relation between $M_{\rm vir}$ and  $M_{\rm mol}$, there is a moderate correlation in {\it Arm} and {\it Bar-end}, while there is no apparent correlation in {\it Bar} (Fig. ~\ref{fig:Lco_Mvir}). We find the $\alpha_{\rm vir}$ decreases with increasing $M_{\rm mol}$, which suggests the high mass GMCs tend to be strongly bound as seen in M51. There is a moderate correlation between $R$ and $M_{\rm mol}$ in each environment (Fig. ~\ref{fig:R_Lco}). Further, we find the $P_{\rm int}/k$ is roughly constant across the different environments in NGC~1300 (Fig. ~\ref{fig:Smol_Scale}).
   
    \item No clear difference is found in the shape of GMC mass spectra  between in {\it Bar} and {\it Arm}.  The slope of the spectrum in the {\it Bar-end} is slightly  flatter than those in {\it Arm} and {\it Bar}, and massive GMCs are seen only in the {\it Bar-end}  (Fig. ~\ref{fig:CMF}). The similarity of the feature between {\it Bar-end} and M51 spiral arms suggests the mechanism regulating formation and destruction of GMCs is similar to that in M51 spiral arms.  

    \item It appears that systematic differences in the physical properties of the GMCs are not the cause for the low SFE in the bar region. Other mechanisms such as fast CCCs may control the SFE of GMCs in NGC 1300 (Section \ref{sec:discussion}).
    
\end{enumerate}

\section*{Acknowledgements}
We would like to thank the referee for useful comments and suggestions. We are grateful to K. Nakanishi, F. Egusa, Y. Miyamoto, K. Saigo, R. Kawabe and the staff at the ALMA Regional Center for their help in data reduction. F.M. is supported by Research Fellowship for Young Scientists from the Japan Society of the Promotion of Science (JSPS). K.O. is supported by Grants-in-Aids for Scientific Research (C) (16K05294 and 19K03928) from JSPS.
A.H. is funded by the JSPS KAKENHI Grant Number JP19K03923.
This paper makes use of the following ALMA data: ADS/JAO.ALMA\#2017.1.00248.S. ALMA is a partnership of ESO (representing its member states), NSF (USA) and NINS (Japan), together with NRC (Canada), MOST and ASIAA (Taiwan), and KASI (Republic of Korea), in cooperation with the Republic of Chile. The Joint ALMA Observatory is operated by ESO, AUI/NRAO and NAOJ.

\bibliographystyle{mnras}
\bibliography{Reference} 

\begin{thebibliography}{}
\makeatletter
\relax
\def\mn@urlcharsother{\let\do\@makeother \do\$\do\&\do\#\do\^\do\_\do\%\do\~}
\def\mn@doi{\begingroup\mn@urlcharsother \@ifnextchar [ {\mn@doi@}
  {\mn@doi@[]}}
\def\mn@doi@[#1]#2{\def\@tempa{#1}\ifx\@tempa\@empty \href
  {http://dx.doi.org/#2} {doi:#2}\else \href {http://dx.doi.org/#2} {#1}\fi
  \endgroup}
\def\mn@eprint#1#2{\mn@eprint@#1:#2::\@nil}
\def\mn@eprint@arXiv#1{\href {http://arxiv.org/abs/#1} {{\tt arXiv:#1}}}
\def\mn@eprint@dblp#1{\href {http://dblp.uni-trier.de/rec/bibtex/#1.xml}
  {dblp:#1}}
\def\mn@eprint@#1:#2:#3:#4\@nil{\def\@tempa {#1}\def\@tempb {#2}\def\@tempc
  {#3}\ifx \@tempc \@empty \let \@tempc \@tempb \let \@tempb \@tempa \fi \ifx
  \@tempb \@empty \def\@tempb {arXiv}\fi \@ifundefined
  {mn@eprint@\@tempb}{\@tempb:\@tempc}{\expandafter \expandafter \csname
  mn@eprint@\@tempb\endcsname \expandafter{\@tempc}}}

\bibitem[\protect\citeauthoryear{{Athanassoula}}{{Athanassoula}}{1992}]{Athanassoula1992MNRAS}
{Athanassoula} E.,  1992, \mn@doi [\mnras] {10.1093/mnras/259.2.345}, \href
  {https://ui.adsabs.harvard.edu/abs/1992MNRAS.259..345A} {259, 345}

\bibitem[\protect\citeauthoryear{{Bertoldi} \& {McKee}}{{Bertoldi} \&
  {McKee}}{1992}]{BertoldiandMckee}
{Bertoldi} F.,  {McKee} C.~F.,  1992, \mn@doi [\apj] {10.1086/171638}, \href
  {https://ui.adsabs.harvard.edu/abs/1992ApJ...395..140B} {395, 140}

\bibitem[\protect\citeauthoryear{{Bigiel}, {Leroy}, {Walter}, {Brinks}, {de
  Blok}, {Madore}  \& {Thornley}}{{Bigiel} et~al.}{2008}]{Bigiel2008AJ}
{Bigiel} F.,  {Leroy} A.,  {Walter} F.,  {Brinks} E.,  {de Blok} W.~J.~G.,
  {Madore} B.,   {Thornley} M.~D.,  2008, \mn@doi [\aj]
  {10.1088/0004-6256/136/6/2846}, \href
  {https://ui.adsabs.harvard.edu/abs/2008AJ....136.2846B} {136, 2846}

\bibitem[\protect\citeauthoryear{{Bolatto}, {Leroy}, {Rosolowsky}, {Walter}  \&
  {Blitz}}{{Bolatto} et~al.}{2008}]{Bolatto2008ApJ}
{Bolatto} A.~D.,  {Leroy} A.~K.,  {Rosolowsky} E.,  {Walter} F.,   {Blitz} L.,
  2008, \mn@doi [\apj] {10.1086/591513}, \href
  {https://ui.adsabs.harvard.edu/abs/2008ApJ...686..948B} {686, 948}

\bibitem[\protect\citeauthoryear{{Chevance} et~al.,}{{Chevance}
  et~al.}{2019}]{Chevance2019MNRAS}
{Chevance} M.,  et~al., 2019, \mn@doi [\mnras] {10.1093/mnras/stz3525}, \href
  {https://ui.adsabs.harvard.edu/abs/2019MNRAS.tmp.3155C} {p.~3155}

\bibitem[\protect\citeauthoryear{{Colombo} et~al.,}{{Colombo}
  et~al.}{2014}]{Colombo:2014ei}
{Colombo} D.,  et~al., 2014, \mn@doi [\apj] {10.1088/0004-637X/784/1/3}, \href
  {https://ui.adsabs.harvard.edu/abs/2014ApJ...784....3C} {784, 3}

\bibitem[\protect\citeauthoryear{{Cornwell}}{{Cornwell}}{2008}]{Cornwell2008}
{Cornwell} T.~J.,  2008, \mn@doi [IEEE Journal of Selected Topics in Signal
  Processing] {10.1109/JSTSP.2008.2006388}, \href
  {https://ui.adsabs.harvard.edu/abs/2008ISTSP...2..793C} {2, 793}

\bibitem[\protect\citeauthoryear{{Dobbs}}{{Dobbs}}{2008}]{Dobbs:2008ez}
{Dobbs} C.~L.,  2008, \mn@doi [\mnras] {10.1111/j.1365-2966.2008.13939.x},
  \href {https://ui.adsabs.harvard.edu/abs/2008MNRAS.391..844D} {391, 844}

\bibitem[\protect\citeauthoryear{{Egusa}, {Hirota}, {Baba}  \&
  {Muraoka}}{{Egusa} et~al.}{2018}]{Egusa:2018hq}
{Egusa} F.,  {Hirota} A.,  {Baba} J.,   {Muraoka} K.,  2018, \mn@doi [\apj]
  {10.3847/1538-4357/aaa76d}, \href
  {https://ui.adsabs.harvard.edu/abs/2018ApJ...854...90E} {854, 90}

\bibitem[\protect\citeauthoryear{{England}}{{England}}{1989}]{England1989a}
{England} M.~N.,  1989, \mn@doi [\apj] {10.1086/167097}, \href
  {https://ui.adsabs.harvard.edu/abs/1989ApJ...337..191E} {337, 191}

\bibitem[\protect\citeauthoryear{{Faesi}, {Lada}  \& {Forbrich}}{{Faesi}
  et~al.}{2018}]{Faesi:2018gg}
{Faesi} C.~M.,  {Lada} C.~J.,   {Forbrich} J.,  2018, \mn@doi [\apj]
  {10.3847/1538-4357/aaad60}, \href
  {https://ui.adsabs.harvard.edu/abs/2018ApJ...857...19F} {857, 19}

\bibitem[\protect\citeauthoryear{{Fujimoto}, {Tasker}  \& {Habe}}{{Fujimoto}
  et~al.}{2014}]{Fujimoto:2014eb}
{Fujimoto} Y.,  {Tasker} E.~J.,   {Habe} A.,  2014, \mn@doi [\mnras]
  {10.1093/mnrasl/slu138}, \href
  {https://ui.adsabs.harvard.edu/abs/2014MNRAS.445L..65F} {445, L65}

\bibitem[\protect\citeauthoryear{{Fujimoto}, {Chevance}, {Haydon}, {Krumholz}
  \& {Kruijssen}}{{Fujimoto} et~al.}{2019}]{Fujimoto2019MNRAS}
{Fujimoto} Y.,  {Chevance} M.,  {Haydon} D.~T.,  {Krumholz} M.~R.,
  {Kruijssen} J.~M.~D.,  2019, \mn@doi [\mnras] {10.1093/mnras/stz641}, \href
  {https://ui.adsabs.harvard.edu/abs/2019MNRAS.487.1717F} {487, 1717}

\bibitem[\protect\citeauthoryear{{Fukui}, {Mizuno}, {Yamaguchi}, {Mizuno}  \&
  {Onishi}}{{Fukui} et~al.}{2001}]{Fukui2001PASJ}
{Fukui} Y.,  {Mizuno} N.,  {Yamaguchi} R.,  {Mizuno} A.,   {Onishi} T.,  2001,
  \mn@doi [\pasj] {10.1093/pasj/53.6.L41}, \href
  {https://ui.adsabs.harvard.edu/abs/2001PASJ...53L..41F} {53, L41}

\bibitem[\protect\citeauthoryear{{Fukui} et~al.,}{{Fukui}
  et~al.}{2008}]{Fukui2008ApJS}
{Fukui} Y.,  et~al., 2008, \mn@doi [\apjs] {10.1086/589833}, \href
  {https://ui.adsabs.harvard.edu/abs/2008ApJS..178...56F} {178, 56}

\bibitem[\protect\citeauthoryear{{Fukui} et~al.,}{{Fukui}
  et~al.}{2014}]{Fukui2014ApJ...780}
{Fukui} Y.,  et~al., 2014, \mn@doi [\apj] {10.1088/0004-637X/780/1/36}, \href
  {https://ui.adsabs.harvard.edu/abs/2014ApJ...780...36F} {780, 36}

\bibitem[\protect\citeauthoryear{{Gallagher} et~al.,}{{Gallagher}
  et~al.}{2018}]{Gallagher:2018ep}
{Gallagher} M.~J.,  et~al., 2018, \mn@doi [\apj] {10.3847/1538-4357/aabad8},
  \href {https://ui.adsabs.harvard.edu/abs/2018ApJ...858...90G} {858, 90}

\bibitem[\protect\citeauthoryear{{Gratier} et~al.,}{{Gratier}
  et~al.}{2012}]{Gratier2012A&A}
{Gratier} P.,  et~al., 2012, \mn@doi [\aap] {10.1051/0004-6361/201116612},
  \href {https://ui.adsabs.harvard.edu/abs/2012A&A...542A.108G} {542, A108}

\bibitem[\protect\citeauthoryear{{Habe} \& {Ohta}}{{Habe} \&
  {Ohta}}{1992}]{Habe1992PASJ}
{Habe} A.,  {Ohta} K.,  1992, \pasj, \href
  {https://ui.adsabs.harvard.edu/abs/1992PASJ...44..203H} {44, 203}

\bibitem[\protect\citeauthoryear{{Heyer}, {Krawczyk}, {Duval}  \&
  {Jackson}}{{Heyer} et~al.}{2009}]{Heyer2009ApJ}
{Heyer} M.,  {Krawczyk} C.,  {Duval} J.,   {Jackson} J.~M.,  2009, \mn@doi
  [\apj] {10.1088/0004-637X/699/2/1092}, \href
  {https://ui.adsabs.harvard.edu/abs/2009ApJ...699.1092H} {699, 1092}

\bibitem[\protect\citeauthoryear{{Hirota} et~al.,}{{Hirota}
  et~al.}{2014}]{Hirota:2014bt}
{Hirota} A.,  et~al., 2014, \mn@doi [\pasj] {10.1093/pasj/psu006}, \href
  {https://ui.adsabs.harvard.edu/abs/2014PASJ...66...46H} {66, 46}

\bibitem[\protect\citeauthoryear{{Hirota} et~al.,}{{Hirota}
  et~al.}{2018}]{Hirota:2018jp}
{Hirota} A.,  et~al., 2018, \mn@doi [\pasj] {10.1093/pasj/psy071}, \href
  {https://ui.adsabs.harvard.edu/abs/2018PASJ...70...73H} {70, 73}

\bibitem[\protect\citeauthoryear{{Hughes} et~al.,}{{Hughes}
  et~al.}{2013}]{Hughes2013ApJ}
{Hughes} A.,  et~al., 2013, \mn@doi [\apj] {10.1088/0004-637X/779/1/46}, \href
  {https://ui.adsabs.harvard.edu/abs/2013ApJ...779...46H} {779, 46}

\bibitem[\protect\citeauthoryear{{Kawamura} et~al.,}{{Kawamura}
  et~al.}{2009}]{Kawamura2009ApJS}
{Kawamura} A.,  et~al., 2009, \mn@doi [\apjs] {10.1088/0067-0049/184/1/1},
  \href {https://ui.adsabs.harvard.edu/abs/2009ApJS..184....1K} {184, 1}

\bibitem[\protect\citeauthoryear{{Kennicutt}}{{Kennicutt}}{1998}]{Kennicutt1998ARA&A}
{Kennicutt} Robert~C. J.,  1998, \mn@doi [\araa]
  {10.1146/annurev.astro.36.1.189}, \href
  {https://ui.adsabs.harvard.edu/abs/1998ARA&A..36..189K} {36, 189}

\bibitem[\protect\citeauthoryear{{Kobayashi}, {Inutsuka}, {Kobayashi}  \&
  {Hasegawa}}{{Kobayashi} et~al.}{2017}]{Kobayashi2017ApJ}
{Kobayashi} M. I.~N.,  {Inutsuka} S.-i.,  {Kobayashi} H.,   {Hasegawa} K.,
  2017, \mn@doi [\apj] {10.3847/1538-4357/836/2/175}, \href
  {https://ui.adsabs.harvard.edu/abs/2017ApJ...836..175K} {836, 175}

\bibitem[\protect\citeauthoryear{{Larson}}{{Larson}}{1981}]{Larson1981}
{Larson} R.~B.,  1981, \mn@doi [\mnras] {10.1093/mnras/194.4.809}, \href
  {https://ui.adsabs.harvard.edu/abs/1981MNRAS.194..809L} {194, 809}

\bibitem[\protect\citeauthoryear{{Leroy} et~al.,}{{Leroy}
  et~al.}{2013}]{Leroy2013AJ.146}
{Leroy} A.~K.,  et~al., 2013, \mn@doi [\aj] {10.1088/0004-6256/146/2/19}, \href
  {https://ui.adsabs.harvard.edu/abs/2013AJ....146...19L} {146, 19}

\bibitem[\protect\citeauthoryear{{Leroy} et~al.,}{{Leroy}
  et~al.}{2015}]{Leroy:2015ds}
{Leroy} A.~K.,  et~al., 2015, \mn@doi [\apj] {10.1088/0004-637X/801/1/25},
  \href {https://ui.adsabs.harvard.edu/abs/2015ApJ...801...25L} {801, 25}

\bibitem[\protect\citeauthoryear{{Longmore} et~al.,}{{Longmore}
  et~al.}{2013}]{Longmore2013MNRAS.429}
{Longmore} S.~N.,  et~al., 2013, \mn@doi [\mnras] {10.1093/mnras/sts376}, \href
  {https://ui.adsabs.harvard.edu/abs/2013MNRAS.429..987L} {429, 987}

\bibitem[\protect\citeauthoryear{{Maeda}, {Ohta}, {Fujimoto}, {Habe}  \&
  {Baba}}{{Maeda} et~al.}{2018}]{Maeda:2018bg}
{Maeda} F.,  {Ohta} K.,  {Fujimoto} Y.,  {Habe} A.,   {Baba} J.,  2018, \mn@doi
  [\pasj] {10.1093/pasj/psy028}, \href
  {https://ui.adsabs.harvard.edu/abs/2018PASJ...70...37M} {70, 37}

\bibitem[\protect\citeauthoryear{{Meidt} et~al.,}{{Meidt}
  et~al.}{2015}]{Meidt2015ApJ}
{Meidt} S.~E.,  et~al., 2015, \mn@doi [\apj] {10.1088/0004-637X/806/1/72},
  \href {https://ui.adsabs.harvard.edu/abs/2015ApJ...806...72M} {806, 72}

\bibitem[\protect\citeauthoryear{{Momose}, {Okumura}, {Koda}  \&
  {Sawada}}{{Momose} et~al.}{2010}]{Momose2010ApJ}
{Momose} R.,  {Okumura} S.~K.,  {Koda} J.,   {Sawada} T.,  2010, \mn@doi [\apj]
  {10.1088/0004-637X/721/1/383}, \href
  {https://ui.adsabs.harvard.edu/abs/2010ApJ...721..383M} {721, 383}

\bibitem[\protect\citeauthoryear{{Morokuma-Matsui}, {Sorai}, {Watanabe}  \&
  {Kuno}}{{Morokuma-Matsui} et~al.}{2015}]{MorokumaMatusi2015PASJ}
{Morokuma-Matsui} K.,  {Sorai} K.,  {Watanabe} Y.,   {Kuno} N.,  2015, \mn@doi
  [\pasj] {10.1093/pasj/psu126}, \href
  {https://ui.adsabs.harvard.edu/abs/2015PASJ...67....2M} {67, 2}

\bibitem[\protect\citeauthoryear{{Mould} et~al.,}{{Mould}
  et~al.}{2000}]{MouldEtAl00}
{Mould} J.~R.,  et~al., 2000, \mn@doi [\apj] {10.1086/308304}, \href
  {https://ui.adsabs.harvard.edu/abs/2000ApJ...529..786M} {529, 786}

\bibitem[\protect\citeauthoryear{{Muraoka} et~al.,}{{Muraoka}
  et~al.}{2016}]{Muraoka:2016ip}
{Muraoka} K.,  et~al., 2016, \mn@doi [\pasj] {10.1093/pasj/psw080}, \href
  {https://ui.adsabs.harvard.edu/abs/2016PASJ...68...89M} {68, 89}

\bibitem[\protect\citeauthoryear{{Nimori}, {Habe}, {Sorai}, {Watanabe},
  {Hirota}  \& {Namekata}}{{Nimori} et~al.}{2013}]{Nimori2013MNRAS}
{Nimori} M.,  {Habe} A.,  {Sorai} K.,  {Watanabe} Y.,  {Hirota} A.,
  {Namekata} D.,  2013, \mn@doi [\mnras] {10.1093/mnras/sts487}, \href
  {https://ui.adsabs.harvard.edu/abs/2013MNRAS.429.2175N} {429, 2175}

\bibitem[\protect\citeauthoryear{{Pan} \& {Kuno}}{{Pan} \&
  {Kuno}}{2017}]{Pan:2017gy}
{Pan} H.-A.,  {Kuno} N.,  2017, \mn@doi [\apj] {10.3847/1538-4357/aa60c2},
  \href {https://ui.adsabs.harvard.edu/abs/2017ApJ...839..133P} {839, 133}

\bibitem[\protect\citeauthoryear{{Pety} et~al.,}{{Pety}
  et~al.}{2013}]{Pety:2013fw}
{Pety} J.,  et~al., 2013, \mn@doi [\apj] {10.1088/0004-637X/779/1/43}, \href
  {https://ui.adsabs.harvard.edu/abs/2013ApJ...779...43P} {779, 43}

\bibitem[\protect\citeauthoryear{{Querejeta} et~al.,}{{Querejeta}
  et~al.}{2019}]{Querejeta:2019jl}
{Querejeta} M.,  et~al., 2019, \mn@doi [\aap] {10.1051/0004-6361/201834915},
  \href {https://ui.adsabs.harvard.edu/abs/2019A&A...625A..19Q} {625, A19}

\bibitem[\protect\citeauthoryear{{Regan}, {Sheth}  \& {Vogel}}{{Regan}
  et~al.}{1999}]{Regan1999}
{Regan} M.~W.,  {Sheth} K.,   {Vogel} S.~N.,  1999, \mn@doi [\apj]
  {10.1086/307960}, \href
  {https://ui.adsabs.harvard.edu/abs/1999ApJ...526...97R} {526, 97}

\bibitem[\protect\citeauthoryear{{Reynaud} \& {Downes}}{{Reynaud} \&
  {Downes}}{1998}]{ReynaudDownes1998A&A}
{Reynaud} D.,  {Downes} D.,  1998, \aap, \href
  {https://ui.adsabs.harvard.edu/abs/1998A&A...337..671R} {337, 671}

\bibitem[\protect\citeauthoryear{{Rich}, {de Blok}, {Cornwell}, {Brinks},
  {Walter}, {Bagetakos}  \& {Kennicutt}}{{Rich} et~al.}{2008}]{Rich2008AJ}
{Rich} J.~W.,  {de Blok} W.~J.~G.,  {Cornwell} T.~J.,  {Brinks} E.,  {Walter}
  F.,  {Bagetakos} I.,   {Kennicutt} R.~C. J.,  2008, \mn@doi [\aj]
  {10.1088/0004-6256/136/6/2897}, \href
  {https://ui.adsabs.harvard.edu/abs/2008AJ....136.2897R} {136, 2897}

\bibitem[\protect\citeauthoryear{{Rosolowsky}}{{Rosolowsky}}{2007}]{Rosolowsky2007ApJ}
{Rosolowsky} E.,  2007, \mn@doi [\apj] {10.1086/509249}, \href
  {https://ui.adsabs.harvard.edu/abs/2007ApJ...654..240R} {654, 240}

\bibitem[\protect\citeauthoryear{{Rosolowsky} \& {Leroy}}{{Rosolowsky} \&
  {Leroy}}{2006}]{RosolowskyLeroy}
{Rosolowsky} E.,  {Leroy} A.,  2006, \mn@doi [\pasp] {10.1086/502982}, \href
  {https://ui.adsabs.harvard.edu/abs/2006PASP..118..590R} {118, 590}

\bibitem[\protect\citeauthoryear{{Sandage} \& {Tammann}}{{Sandage} \&
  {Tammann}}{1981}]{Sandae_Tammann}
{Sandage} A.,  {Tammann} G.~A.,  1981, {A Revised Shapley-Ames Catalog of
  Bright Galaxies}

\bibitem[\protect\citeauthoryear{{Schinnerer} et~al.,}{{Schinnerer}
  et~al.}{2013}]{Schinnerer:2013jy}
{Schinnerer} E.,  et~al., 2013, \mn@doi [\apj] {10.1088/0004-637X/779/1/42},
  \href {https://ui.adsabs.harvard.edu/abs/2013ApJ...779...42S} {779, 42}

\bibitem[\protect\citeauthoryear{{Schruba} et~al.,}{{Schruba}
  et~al.}{2011}]{Schruba2011AJ}
{Schruba} A.,  et~al., 2011, \mn@doi [\aj] {10.1088/0004-6256/142/2/37}, \href
  {https://ui.adsabs.harvard.edu/abs/2011AJ....142...37S} {142, 37}

\bibitem[\protect\citeauthoryear{{Solomon}, {Rivolo}, {Barrett}  \&
  {Yahil}}{{Solomon} et~al.}{1987}]{Solomon87}
{Solomon} P.~M.,  {Rivolo} A.~R.,  {Barrett} J.,   {Yahil} A.,  1987, \mn@doi
  [\apj] {10.1086/165493}, \href
  {https://ui.adsabs.harvard.edu/abs/1987ApJ...319..730S} {319, 730}

\bibitem[\protect\citeauthoryear{{Sorai} et~al.,}{{Sorai}
  et~al.}{2012}]{Sorai2012PASJ}
{Sorai} K.,  et~al., 2012, \mn@doi [\pasj] {10.1093/pasj/64.3.51}, \href
  {https://ui.adsabs.harvard.edu/abs/2012PASJ...64...51S} {64, 51}

\bibitem[\protect\citeauthoryear{Stephens}{Stephens}{1970}]{Stephens:1970ic}
Stephens M.~A.,  1970, Journal of the Royal Statistical Society: Series B
  (Methodological), 32, 115

\bibitem[\protect\citeauthoryear{{Sun} et~al.,}{{Sun}
  et~al.}{2018}]{Sun:2018ib}
{Sun} J.,  et~al., 2018, \mn@doi [\apj] {10.3847/1538-4357/aac326}, \href
  {https://ui.adsabs.harvard.edu/abs/2018ApJ...860..172S} {860, 172}

\bibitem[\protect\citeauthoryear{{Takahira}, {Tasker}  \& {Habe}}{{Takahira}
  et~al.}{2014}]{Takahira:2014cq}
{Takahira} K.,  {Tasker} E.~J.,   {Habe} A.,  2014, \mn@doi [\apj]
  {10.1088/0004-637X/792/1/63}, \href
  {https://ui.adsabs.harvard.edu/abs/2014ApJ...792...63T} {792, 63}

\bibitem[\protect\citeauthoryear{{Takahira}, {Shima}, {Habe}  \&
  {Tasker}}{{Takahira} et~al.}{2018}]{Takahira:2018gx}
{Takahira} K.,  {Shima} K.,  {Habe} A.,   {Tasker} E.~J.,  2018, \mn@doi
  [\pasj] {10.1093/pasj/psy011}, \href
  {https://ui.adsabs.harvard.edu/abs/2018PASJ...70S..58T} {70, S58}

\bibitem[\protect\citeauthoryear{{Tubbs}}{{Tubbs}}{1982}]{Tubbs1982ApJ}
{Tubbs} A.~D.,  1982, \mn@doi [\apj] {10.1086/159846}, \href
  {https://ui.adsabs.harvard.edu/abs/1982ApJ...255..458T} {255, 458}

\bibitem[\protect\citeauthoryear{{Usero} et~al.,}{{Usero}
  et~al.}{2015}]{Usero2015AJ}
{Usero} A.,  et~al., 2015, \mn@doi [\aj] {10.1088/0004-6256/150/4/115}, \href
  {https://ui.adsabs.harvard.edu/abs/2015AJ....150..115U} {150, 115}

\bibitem[\protect\citeauthoryear{{Watanabe}, {Sorai}, {Kuno}  \&
  {Habe}}{{Watanabe} et~al.}{2011}]{Watanabe2011}
{Watanabe} Y.,  {Sorai} K.,  {Kuno} N.,   {Habe} A.,  2011, \mn@doi [\mnras]
  {10.1111/j.1365-2966.2010.17746.x}, \href
  {https://ui.adsabs.harvard.edu/abs/2011MNRAS.411.1409W} {411, 1409}

\bibitem[\protect\citeauthoryear{{Williams} \& {McKee}}{{Williams} \&
  {McKee}}{1997}]{WilliamsMcKee1997ApJ}
{Williams} J.~P.,  {McKee} C.~F.,  1997, \mn@doi [\apj] {10.1086/303588}, \href
  {https://ui.adsabs.harvard.edu/abs/1997ApJ...476..166W} {476, 166}

\bibitem[\protect\citeauthoryear{{Yajima} et~al.,}{{Yajima}
  et~al.}{2019}]{Yajima:2019do}
{Yajima} Y.,  et~al., 2019, \mn@doi [\pasj] {10.1093/pasj/psz022}, \href
  {https://ui.adsabs.harvard.edu/abs/2019PASJ..tmp...41Y} {p.~41}

\makeatother
\end{thebibliography}

\appendix

\section{CPROPS bias}\label{apx:CPROPS}
CPROPS corrects for the sensitivity by extrapolating GMC properties to those we would expect to measure with perfect sensitivity (i.e., 0 K). As described in section~\ref{sec:derivation of R}, the extrapolated $R$, $\sigma_v$ and $L_{\rm CO}$ is typically $1.5 \sim 2.0$ times higher than  the directly measured values. CPROPS also corrects for the resolution by  deconvolution for beam and channel width. The corrected $R$, $\sigma_v$ is typically by a factor of $0.7 \sim 0.8$ lower than the extrapolated values. Since the accuracy of these corrections (extrapolation and deconvolution) depends on the sensitivity, spatial resolution and velocity resolution \citep{RosolowskyLeroy}, it is necessary to  assess the reliability of the measurements of GMC properties.

Using CASA, we simulated ALMA observation of mock GMCs, which  are three-dimensional Gaussian clouds with a given $M_{\rm mol}$, $R$, and $\sigma_v$ in a position-position-velocity space. We create a fits image with 100 mock GMCs centered in fixed positions. The $M_{\rm mol}$, $R$, and $\sigma_v$ are randomly determined in the range of $5.3 \leq \log(M_{\rm mol}/[M_\odot]) \leq 6.8$, $1.5\leq \log(R/[{\rm pc}]) \leq 2.2$, and  $0.3 \leq \log(\sigma_v/[{\rm km~s^{-1}}]) \leq 1.1$, respectively. The GMCs are round clouds (i.e., axis ratio = 1.0). Here, we assume the CO-to-H$_2$ conversion factor of $4.4~M_\odot~(\rm K~km~s^{-1}~pc^2)^{-1}$. Then, using the task of \verb|simobserve| in CASA, we simulate observation under the same configuration, pointings, and noise condition of our ALMA observations (see Section~\ref{sec:obs}). After reconstructing the image with the same \verb|tclean| parameters (see Section~\ref{sec:obs}), we identify the GMCs using CPROPS with the same settings described in section~\ref{sec:GMC identification and characterization}. We repeated this procedure 12 times, and then we extracted the GMCs with $4 \leq {\rm S/N} \leq 10$, corresponding to the observed value.

\begin{figure*}
\begin{center}
 \includegraphics[width=150mm]{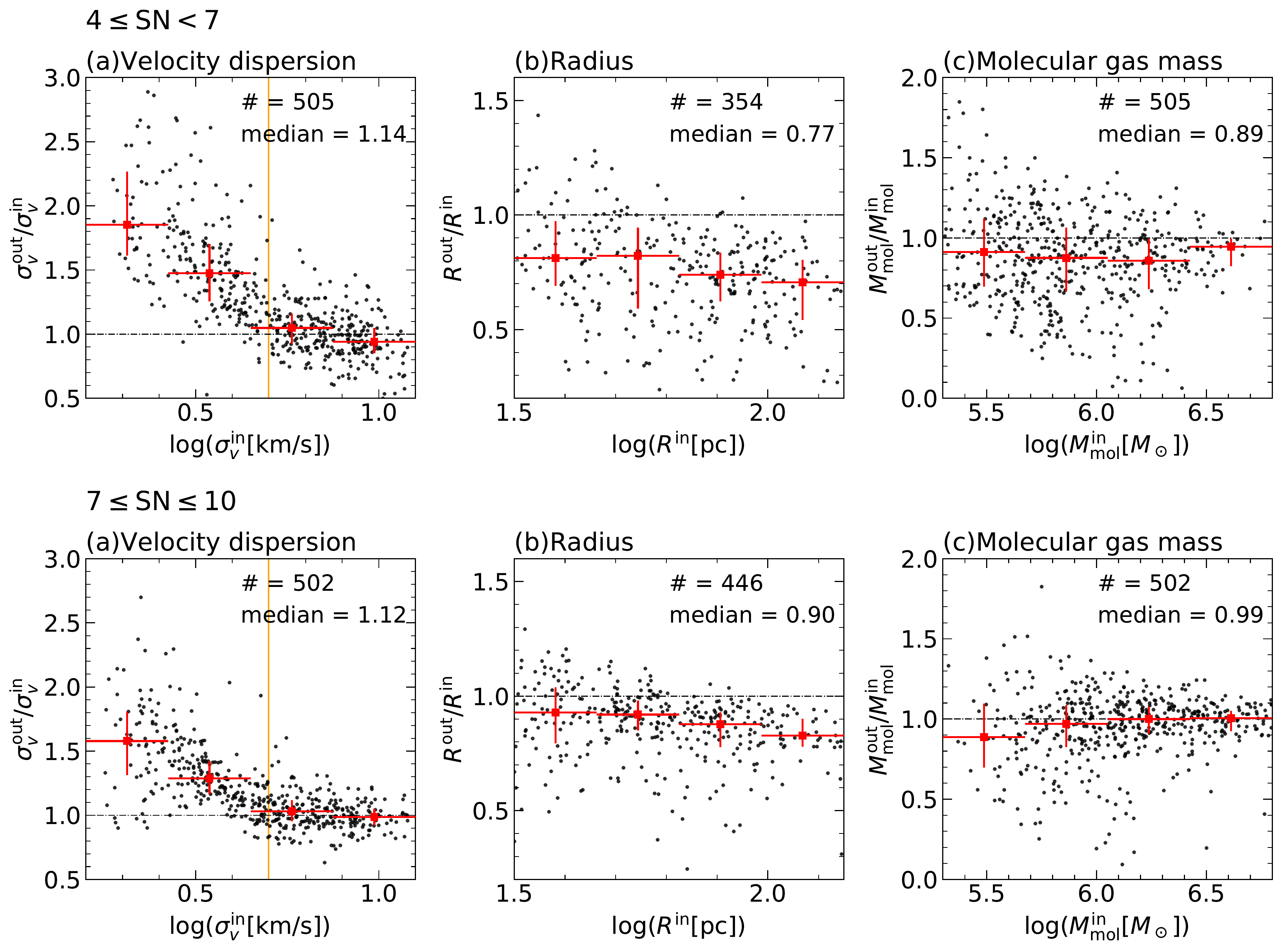}
 \caption{ Results of the  ALMA observation simulation of the mock GMCs. (a): Ratio of the output velocity dispersion by CPROPS ($\sigma_v^{\rm out}$) to the input (model) velocity dispersion ($\sigma_v^{\rm in}$) as a function of the $\sigma_v^{\rm in}$. We shows the GMCs with lower S/N ($4 \leq {\rm S/N} < 7$) and higher S/N  ($7 \leq {\rm S/N} \leq 10$) in upper and lower panel, respectively. Red square shows the median value in  a bin whose range is shown as error bar in x-axis, and a error bar in y-axis shows the $\Delta Q$ in the bin. Orange solid line shows the channel width of $5.0~\rm km~s^{-1}$. Black dash-dotted line shows $\sigma_v^{\rm out}/\sigma_v^{\rm in} =  1.0$. (b),(c): Same as panel (a) but for radius and molecular gas mass.}
 \label{fig:NGC1300_CASA_simulation}
\end{center}
\end{figure*}

Fig. ~\ref{fig:NGC1300_CASA_simulation} shows the results of this simulation. The superscript {\it in} denotes the input value of the mock GMC and  {\it out} denotes the  output value CPROPS measured and corrected. We plot the ratio of output value to input value as a function of input value for $\sigma_v$, $R$, and $M_{\rm mol}$, dividing GMCs into lower  S/N ($4 \leq {\rm S/N} < 7$) and higher S/N  ($7 \leq {\rm S/N} \leq 10$). In panel (a), we find CPROPS measurements work well for the GMCs whose $\sigma_v^{\rm in}$ is larger than $5.0~ \rm km~s^{-1}$. However, for the GMCs with $\sigma_v^{\rm in} \leq 5.0~\rm km~s^{-1}$, CPROPS overestimates $\sigma_v$ by a factor of $\sim 1.5$  regardless of S/N.

\begin{figure*}
\begin{center}
 \includegraphics[width=65mm]{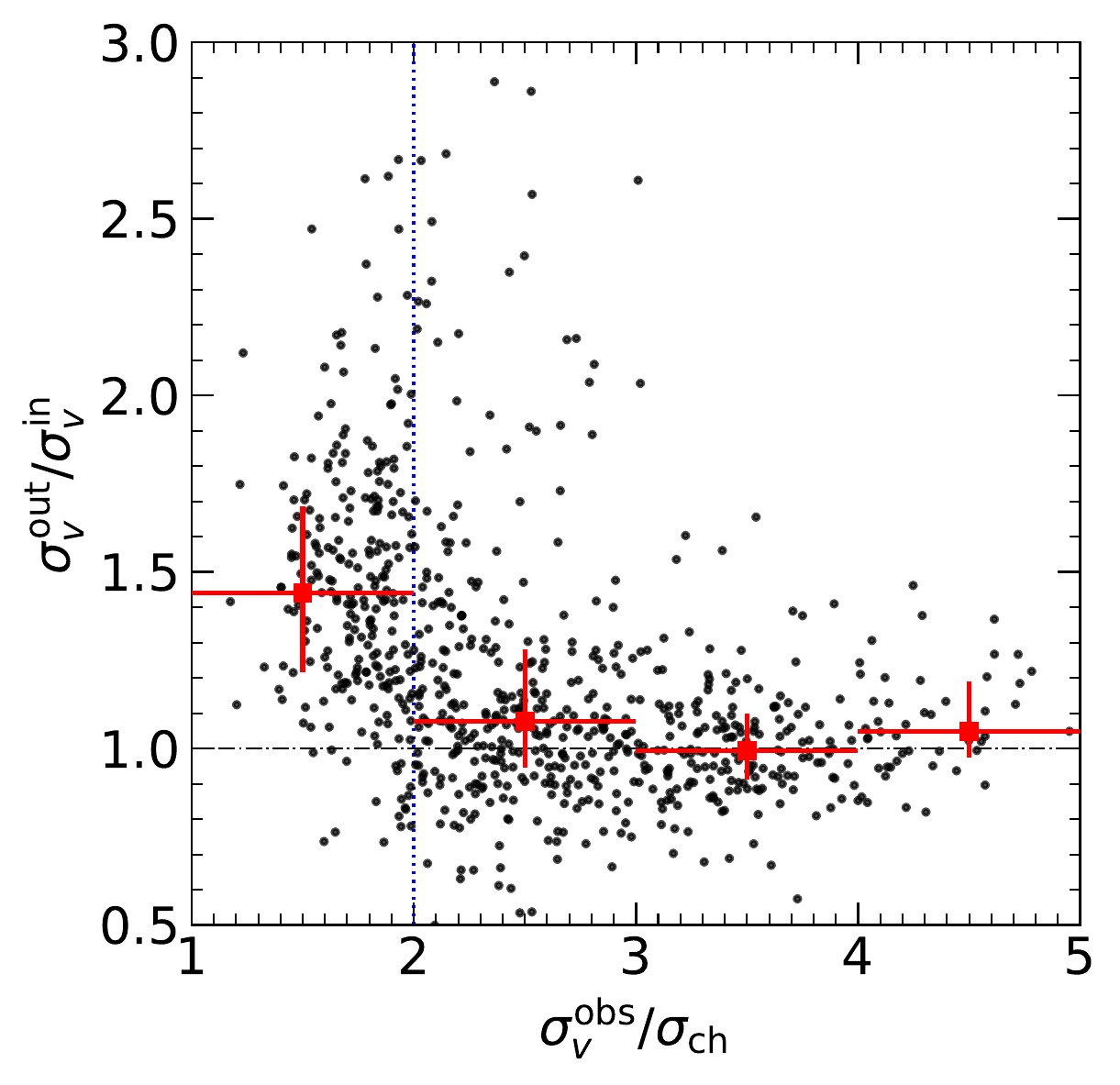}
 \caption{Ratio of the output velocity dispersion by CPROPS ($\sigma_v^{\rm out}$) to the input (model) velocity dispersion ($\sigma_v^{\rm in}$) as a function of the ratio of the velocity dispersion without CPROPS correction ($\sigma_v^{\rm obs}$)  to channel width  ($\sigma_{\rm ch}=\sqrt{\Delta V_{\rm chan}^2/2 \pi} =〜2.0 \rm km~s^{-1}$). Red square shows the median value in  a bin whose range is shown as error bar in x-axis, and a error bar in y-axis shows the $\Delta Q$ in the bin.  Black dash-dotted line shows $\sigma_v^{\rm out}/\sigma_v^{\rm in} =  1.0$. Blue dotted line shows  $\sigma_v^{\rm obs}/\sigma_{\rm ch} =  2.0$.}
 \label{fig:NGC1300_CASA_simulation_vel}
\end{center}
\end{figure*}

According to \citet{RosolowskyLeroy}, the extrapolation works well regardless of the S/N, if the line width of the identified GMC is at least twice the channel width. In Fig. ~\ref{fig:NGC1300_CASA_simulation_vel}, we plot the $\sigma_v^{\rm out}/\sigma_v^{\rm in}$ as a function of the ratio of the velocity dispersion without CPROPS correction ($\sigma_v^{\rm obs}$)  to channel width  ($\sigma_{\rm ch}=\sqrt{\Delta V_{\rm chan}^2/2 \pi} = 2.0~\rm km~s^{-1}$). We find that the $\sigma_v^{\rm out}$ is overestimated by a factor of $\sim 1.5$ if the $\sigma_v^{\rm obs}$ is less than half the channel width. Such GMC accounts for 56~\%, 64~\%, and 64~\% in {\it Bar}, {\it Arm}, and {\it Bar-end}, respectively. Therefore, a large number of the cataloged GMCs may be overestimated in the $\sigma_v$ by a factor of $\sim$ 1.5.

The CPROPS algorithm performs relatively well for the measurements of  the $M_{\rm mol}$ and $R$  in comparison to $\sigma_v$. In panel (b), we find CPROPS slightly  underestimate radius: the $R^{\rm out}$ is typically underestimated by a factor of $\sim 0.8$ and  $\sim 0.9$ for the  GMC with lower S/N  and higher S/N, respectively. In panel (c), the corrected $M_{\rm mol}$ is mostly equal to the input $M_{\rm mol}$ for the GMC with higher S/N, but the corrected $M_{\rm mol}$ is typically  underestimated by a factor of  $\sim 0.9$ for the  GMC with lower S/N. Because about 70~\% of the cataloged GMCs were detected  with $4 \leq {\rm S/N} \leq 7$, $R$ and  $M_{\rm mol}$ may be slightly underestimated by a factor of $\sim 0.8$ and $\sim 0.9$.
Note that the factors of over/underestimation do not change if the GMC's minor-to-major axis ratio is set to be 0.5.

These over/underestimation can propagate to the measurements of $M_{\rm vir}$,$\Sigma_{\rm mol}$,$\alpha_{\rm vir}$, and $c$, which are a combination of $\sigma_v$, $R$, and $M_{\rm mol}$. Thus, we recalculated the GMC properties. We corrected the cataloged $\sigma_v$ by dividing by 1.5 if $\sigma_v^{\rm obs} \geq 2\sigma_{\rm ch}$. The cataloged $R$ is corrected by dividing by 0.8 and 0.9 for the GMCs with lower and higher S/N. For the $M_{\rm mol}$, we corrected by dividing by 0.9 for the GMCs with lower S/N. Then, we recalculated other properties based on the corrected $\sigma_v$, $R$, and $M_{\rm mol}$. Table~\ref{tab:re-GMCproperties} is the same as Tabel~\ref{tab:GMCproperties} but corrected for the over/underestimation. We find the recalculated median $M_{\rm vir}$, $\Sigma_{\rm mol}$, and $c$ become by a factor of $0.7 \sim 0.9$ smaller and the correction factor is roughly comparable in the different environments. The  $\alpha_{\rm vir}$ becomes by a factor of 0.8, 0.5, and 0.8 smaller in {\it Bar}, {\it Arm}, and {\it Bar-end}, respectively. This suggests we overestimate the fraction of gravitationally unbound clouds. Based on the corrected (uncorrected) values, the number fraction is 50~\% (58~\%), 29~\% (33~\%), and 23~\% (29~\%) in  {\it Bar}, {\it Arm}, and {\it Bar-end}, respectively.

We retested the environmental variation of the GMC properties using the corrected catalog. Table~\ref{tab:re-KStest} shows the result of the two-sided K-S test. Comparing the Table~\ref{tab:KStest}, $p$-values do not change much. Therefore, the over/underestimation of the measurements  does not influence on the discussion about the environmental variation described in Section~\ref{sec:Basic properties}. It is notable that $\sim 35$~\% and $\sim 60$~\% of the GMCs in M51 measured by \citet{Colombo:2014ei} may be underestimated and overestimated by a factor of $\sim 1.2$. Thus, the over/underestimation seems not to influence on the discussion about the comparison with the GMCs in M51.

\begin{table*}
 \caption{GMC properties in the different environments of NGC~1300 corrected for the CPROPS bias}
 \label{tab:re-GMCproperties}
 \begin{tabular}{lccccccc}
 \hline
 Envir. & $\sigma_{\rm v}$  & $R$  & $M_{\rm mol}$  & $M_{\rm vir}$   & $\Sigma_{\rm mol}$   & $\alpha_{\rm vir}$   & $c$   \\
    & ($\rm km~s^{-1}$) & (pc) & ($10^5~M_\odot$) &  ($10^5~M_\odot$) &  ($M_\odot \rm ~pc^{-2}$) &  & ($\rm km~s^{-1}~pc^{-0.5}$) \\
    \hline
All      & $4.0^{+1.8}_{-0.9}$ & $56.4^{+12.1}_{-13.7}$ & $5.7^{+4.7}_{-1.9}$ & $11.0^{+12.6}_{-4.78}$ & $127.8^{+45.6}_{-45.9}$ & $1.1^{+1.2}_{-0.6}$  & $0.6^{+0.2}_{-0.2}$  \\
 \hline
Bar      & $4.8^{+1.3}_{-1.5}$ & $58.4^{+22.2}_{-19.3}$ & $6.1^{+2.0}_{-2.1}$ & $14.6^{+8.5}_{-8.77}$  & $91.7^{+64.6}_{-28.6}$  & $2.0^{+0.9}_{-1.4}$  & $0.7^{+0.1}_{-0.2}$ \\
Arm      & $3.7^{+1.5}_{-0.8}$ & $55.0^{+10.7}_{-11.8}$ & $5.5^{+4.2}_{-1.8}$ & $ 9.8^{+6.4}_{-4.14}$  & $111.4^{+55.7}_{-29.3}$ & $0.9^{+1.2}_{-0.4}$  & $0.6^{+0.2}_{-0.2}$ \\
Bar-end  & $3.9^{+2.6}_{-0.7}$ & $57.0^{+21.4}_{-15.7}$ & $9.1^{+6.4}_{-3.6}$ & $14.0^{+17.9}_{-7.69}$ & $143.9^{+77.8}_{-33.4}$ & $1.1^{+0.6}_{-0.5}$  & $0.7^{+0.2}_{-0.2}$ \\
\hline
 \end{tabular}
\end{table*}

\begin{table*}
 \caption{Kolmogorov-Smirnov Test for GMC properties corrected for the CPROPS bias}
 \label{tab:re-KStest}
 \begin{tabular}{lccc}
 \hline
  Prop. & Bar vs. Arm & Bar vs. Bar-end & Arm vs. Bar-end \\
\hline
 $\sigma_{\rm v}$          & $0.138 \pm 0.062 (0.169)$ &$0.538 \pm 0.166 (0.538)$ & $0.422 \pm 0.123 (0.460)$  \\
 $R$                 & $0.746 \pm 0.100 (0.914)$ &$0.896 \pm 0.078 (0.952)$ & $0.556 \pm 0.149 (0.450)$  \\
 $M_{\rm mol}$             & $0.841 \pm 0.103 (0.938)$ &$0.028 \pm 0.018 (0.027)$ & $0.013 \pm 0.007 (0.006)$  \\
 $M_{\rm vir}$       & $0.571 \pm 0.199 (0.438)$ &$0.865 \pm 0.099 (0.865)$ & $0.242 \pm 0.092 (0.261)$  \\
 $\Sigma_{\rm mol}$  & $0.547 \pm 0.183 (0.530)$ &$0.142 \pm 0.077 (0.142)$ & $0.093 \pm 0.055 (0.139)$  \\
 $\alpha_{\rm vir}$  & $0.312 \pm 0.120 (0.283)$ &$0.185 \pm 0.077 (0.221)$ & $0.669 \pm 0.148 (0.664)$  \\
 $c$                 & $0.409 \pm 0.092 (0.409)$ &$0.739 \pm 0.166 (0.712)$ & $0.359 \pm 0.111 (0.430)$  \\
 \hline
 \end{tabular}
\end{table*}

\section{GMC catalog}\label{apx:catalog}
Table~\ref{tab:catalog} presents the GMC catalog in NGC~1300, which contains columns as follws:
\begin{enumerate}
  \item \verb|Column 1|: ID, GMC identification number
  \item \verb|Column 2|: RA, GMC right ascension as measured by the
intensity-weighted 1st moment along this direction
  \item \verb|Column 3|: Dec, GMC declination measured as above
  \item \verb|Column 4|: $v_{\rm LSR}$, GMC central velocity as measured by the intensity-weighted first moment along the velocity axis
  \item \verb|Column 5|: $T_{\rm peak}$, GMC's peak brightness temperature in K
  \item \verb|Column 6|: S/N, GMC's peak signal-to-noise ratio
  \item \verb|Column 7|: $\sigma_v$, GMC's deconvolved, extrapolated velocity dispersion in $\rm km~s^{-1}$ with uncertainty
  \item \verb|Column 8|: $R$, GMC's deconvolved, extrapolated effective radius in pc with uncertainty
  \item \verb|Column 9|: $M_{\rm mol}$, GMC's mass in $10^5~M_\odot$ calculated from CO luminosity and $\alpha_{\rm CO} = 4.4~M_\odot~(\rm K~km~s^{-1}~pc^2)^{-1}$ with uncertainty
  \item \verb|Column 10|: $M_{\rm vir}$, GMC's mass inferred from virial theorem in $10^5~M_\odot$ with uncertainty
  \item \verb|Column 11|: $\alpha_{\rm vir}$, GMC's virial parameter with uncertainty
  \item \verb|Column 12|: Region where a given GMC has been identified,
  i.e.,  Bar, Arm, Bar-end(BE), and "other", which represents the GMC outside the three regions
  \item \verb|Column 13|: Flag for the measurement of radius: $0 =$ actual measurement of radius, $1 =$ upper limit
  
\end{enumerate}

 \small
\begin{table*}
 \caption{GMC catalog}
 \label{tab:catalog}
 \begin{tabular}{lcccccccccccc}
 \hline
ID & RA  & Dec & $v_{\rm LSR}$ & $T_{\rm peak}$ & S/N & $\sigma_v$ & $R$ & $M_{\rm mol}$ & $M_{\rm vir}$ & $\alpha_{\rm vir}$ & Reg & Flag \\
   & (J2000) & (J2000) & ($\rm km~s^{-1}$) & (K) & & ($\rm km~s^{-1}$) & (pc) & ($10^5~M_\odot$) & ($10^5~M_\odot$) &  & &\\
\hline
   1 &$ 3^{\rm h}19^{\rm m}35.56^{\rm s}$ & $-19^\circ24^{\prime}25.1^{\prime\prime}$ &1675.8 &2.8 &5.5 & 7.9 $\pm$ 4.8 &43.2 $\pm$ 34.9 & 10.4 $\pm$ 6.4 & 28.3 $\pm$ 53.3 & 3.0 $\pm$ 4.9 & BE & 0 \\
   2 &$ 3^{\rm h}19^{\rm m}35.53^{\rm s}$ & $-19^\circ24^{\prime}24.6^{\prime\prime}$ &1670.1 &2.8 &5.3 & 7.9 $\pm$ 5.2 &$<$29.6 & 2.9 $\pm$ 4.1 & - & - & BE & 1 \\
   3 &$ 3^{\rm h}19^{\rm m}35.53^{\rm s}$ & $-19^\circ24^{\prime}25.7^{\prime\prime}$ &1666.2 &4.2 &7.9 & 7.7 $\pm$ 2.4 &77.5 $\pm$ 16.8 & 26.5 $\pm$ 6.9 & 47.9 $\pm$ 30.1 & 2.0 $\pm$ 1.2 & BE & 0 \\
   4 &$ 3^{\rm h}19^{\rm m}35.50^{\rm s}$ & $-19^\circ24^{\prime}04.8^{\prime\prime}$ &1667.9 &1.8 &4.7 & 4.4 $\pm$ 5.1 &$<$29.6 & 1.1 $\pm$ 1.3 & - & - & Arm & 1 \\
   5 &$ 3^{\rm h}19^{\rm m}35.46^{\rm s}$ & $-19^\circ24^{\prime}04.3^{\prime\prime}$ &1666.6 &2.1 &5.3 & 4.2 $\pm$ 2.6 &24.7 $\pm$ 14.9 & 2.8 $\pm$ 1.7 & 4.5 $\pm$ 6.9 & 1.8 $\pm$ 2.2 & Arm & 0 \\
   6 &$ 3^{\rm h}19^{\rm m}35.74^{\rm s}$ & $-19^\circ24^{\prime}28.2^{\prime\prime}$ &1662.4 &4.2 &8.9 & 5.1 $\pm$ 2.0 &61.1 $\pm$ 26.4 & 15.5 $\pm$ 11.0 & 16.3 $\pm$ 16.0 & 1.2 $\pm$ 1.2 & BE & 0 \\
   7 &$ 3^{\rm h}19^{\rm m}35.72^{\rm s}$ & $-19^\circ24^{\prime}19.8^{\prime\prime}$ &1662.1 &4.0 &8.4 & 6.6 $\pm$ 2.2 &56.4 $\pm$ 20.5 & 24.1 $\pm$ 9.5 & 25.7 $\pm$ 17.6 & 1.2 $\pm$ 0.9 & BE & 0 \\
   8 &$ 3^{\rm h}19^{\rm m}35.66^{\rm s}$ & $-19^\circ24^{\prime}18.6^{\prime\prime}$ &1666.7 &2.9 &6.8 & 6.4 $\pm$ 6.4 &14.4 $\pm$ 21.0 & 5.1 $\pm$ 4.2 & 6.1 $\pm$ 14.6 & 1.3 $\pm$ 3.3 & BE & 0 \\
   9 &$ 3^{\rm h}19^{\rm m}35.73^{\rm s}$ & $-19^\circ24^{\prime}17.7^{\prime\prime}$ &1663.0 &3.5 &8.3 & 5.7 $\pm$ 3.8 &35.4 $\pm$ 26.1 & 6.6 $\pm$ 5.4 & 11.8 $\pm$ 19.4 & 2.0 $\pm$ 2.8 & BE & 0 \\
  10 &$ 3^{\rm h}19^{\rm m}35.51^{\rm s}$ & $-19^\circ24^{\prime}03.7^{\prime\prime}$ &1659.9 &2.2 &5.4 & 3.1 $\pm$ 2.3 &$<$29.6 & 2.7 $\pm$ 1.0 & - & - & Arm & 1 \\
  11 &$ 3^{\rm h}19^{\rm m}35.62^{\rm s}$ & $-19^\circ24^{\prime}25.2^{\prime\prime}$ &1658.6 &2.8 &5.5 & 6.9 $\pm$ 2.7 &31.7 $\pm$ 14.6 & 9.8 $\pm$ 3.2 & 15.6 $\pm$ 15.3 & 1.8 $\pm$ 1.5 & BE & 0 \\
  12 &$ 3^{\rm h}19^{\rm m}35.59^{\rm s}$ & $-19^\circ24^{\prime}22.8^{\prime\prime}$ &1660.7 &7.5 &15.7 & 8.8 $\pm$ 1.3 &86.7 $\pm$ 11.9 & 79.5 $\pm$ 12.4 & 70.5 $\pm$ 22.8 & 1.0 $\pm$ 0.3 & BE & 0 \\
  13 &$ 3^{\rm h}19^{\rm m}35.52^{\rm s}$ & $-19^\circ24^{\prime}22.3^{\prime\prime}$ &1654.1 &4.0 &8.5 & 3.0 $\pm$ 3.8 &21.6 $\pm$ 20.4 & 7.5 $\pm$ 14.0 & 2.0 $\pm$ 5.5 & 0.3 $\pm$ 0.8 & BE & 0 \\
  14 &$ 3^{\rm h}19^{\rm m}35.71^{\rm s}$ & $-19^\circ24^{\prime}12.7^{\prime\prime}$ &1657.1 &4.0 &10.4 & 8.0 $\pm$ 2.3 &57.4 $\pm$ 19.9 & 19.7 $\pm$ 9.6 & 37.9 $\pm$ 30.4 & 2.1 $\pm$ 1.8 & Arm & 0 \\
  15 &$ 3^{\rm h}19^{\rm m}35.69^{\rm s}$ & $-19^\circ24^{\prime}08.6^{\prime\prime}$ &1655.6 &3.8 &10.5 & 8.1 $\pm$ 3.6 &46.7 $\pm$ 64.8 & 14.7 $\pm$ 11.0 & 31.7 $\pm$ 54.5 & 2.4 $\pm$ 4.0 & Arm & 0 \\
  16 &$ 3^{\rm h}19^{\rm m}35.58^{\rm s}$ & $-19^\circ24^{\prime}06.8^{\prime\prime}$ &1655.4 &3.9 &10.3 & 5.7 $\pm$ 1.5 &56.5 $\pm$ 26.5 & 16.0 $\pm$ 9.0 & 18.9 $\pm$ 13.6 & 1.3 $\pm$ 1.0 & Arm & 0 \\
  17 &$ 3^{\rm h}19^{\rm m}35.79^{\rm s}$ & $-19^\circ24^{\prime}27.9^{\prime\prime}$ &1649.6 &5.0 &10.5 & 4.6 $\pm$ 3.0 &30.5 $\pm$ 13.0 & 10.9 $\pm$ 6.4 & 6.8 $\pm$ 9.7 & 0.7 $\pm$ 0.9 & BE & 0 \\
  18 &$ 3^{\rm h}19^{\rm m}35.51^{\rm s}$ & $-19^\circ24^{\prime}26.5^{\prime\prime}$ &1654.4 &3.7 &7.0 & 5.3 $\pm$ 2.4 &18.2 $\pm$ 7.9 & 5.5 $\pm$ 2.9 & 5.3 $\pm$ 5.6 & 1.1 $\pm$ 1.2 & BE & 0 \\
  19 &$ 3^{\rm h}19^{\rm m}35.72^{\rm s}$ & $-19^\circ24^{\prime}15.8^{\prime\prime}$ &1646.2 &9.1 &22.0 & 10.8 $\pm$ 1.0 &69.9 $\pm$ 8.2 & 102.9 $\pm$ 5.9 & 84.4 $\pm$ 18.4 & 0.9 $\pm$ 0.2 & BE & 0 \\
  20 &$ 3^{\rm h}19^{\rm m}35.66^{\rm s}$ & $-19^\circ24^{\prime}07.2^{\prime\prime}$ &1653.2 &4.6 &12.4 & 6.8 $\pm$ 2.7 &53.1 $\pm$ 22.0 & 21.5 $\pm$ 10.8 & 25.9 $\pm$ 28.0 & 1.3 $\pm$ 1.4 & Arm & 0 \\
  21 &$ 3^{\rm h}19^{\rm m}35.62^{\rm s}$ & $-19^\circ24^{\prime}04.9^{\prime\prime}$ &1653.1 &4.1 &12.7 & 6.8 $\pm$ 4.5 &$<$29.6 & 9.6 $\pm$ 16.1 & - & - & Arm & 1 \\
  22 &$ 3^{\rm h}19^{\rm m}35.62^{\rm s}$ & $-19^\circ24^{\prime}04.1^{\prime\prime}$ &1653.7 &5.2 &15.5 & 6.5 $\pm$ 2.8 &26.8 $\pm$ 19.1 & 12.1 $\pm$ 9.9 & 11.7 $\pm$ 13.9 & 1.1 $\pm$ 1.1 & Arm & 0 \\
  23 &$ 3^{\rm h}19^{\rm m}35.63^{\rm s}$ & $-19^\circ24^{\prime}24.1^{\prime\prime}$ &1648.1 &3.6 &7.4 & 5.4 $\pm$ 3.2 &92.9 $\pm$ 39.1 & 14.7 $\pm$ 10.4 & 28.4 $\pm$ 35.2 & 2.2 $\pm$ 2.3 & BE & 0 \\
  24 &$ 3^{\rm h}19^{\rm m}35.66^{\rm s}$ & $-19^\circ24^{\prime}19.3^{\prime\prime}$ &1645.9 &4.5 &9.8 & 11.0 $\pm$ 2.9 &84.1 $\pm$ 17.1 & 39.5 $\pm$ 13.0 & 106.0 $\pm$ 66.4 & 3.0 $\pm$ 1.8 & BE & 0 \\
  25 &$ 3^{\rm h}19^{\rm m}35.68^{\rm s}$ & $-19^\circ24^{\prime}17.8^{\prime\prime}$ &1643.8 &6.3 &15.5 & 9.6 $\pm$ 1.3 &71.3 $\pm$ 8.7 & 58.9 $\pm$ 7.6 & 67.6 $\pm$ 21.4 & 1.3 $\pm$ 0.4 & BE & 0 \\
  26 &$ 3^{\rm h}19^{\rm m}35.73^{\rm s}$ & $-19^\circ24^{\prime}10.3^{\prime\prime}$ &1638.3 &4.2 &11.8 & 7.5 $\pm$ 4.3 &47.4 $\pm$ 19.2 & 19.4 $\pm$ 19.0 & 28.0 $\pm$ 35.4 & 1.6 $\pm$ 1.9 & Arm & 0 \\
  27 &$ 3^{\rm h}19^{\rm m}35.84^{\rm s}$ & $-19^\circ24^{\prime}07.7^{\prime\prime}$ &1639.8 &2.0 &5.8 & 1.9 $\pm$ 2.2 &$<$29.6 & 3.3 $\pm$ 5.0 & - & - & Arm & 1 \\
  28 &$ 3^{\rm h}19^{\rm m}35.71^{\rm s}$ & $-19^\circ24^{\prime}06.0^{\prime\prime}$ &1642.7 &3.1 &7.9 & 4.1 $\pm$ 4.7 &52.5 $\pm$ 25.0 & 5.3 $\pm$ 2.7 & 9.3 $\pm$ 22.7 & 2.0 $\pm$ 4.6 & Arm & 0 \\
  29 &$ 3^{\rm h}19^{\rm m}35.72^{\rm s}$ & $-19^\circ24^{\prime}05.1^{\prime\prime}$ &1642.6 &2.4 &6.3 & 5.7 $\pm$ 2.2 &37.5 $\pm$ 15.2 & 4.3 $\pm$ 1.4 & 12.7 $\pm$ 10.5 & 3.3 $\pm$ 2.7 & Arm & 0 \\
  30 &$ 3^{\rm h}19^{\rm m}35.88^{\rm s}$ & $-19^\circ24^{\prime}26.7^{\prime\prime}$ &1632.2 &3.0 &6.6 & 3.7 $\pm$ 4.8 &70.8 $\pm$ 48.9 & 1.8 $\pm$ 1.6 & 10.3 $\pm$ 24.6 & 6.5 $\pm$ 16.2 & BE & 0 \\
  31 &$ 3^{\rm h}19^{\rm m}35.77^{\rm s}$ & $-19^\circ24^{\prime}24.3^{\prime\prime}$ &1634.8 &4.9 &10.0 & 5.8 $\pm$ 3.2 &47.3 $\pm$ 19.3 & 25.5 $\pm$ 15.7 & 16.6 $\pm$ 19.2 & 0.7 $\pm$ 0.7 & BE & 0 \\
  32 &$ 3^{\rm h}19^{\rm m}35.69^{\rm s}$ & $-19^\circ24^{\prime}22.6^{\prime\prime}$ &1635.1 &4.7 &10.4 & 5.5 $\pm$ 1.3 &75.7 $\pm$ 14.9 & 32.6 $\pm$ 5.9 & 23.9 $\pm$ 13.2 & 0.8 $\pm$ 0.4 & BE & 0 \\
  33 &$ 3^{\rm h}19^{\rm m}35.63^{\rm s}$ & $-19^\circ24^{\prime}20.6^{\prime\prime}$ &1640.7 &3.4 &8.1 & 3.5 $\pm$ 4.8 &35.5 $\pm$ 39.6 & 6.8 $\pm$ 10.9 & 4.5 $\pm$ 13.8 & 0.7 $\pm$ 2.4 & BE & 0 \\
  34 &$ 3^{\rm h}19^{\rm m}35.69^{\rm s}$ & $-19^\circ24^{\prime}11.3^{\prime\prime}$ &1644.7 &4.1 &10.7 & 9.3 $\pm$ 4.2 &30.8 $\pm$ 12.5 & 14.5 $\pm$ 9.8 & 27.8 $\pm$ 26.9 & 2.1 $\pm$ 1.7 & Arm & 0 \\
  35 &$ 3^{\rm h}19^{\rm m}35.71^{\rm s}$ & $-19^\circ24^{\prime}09.6^{\prime\prime}$ &1635.4 &3.9 &9.2 & 4.6 $\pm$ 6.3 &45.6 $\pm$ 47.0 & 24.7 $\pm$ 31.8 & 9.9 $\pm$ 27.2 & 0.4 $\pm$ 1.1 & Arm & 0 \\
  36 &$ 3^{\rm h}19^{\rm m}35.84^{\rm s}$ & $-19^\circ24^{\prime}09.2^{\prime\prime}$ &1639.0 &3.5 &10.1 & 8.9 $\pm$ 3.0 &76.7 $\pm$ 20.4 & 38.3 $\pm$ 14.6 & 63.0 $\pm$ 44.0 & 1.8 $\pm$ 1.1 & Arm & 0 \\
  37 &$ 3^{\rm h}19^{\rm m}35.71^{\rm s}$ & $-19^\circ24^{\prime}04.2^{\prime\prime}$ &1640.5 &2.3 &5.9 & 4.7 $\pm$ 3.2 &$<$29.6 & 3.0 $\pm$ 1.7 & - & - & Arm & 1 \\
  38 &$ 3^{\rm h}19^{\rm m}36.28^{\rm s}$ & $-19^\circ24^{\prime}32.7^{\prime\prime}$ &1632.9 &3.0 &7.6 & 4.3 $\pm$ 2.4 &$<$29.6 & 2.3 $\pm$ 1.2 & - & - & BE & 1 \\
  39 &$ 3^{\rm h}19^{\rm m}35.97^{\rm s}$ & $-19^\circ24^{\prime}27.8^{\prime\prime}$ &1631.1 &4.2 &9.6 & 7.2 $\pm$ 1.8 &38.7 $\pm$ 14.4 & 12.3 $\pm$ 7.0 & 20.8 $\pm$ 11.9 & 1.9 $\pm$ 1.0 & BE & 0 \\
  40 &$ 3^{\rm h}19^{\rm m}35.90^{\rm s}$ & $-19^\circ24^{\prime}27.1^{\prime\prime}$ &1632.8 &3.0 &6.7 & 1.8 $\pm$ 3.2 &64.2 $\pm$ 73.9 & 8.4 $\pm$ 11.5 & 2.2 $\pm$ 7.4 & 0.3 $\pm$ 1.0 & BE & 0 \\
  41 &$ 3^{\rm h}19^{\rm m}36.02^{\rm s}$ & $-19^\circ24^{\prime}25.8^{\prime\prime}$ &1631.1 &3.2 &8.0 & 3.1 $\pm$ 2.9 &42.6 $\pm$ 20.5 & 4.2 $\pm$ 2.0 & 4.3 $\pm$ 8.7 & 1.2 $\pm$ 2.2 & BE & 0 \\
  42 &$ 3^{\rm h}19^{\rm m}35.85^{\rm s}$ & $-19^\circ24^{\prime}25.4^{\prime\prime}$ &1634.4 &4.7 &10.2 & 7.3 $\pm$ 3.1 &89.5 $\pm$ 22.1 & 47.3 $\pm$ 27.5 & 50.0 $\pm$ 45.4 & 1.2 $\pm$ 1.0 & BE & 0 \\
  43 &$ 3^{\rm h}19^{\rm m}35.87^{\rm s}$ & $-19^\circ24^{\prime}24.5^{\prime\prime}$ &1631.5 &3.1 &6.8 & 2.2 $\pm$ 3.7 &28.8 $\pm$ 38.5 & 3.7 $\pm$ 4.9 & 1.5 $\pm$ 5.0 & 0.5 $\pm$ 1.4 & BE & 0 \\
  44 &$ 3^{\rm h}19^{\rm m}35.67^{\rm s}$ & $-19^\circ24^{\prime}21.1^{\prime\prime}$ &1632.0 &5.0 &11.2 & 7.8 $\pm$ 2.9 &32.5 $\pm$ 16.9 & 9.1 $\pm$ 7.9 & 20.3 $\pm$ 19.6 & 2.5 $\pm$ 2.3 & BE & 0 \\
  45 &$ 3^{\rm h}19^{\rm m}35.71^{\rm s}$ & $-19^\circ24^{\prime}12.9^{\prime\prime}$ &1635.8 &3.9 &9.7 & 6.0 $\pm$ 2.2 &57.6 $\pm$ 18.6 & 23.0 $\pm$ 10.5 & 21.2 $\pm$ 18.0 & 1.0 $\pm$ 0.7 & Arm & 0 \\
  46 &$ 3^{\rm h}19^{\rm m}36.25^{\rm s}$ & $-19^\circ24^{\prime}30.5^{\prime\prime}$ &1622.9 &4.4 &11.2 & 5.1 $\pm$ 3.7 &46.3 $\pm$ 38.7 & 15.6 $\pm$ 27.9 & 12.7 $\pm$ 20.7 & 0.9 $\pm$ 1.7 & BE & 0 \\
  47 &$ 3^{\rm h}19^{\rm m}36.06^{\rm s}$ & $-19^\circ24^{\prime}29.6^{\prime\prime}$ &1626.1 &3.3 &8.1 & 5.6 $\pm$ 4.0 &46.8 $\pm$ 26.7 & 8.1 $\pm$ 4.1 & 15.1 $\pm$ 21.4 & 2.1 $\pm$ 3.4 & BE & 0 \\
  48 &$ 3^{\rm h}19^{\rm m}36.00^{\rm s}$ & $-19^\circ24^{\prime}29.4^{\prime\prime}$ &1625.1 &2.0 &4.9 & 7.7 $\pm$ 8.5 &$<$29.6 & 3.4 $\pm$ 4.2 & - & - & BE & 1 \\
  49 &$ 3^{\rm h}19^{\rm m}36.00^{\rm s}$ & $-19^\circ24^{\prime}28.7^{\prime\prime}$ &1625.5 &3.7 &8.3 & 8.1 $\pm$ 5.5 &51.3 $\pm$ 23.1 & 14.4 $\pm$ 11.6 & 35.1 $\pm$ 48.4 & 2.7 $\pm$ 3.6 & BE & 0 \\
  50 &$ 3^{\rm h}19^{\rm m}35.97^{\rm s}$ & $-19^\circ24^{\prime}26.4^{\prime\prime}$ &1623.5 &2.7 &5.6 & 4.2 $\pm$ 2.6 &39.5 $\pm$ 22.4 & 9.8 $\pm$ 7.6 & 7.3 $\pm$ 10.8 & 0.8 $\pm$ 1.4 & BE & 0 \\
  51 &$ 3^{\rm h}19^{\rm m}35.73^{\rm s}$ & $-19^\circ24^{\prime}17.2^{\prime\prime}$ &1627.0 &2.0 &5.0 & 3.3 $\pm$ 5.0 &27.4 $\pm$ 28.8 & 2.3 $\pm$ 3.5 & 3.1 $\pm$ 9.8 & 1.5 $\pm$ 4.6 & BE & 0 \\
  52 &$ 3^{\rm h}19^{\rm m}36.12^{\rm s}$ & $-19^\circ24^{\prime}29.9^{\prime\prime}$ &1622.7 &3.1 &7.8 & 6.0 $\pm$ 10.5 &49.7 $\pm$ 33.9 & 8.4 $\pm$ 8.7 & 18.4 $\pm$ 57.9 & 2.4 $\pm$ 6.9 & BE & 0 \\
  53 &$ 3^{\rm h}19^{\rm m}36.17^{\rm s}$ & $-19^\circ24^{\prime}29.2^{\prime\prime}$ &1620.5 &2.4 &6.3 & 5.5 $\pm$ 3.0 &$<$29.6 & 3.3 $\pm$ 2.9 & - & - & BE & 1 \\
  54 &$ 3^{\rm h}19^{\rm m}36.06^{\rm s}$ & $-19^\circ24^{\prime}28.0^{\prime\prime}$ &1622.7 &4.2 &10.5 & 7.0 $\pm$ 2.7 &63.2 $\pm$ 18.9 & 11.9 $\pm$ 8.5 & 32.3 $\pm$ 29.7 & 3.0 $\pm$ 3.1 & BE & 0 \\
  55 &$ 3^{\rm h}19^{\rm m}35.64^{\rm s}$ & $-19^\circ24^{\prime}22.2^{\prime\prime}$ &1623.7 &2.5 &5.9 & 5.0 $\pm$ 3.4 &$<$29.6 & 1.8 $\pm$ 2.9 & - & - & BE & 1 \\
  56 &$ 3^{\rm h}19^{\rm m}35.71^{\rm s}$ & $-19^\circ24^{\prime}21.6^{\prime\prime}$ &1624.8 &2.3 &5.9 & 4.5 $\pm$ 3.8 &$<$29.6 & 1.8 $\pm$ 1.2 & - & - & BE & 1 \\
  57 &$ 3^{\rm h}19^{\rm m}35.73^{\rm s}$ & $-19^\circ24^{\prime}16.5^{\prime\prime}$ &1623.0 &3.4 &8.0 & 4.8 $\pm$ 4.4 &$<$29.6 & 2.3 $\pm$ 3.0 & - & - & BE & 1 \\
  58 &$ 3^{\rm h}19^{\rm m}35.72^{\rm s}$ & $-19^\circ24^{\prime}14.4^{\prime\prime}$ &1622.1 &2.2 &6.1 & 7.9 $\pm$ 5.0 &$<$29.6 & 2.8 $\pm$ 2.5 & - & - & Arm & 1 \\
  59 &$ 3^{\rm h}19^{\rm m}36.40^{\rm s}$ & $-19^\circ24^{\prime}32.9^{\prime\prime}$ &1619.1 &1.6 &4.3 & 8.0 $\pm$ 5.7 &16.4 $\pm$ 17.5 & 1.4 $\pm$ 1.2 & 11.0 $\pm$ 21.7 & 8.7 $\pm$ 16.5 & BE & 0 \\
  60 &$ 3^{\rm h}19^{\rm m}36.44^{\rm s}$ & $-19^\circ24^{\prime}32.5^{\prime\prime}$ &1615.7 &2.5 &6.9 & 4.7 $\pm$ 1.9 &$<$29.6 & 4.1 $\pm$ 1.3 & - & - & BE & 1 \\
  61 &$ 3^{\rm h}19^{\rm m}36.31^{\rm s}$ & $-19^\circ24^{\prime}31.2^{\prime\prime}$ &1615.3 &3.8 &9.9 & 9.5 $\pm$ 8.8 &31.5 $\pm$ 40.0 & 6.4 $\pm$ 3.8 & 29.8 $\pm$ 62.8 & 5.2 $\pm$ 11.3 & BE & 0 \\
  62 &$ 3^{\rm h}19^{\rm m}35.51^{\rm s}$ & $-19^\circ24^{\prime}18.1^{\prime\prime}$ &1673.3 &4.9 &10.5 & 4.6 $\pm$ 1.1 &59.9 $\pm$ 10.0 & 17.9 $\pm$ 3.6 & 13.0 $\pm$ 6.9 & 0.8 $\pm$ 0.4 & BE & 0 \\
  \hline
\end{tabular}
\end{table*}

\begin{table*}
    \contcaption{GMC catalog}
    \label{tab:catalog}
    \begin{tabular}{lcccccccccccc}
    \hline
   ID & RA  & Dec & $v_{\rm LSR}$ & $T_{\rm peak}$ & S/N & $\sigma_v$ & $R$ & $M_{\rm mol}$ & $M_{\rm vir}$ & $\alpha_{\rm vir}$ & Reg & Flag \\
      & (J2000) & (J2000) & ($\rm km~s^{-1}$) & (K) & & ($\rm km~s^{-1}$) & (pc) & ($10^5~M_\odot$) & ($10^5~M_\odot$) &  & &\\
   \hline
  63 &$ 3^{\rm h}19^{\rm m}35.55^{\rm s}$ & $-19^\circ24^{\prime}18.5^{\prime\prime}$ &1651.5 &2.8 &6.4 & 5.8 $\pm$ 2.0 &47.3 $\pm$ 17.5 & 8.1 $\pm$ 4.4 & 16.4 $\pm$ 15.8 & 2.2 $\pm$ 2.0 & BE & 0 \\
  64 &$ 3^{\rm h}19^{\rm m}35.54^{\rm s}$ & $-19^\circ24^{\prime}13.3^{\prime\prime}$ &1677.4 &2.6 &6.8 & 4.7 $\pm$ 1.7 &$<$29.6 & 2.4 $\pm$ 0.6 & - & - & Arm & 1 \\
  65 &$ 3^{\rm h}19^{\rm m}36.45^{\rm s}$ & $-19^\circ24^{\prime}51.3^{\prime\prime}$ &1672.6 &2.6 &4.5 & 6.8 $\pm$ 4.3 &$<$29.6 & 2.3 $\pm$ 0.9 & - & - & other & 1 \\
  66 &$ 3^{\rm h}19^{\rm m}35.01^{\rm s}$ & $-19^\circ24^{\prime}04.5^{\prime\prime}$ &1672.0 &2.4 &5.2 & 7.6 $\pm$ 3.3 &14.4 $\pm$ 15.6 & 2.7 $\pm$ 1.1 & 8.7 $\pm$ 11.3 & 3.5 $\pm$ 4.5 & Arm & 0 \\
  67 &$ 3^{\rm h}19^{\rm m}35.69^{\rm s}$ & $-19^\circ23^{\prime}53.3^{\prime\prime}$ &1651.9 &2.3 &5.6 & 5.7 $\pm$ 2.0 &33.2 $\pm$ 11.4 & 6.4 $\pm$ 2.0 & 11.3 $\pm$ 9.3 & 2.0 $\pm$ 1.5 & Arm & 0 \\
  68 &$ 3^{\rm h}19^{\rm m}35.84^{\rm s}$ & $-19^\circ23^{\prime}52.8^{\prime\prime}$ &1649.8 &3.1 &7.7 & 4.1 $\pm$ 7.5 &37.3 $\pm$ 49.5 & 8.2 $\pm$ 19.6 & 6.6 $\pm$ 24.4 & 0.9 $\pm$ 3.3 & Arm & 0 \\
  69 &$ 3^{\rm h}19^{\rm m}36.00^{\rm s}$ & $-19^\circ23^{\prime}50.6^{\prime\prime}$ &1650.7 &2.5 &6.4 & 4.5 $\pm$ 2.2 &49.0 $\pm$ 19.0 & 4.5 $\pm$ 3.7 & 10.2 $\pm$ 10.5 & 2.5 $\pm$ 2.4 & Arm & 0 \\
  70 &$ 3^{\rm h}19^{\rm m}35.79^{\rm s}$ & $-19^\circ23^{\prime}52.7^{\prime\prime}$ &1649.3 &4.1 &10.1 & 4.5 $\pm$ 3.9 &72.9 $\pm$ 33.8 & 10.4 $\pm$ 9.9 & 15.6 $\pm$ 30.7 & 1.7 $\pm$ 3.3 & Arm & 0 \\
  71 &$ 3^{\rm h}19^{\rm m}35.86^{\rm s}$ & $-19^\circ23^{\prime}51.6^{\prime\prime}$ &1652.5 &2.0 &5.0 & 6.7 $\pm$ 4.0 &$<$29.6 & 4.3 $\pm$ 2.7 & - & - & Arm & 1 \\
  72 &$ 3^{\rm h}19^{\rm m}35.97^{\rm s}$ & $-19^\circ23^{\prime}51.4^{\prime\prime}$ &1644.5 &3.8 &9.4 & 5.1 $\pm$ 1.8 &74.2 $\pm$ 15.5 & 19.1 $\pm$ 4.1 & 19.8 $\pm$ 15.7 & 1.2 $\pm$ 0.9 & Arm & 0 \\
  73 &$ 3^{\rm h}19^{\rm m}35.00^{\rm s}$ & $-19^\circ24^{\prime}09.3^{\prime\prime}$ &1668.6 &2.2 &4.1 & 4.9 $\pm$ 3.5 &37.0 $\pm$ 13.2 & 3.3 $\pm$ 1.8 & 9.4 $\pm$ 15.0 & 3.1 $\pm$ 4.7 & Arm & 0 \\
  74 &$ 3^{\rm h}19^{\rm m}35.01^{\rm s}$ & $-19^\circ24^{\prime}08.9^{\prime\prime}$ &1665.4 &4.1 &7.6 & 5.6 $\pm$ 2.2 &$<$29.6 & 3.0 $\pm$ 0.6 & - & - & Arm & 1 \\
  75 &$ 3^{\rm h}19^{\rm m}35.43^{\rm s}$ & $-19^\circ24^{\prime}08.2^{\prime\prime}$ &1664.7 &3.5 &8.9 & 4.6 $\pm$ 2.8 &$<$29.6 & 3.2 $\pm$ 0.6 & - & - & Arm & 1 \\
  76 &$ 3^{\rm h}19^{\rm m}34.93^{\rm s}$ & $-19^\circ24^{\prime}05.5^{\prime\prime}$ &1663.5 &2.7 &5.2 & 5.2 $\pm$ 1.5 &45.8 $\pm$ 12.9 & 7.8 $\pm$ 2.0 & 13.1 $\pm$ 8.9 & 1.9 $\pm$ 1.4 & Arm & 0 \\
  77 &$ 3^{\rm h}19^{\rm m}34.91^{\rm s}$ & $-19^\circ24^{\prime}04.4^{\prime\prime}$ &1663.8 &2.3 &4.3 & 6.5 $\pm$ 2.2 &33.1 $\pm$ 25.4 & 8.4 $\pm$ 2.4 & 14.6 $\pm$ 15.7 & 1.9 $\pm$ 2.1 & Arm & 0 \\
  78 &$ 3^{\rm h}19^{\rm m}34.96^{\rm s}$ & $-19^\circ24^{\prime}02.3^{\prime\prime}$ &1670.6 &2.5 &5.2 & 8.3 $\pm$ 2.9 &18.2 $\pm$ 10.5 & 4.4 $\pm$ 1.2 & 12.9 $\pm$ 13.2 & 3.3 $\pm$ 3.1 & Arm & 0 \\
  79 &$ 3^{\rm h}19^{\rm m}35.49^{\rm s}$ & $-19^\circ23^{\prime}56.6^{\prime\prime}$ &1657.2 &5.3 &12.6 & 4.5 $\pm$ 0.7 &74.8 $\pm$ 10.1 & 24.2 $\pm$ 2.0 & 15.6 $\pm$ 5.5 & 0.7 $\pm$ 0.2 & Arm & 0 \\
  80 &$ 3^{\rm h}19^{\rm m}35.64^{\rm s}$ & $-19^\circ23^{\prime}54.7^{\prime\prime}$ &1650.2 &2.5 &6.5 & 2.9 $\pm$ 1.6 &38.2 $\pm$ 15.3 & 5.6 $\pm$ 2.6 & 3.3 $\pm$ 4.3 & 0.7 $\pm$ 0.9 & Arm & 0 \\
  81 &$ 3^{\rm h}19^{\rm m}35.60^{\rm s}$ & $-19^\circ23^{\prime}55.7^{\prime\prime}$ &1648.8 &2.9 &7.1 & 5.1 $\pm$ 2.5 &50.4 $\pm$ 13.9 & 8.2 $\pm$ 3.9 & 13.8 $\pm$ 14.1 & 1.9 $\pm$ 1.9 & Arm & 0 \\
  82 &$ 3^{\rm h}19^{\rm m}35.60^{\rm s}$ & $-19^\circ23^{\prime}52.5^{\prime\prime}$ &1651.7 &2.3 &5.0 & 6.5 $\pm$ 4.9 &$<$29.6 & 5.1 $\pm$ 4.1 & - & - & Arm & 1 \\
  83 &$ 3^{\rm h}19^{\rm m}35.60^{\rm s}$ & $-19^\circ23^{\prime}51.9^{\prime\prime}$ &1648.8 &2.4 &5.4 & 5.0 $\pm$ 2.9 &30.7 $\pm$ 16.8 & 4.0 $\pm$ 2.1 & 8.1 $\pm$ 11.3 & 2.3 $\pm$ 3.4 & Arm & 0 \\
  84 &$ 3^{\rm h}19^{\rm m}35.63^{\rm s}$ & $-19^\circ23^{\prime}50.7^{\prime\prime}$ &1642.4 &2.5 &6.1 & 6.6 $\pm$ 5.0 &33.8 $\pm$ 23.4 & 4.4 $\pm$ 4.9 & 15.3 $\pm$ 27.1 & 3.9 $\pm$ 7.6 & Arm & 0 \\
  85 &$ 3^{\rm h}19^{\rm m}35.64^{\rm s}$ & $-19^\circ23^{\prime}48.9^{\prime\prime}$ &1642.7 &2.2 &5.2 & 3.9 $\pm$ 4.9 &$<$29.6 & 1.1 $\pm$ 0.8 & - & - & Arm & 1 \\
  86 &$ 3^{\rm h}19^{\rm m}35.65^{\rm s}$ & $-19^\circ23^{\prime}48.3^{\prime\prime}$ &1648.1 &2.4 &5.3 & 7.2 $\pm$ 3.3 &48.0 $\pm$ 14.1 & 8.4 $\pm$ 3.3 & 26.1 $\pm$ 25.4 & 3.5 $\pm$ 2.9 & Arm & 0 \\
  87 &$ 3^{\rm h}19^{\rm m}35.64^{\rm s}$ & $-19^\circ23^{\prime}50.1^{\prime\prime}$ &1642.3 &3.2 &7.4 & 3.9 $\pm$ 3.9 &$<$29.6 & 4.3 $\pm$ 3.8 & - & - & Arm & 1 \\
  88 &$ 3^{\rm h}19^{\rm m}35.00^{\rm s}$ & $-19^\circ24^{\prime}08.2^{\prime\prime}$ &1665.0 &2.6 &5.0 & 4.7 $\pm$ 2.5 &$<$29.6 & 2.1 $\pm$ 0.8 & - & - & Arm & 1 \\
  89 &$ 3^{\rm h}19^{\rm m}35.42^{\rm s}$ & $-19^\circ23^{\prime}54.8^{\prime\prime}$ &1664.9 &1.7 &4.2 & 5.5 $\pm$ 3.4 &$<$29.6 & 2.3 $\pm$ 1.2 & - & - & Arm & 1 \\
  90 &$ 3^{\rm h}19^{\rm m}35.56^{\rm s}$ & $-19^\circ24^{\prime}15.9^{\prime\prime}$ &1659.2 &2.8 &6.2 & 5.2 $\pm$ 1.6 &41.0 $\pm$ 10.8 & 5.0 $\pm$ 1.0 & 11.6 $\pm$ 8.4 & 2.6 $\pm$ 1.8 & BE & 0 \\
  91 &$ 3^{\rm h}19^{\rm m}35.30^{\rm s}$ & $-19^\circ24^{\prime}03.8^{\prime\prime}$ &1663.4 &2.3 &5.6 & 4.1 $\pm$ 2.9 &$<$29.6 & 2.8 $\pm$ 0.6 & - & - & Arm & 1 \\
  92 &$ 3^{\rm h}19^{\rm m}35.70^{\rm s}$ & $-19^\circ24^{\prime}03.4^{\prime\prime}$ &1646.3 &3.8 &9.9 & 9.7 $\pm$ 2.6 &29.2 $\pm$ 12.6 & 14.0 $\pm$ 3.7 & 28.4 $\pm$ 18.8 & 2.3 $\pm$ 1.3 & Arm & 0 \\
  93 &$ 3^{\rm h}19^{\rm m}35.79^{\rm s}$ & $-19^\circ24^{\prime}02.6^{\prime\prime}$ &1646.3 &5.6 &15.5 & 5.8 $\pm$ 1.0 &87.2 $\pm$ 11.1 & 43.1 $\pm$ 4.0 & 30.5 $\pm$ 10.4 & 0.8 $\pm$ 0.3 & Arm & 0 \\
  94 &$ 3^{\rm h}19^{\rm m}35.82^{\rm s}$ & $-19^\circ24^{\prime}00.5^{\prime\prime}$ &1645.7 &3.0 &8.7 & 4.1 $\pm$ 1.1 &73.2 $\pm$ 16.1 & 14.5 $\pm$ 3.3 & 12.6 $\pm$ 7.2 & 1.0 $\pm$ 0.6 & Arm & 0 \\
  95 &$ 3^{\rm h}19^{\rm m}35.93^{\rm s}$ & $-19^\circ24^{\prime}02.5^{\prime\prime}$ &1639.8 &4.0 &11.9 & 3.2 $\pm$ 2.0 &47.9 $\pm$ 15.9 & 14.2 $\pm$ 5.4 & 5.2 $\pm$ 7.3 & 0.4 $\pm$ 0.6 & Arm & 0 \\
  96 &$ 3^{\rm h}19^{\rm m}35.84^{\rm s}$ & $-19^\circ23^{\prime}59.4^{\prime\prime}$ &1645.5 &1.7 &4.6 & 5.6 $\pm$ 4.8 &$<$29.6 & 3.8 $\pm$ 3.5 & - & - & Arm & 1 \\
  97 &$ 3^{\rm h}19^{\rm m}35.87^{\rm s}$ & $-19^\circ24^{\prime}02.6^{\prime\prime}$ &1634.8 &2.7 &8.4 & 5.0 $\pm$ 3.9 &34.8 $\pm$ 20.5 & 7.6 $\pm$ 6.9 & 9.0 $\pm$ 15.2 & 1.3 $\pm$ 2.2 & Arm & 0 \\
  98 &$ 3^{\rm h}19^{\rm m}35.99^{\rm s}$ & $-19^\circ24^{\prime}03.0^{\prime\prime}$ &1630.0 &1.5 &4.0 & 6.7 $\pm$ 5.9 &$<$29.6 & 0.7 $\pm$ 0.6 & - & - & Arm & 1 \\
  99 &$ 3^{\rm h}19^{\rm m}35.39^{\rm s}$ & $-19^\circ23^{\prime}57.2^{\prime\prime}$ &1660.2 &2.0 &4.7 & 3.7 $\pm$ 1.7 &$<$29.6 & 4.1 $\pm$ 1.3 & - & - & Arm & 1 \\
 100 &$ 3^{\rm h}19^{\rm m}36.42^{\rm s}$ & $-19^\circ24^{\prime}39.2^{\prime\prime}$ &1646.1 &1.9 &4.2 & 5.0 $\pm$ 2.8 &25.0 $\pm$ 14.5 & 4.1 $\pm$ 2.5 & 6.5 $\pm$ 7.8 & 1.8 $\pm$ 2.3 & other & 0 \\
 101 &$ 3^{\rm h}19^{\rm m}36.34^{\rm s}$ & $-19^\circ24^{\prime}38.7^{\prime\prime}$ &1647.4 &2.3 &5.0 & 6.6 $\pm$ 2.5 &51.5 $\pm$ 16.1 & 7.3 $\pm$ 1.9 & 23.0 $\pm$ 19.5 & 3.5 $\pm$ 2.8 & other & 0 \\
 102 &$ 3^{\rm h}19^{\rm m}36.04^{\rm s}$ & $-19^\circ24^{\prime}34.5^{\prime\prime}$ &1651.9 &2.3 &4.6 & 6.8 $\pm$ 2.1 &32.7 $\pm$ 10.7 & 6.1 $\pm$ 1.6 & 15.8 $\pm$ 12.8 & 2.9 $\pm$ 2.4 & other & 0 \\
 103 &$ 3^{\rm h}19^{\rm m}38.35^{\rm s}$ & $-19^\circ24^{\prime}22.5^{\prime\prime}$ &1649.7 &1.7 &4.2 & 6.0 $\pm$ 4.1 &$<$29.6 & 3.6 $\pm$ 1.8 & - & - & other & 1 \\
 104 &$ 3^{\rm h}19^{\rm m}35.83^{\rm s}$ & $-19^\circ23^{\prime}58.4^{\prime\prime}$ &1651.1 &2.5 &6.9 & 5.7 $\pm$ 4.7 &38.3 $\pm$ 21.5 & 5.3 $\pm$ 5.9 & 13.0 $\pm$ 24.8 & 2.7 $\pm$ 6.1 & Arm & 0 \\
 105 &$ 3^{\rm h}19^{\rm m}35.88^{\rm s}$ & $-19^\circ23^{\prime}58.5^{\prime\prime}$ &1642.2 &2.4 &6.7 & 2.2 $\pm$ 4.1 &35.8 $\pm$ 38.8 & 1.2 $\pm$ 1.4 & 1.8 $\pm$ 5.6 & 1.6 $\pm$ 5.1 & Arm & 0 \\
 106 &$ 3^{\rm h}19^{\rm m}35.90^{\rm s}$ & $-19^\circ24^{\prime}33.5^{\prime\prime}$ &1650.4 &2.9 &5.5 & 4.4 $\pm$ 1.8 &$<$29.6 & 3.3 $\pm$ 1.0 & - & - & other & 1 \\
 107 &$ 3^{\rm h}19^{\rm m}35.93^{\rm s}$ & $-19^\circ23^{\prime}57.4^{\prime\prime}$ &1649.9 &1.8 &5.3 & 3.8 $\pm$ 1.5 &40.7 $\pm$ 18.3 & 4.0 $\pm$ 2.3 & 6.3 $\pm$ 5.7 & 1.7 $\pm$ 1.5 & Arm & 0 \\
 108 &$ 3^{\rm h}19^{\rm m}35.92^{\rm s}$ & $-19^\circ23^{\prime}58.1^{\prime\prime}$ &1646.3 &2.2 &6.6 & 3.9 $\pm$ 3.2 &$<$29.6 & 1.7 $\pm$ 1.0 & - & - & Arm & 1 \\
 109 &$ 3^{\rm h}19^{\rm m}35.82^{\rm s}$ & $-19^\circ24^{\prime}05.9^{\prime\prime}$ &1641.5 &3.3 &8.9 & 5.2 $\pm$ 1.1 &38.8 $\pm$ 7.5 & 9.7 $\pm$ 0.8 & 11.1 $\pm$ 5.1 & 1.3 $\pm$ 0.6 & Arm & 0 \\
 110 &$ 3^{\rm h}19^{\rm m}36.03^{\rm s}$ & $-19^\circ24^{\prime}02.2^{\prime\prime}$ &1639.0 &2.6 &7.3 & 5.8 $\pm$ 1.9 &37.7 $\pm$ 8.7 & 5.5 $\pm$ 1.1 & 13.4 $\pm$ 9.3 & 2.7 $\pm$ 1.9 & Arm & 0 \\
 111 &$ 3^{\rm h}19^{\rm m}35.97^{\rm s}$ & $-19^\circ23^{\prime}54.9^{\prime\prime}$ &1641.9 &2.4 &6.9 & 2.0 $\pm$ 2.9 &$<$29.6 & 3.6 $\pm$ 3.5 & - & - & Arm & 1 \\
 112 &$ 3^{\rm h}19^{\rm m}36.02^{\rm s}$ & $-19^\circ23^{\prime}52.6^{\prime\prime}$ &1639.5 &2.2 &5.7 & 3.9 $\pm$ 3.7 &13.6 $\pm$ 27.1 & 2.9 $\pm$ 2.9 & 2.2 $\pm$ 5.8 & 0.8 $\pm$ 2.2 & Arm & 0 \\
 113 &$ 3^{\rm h}19^{\rm m}36.03^{\rm s}$ & $-19^\circ23^{\prime}54.5^{\prime\prime}$ &1630.5 &3.6 &10.3 & 7.8 $\pm$ 2.8 &42.6 $\pm$ 12.2 & 12.2 $\pm$ 2.8 & 26.7 $\pm$ 18.8 & 2.4 $\pm$ 1.6 & Arm & 0 \\
 114 &$ 3^{\rm h}19^{\rm m}36.19^{\rm s}$ & $-19^\circ23^{\prime}52.5^{\prime\prime}$ &1628.2 &3.0 &7.9 & 6.9 $\pm$ 2.3 &59.3 $\pm$ 16.7 & 17.1 $\pm$ 7.1 & 29.5 $\pm$ 21.4 & 1.9 $\pm$ 1.2 & Arm & 0 \\
 115 &$ 3^{\rm h}19^{\rm m}36.21^{\rm s}$ & $-19^\circ23^{\prime}51.9^{\prime\prime}$ &1632.7 &2.5 &6.7 & 8.4 $\pm$ 4.1 &46.3 $\pm$ 29.0 & 6.6 $\pm$ 3.8 & 34.0 $\pm$ 44.1 & 5.8 $\pm$ 6.9 & Arm & 0 \\
 116 &$ 3^{\rm h}19^{\rm m}36.46^{\rm s}$ & $-19^\circ23^{\prime}50.6^{\prime\prime}$ &1647.3 &1.5 &4.1 & 3.6 $\pm$ 6.0 &33.3 $\pm$ 44.3 & 1.6 $\pm$ 2.5 & 4.4 $\pm$ 15.8 & 3.0 $\pm$ 9.7 & Arm & 0 \\
 117 &$ 3^{\rm h}19^{\rm m}36.43^{\rm s}$ & $-19^\circ23^{\prime}50.4^{\prime\prime}$ &1635.5 &2.0 &5.2 & 6.5 $\pm$ 2.5 &23.2 $\pm$ 11.2 & 5.0 $\pm$ 2.0 & 10.1 $\pm$ 9.2 & 2.2 $\pm$ 1.9 & Arm & 0 \\
 118 &$ 3^{\rm h}19^{\rm m}36.59^{\rm s}$ & $-19^\circ23^{\prime}50.0^{\prime\prime}$ &1637.3 &1.7 &4.7 & 7.4 $\pm$ 4.8 &29.2 $\pm$ 18.3 & 2.6 $\pm$ 1.3 & 16.4 $\pm$ 24.6 & 7.0 $\pm$ 9.5 & Arm & 0 \\
 119 &$ 3^{\rm h}19^{\rm m}36.22^{\rm s}$ & $-19^\circ23^{\prime}49.1^{\prime\prime}$ &1638.1 &2.5 &6.0 & 4.6 $\pm$ 1.8 &32.5 $\pm$ 11.4 & 4.9 $\pm$ 0.9 & 7.3 $\pm$ 6.6 & 1.6 $\pm$ 1.4 & Arm & 0 \\
 120 &$ 3^{\rm h}19^{\rm m}37.13^{\rm s}$ & $-19^\circ24^{\prime}40.6^{\prime\prime}$ &1631.2 &2.2 &6.1 & 5.8 $\pm$ 2.3 &35.8 $\pm$ 9.1 & 4.8 $\pm$ 0.8 & 12.4 $\pm$ 11.1 & 2.9 $\pm$ 2.6 & other & 0 \\
 121 &$ 3^{\rm h}19^{\rm m}36.14^{\rm s}$ & $-19^\circ24^{\prime}00.2^{\prime\prime}$ &1638.3 &1.7 &4.9 & 4.4 $\pm$ 2.2 &$<$29.6 & 2.2 $\pm$ 1.3 & - & - & Arm & 1 \\
 122 &$ 3^{\rm h}19^{\rm m}36.57^{\rm s}$ & $-19^\circ23^{\prime}48.2^{\prime\prime}$ &1635.9 &2.7 &6.4 & 5.0 $\pm$ 4.7 &76.5 $\pm$ 47.6 & 13.1 $\pm$ 10.2 & 20.0 $\pm$ 40.4 & 1.7 $\pm$ 3.6 & Arm & 0 \\
 123 &$ 3^{\rm h}19^{\rm m}36.52^{\rm s}$ & $-19^\circ23^{\prime}47.7^{\prime\prime}$ &1629.3 &3.3 &8.6 & 2.2 $\pm$ 1.1 &49.0 $\pm$ 16.7 & 8.9 $\pm$ 5.7 & 2.5 $\pm$ 2.6 & 0.3 $\pm$ 0.3 & Arm & 0 \\
 124 &$ 3^{\rm h}19^{\rm m}36.00^{\rm s}$ & $-19^\circ23^{\prime}47.0^{\prime\prime}$ &1635.2 &2.1 &5.1 & 4.9 $\pm$ 1.9 &35.6 $\pm$ 14.4 & 5.2 $\pm$ 1.7 & 8.8 $\pm$ 8.1 & 1.9 $\pm$ 1.8 & Arm & 0 \\
 \hline
\end{tabular}
\end{table*}

\begin{table*}
    \contcaption{GMC catalog}
    \label{tab:catalog}
    \begin{tabular}{lcccccccccccc}
    \hline
   ID & RA  & Dec & $v_{\rm LSR}$ & $T_{\rm peak}$ & S/N & $\sigma_v$ & $R$ & $M_{\rm mol}$ & $M_{\rm vir}$ & $\alpha_{\rm vir}$ & Reg & Flag \\
      & (J2000) & (J2000) & ($\rm km~s^{-1}$) & (K) & & ($\rm km~s^{-1}$) & (pc) & ($10^5~M_\odot$) & ($10^5~M_\odot$) &  & &\\
   \hline
 125 &$ 3^{\rm h}19^{\rm m}36.79^{\rm s}$ & $-19^\circ23^{\prime}43.8^{\prime\prime}$ &1631.4 &2.2 &4.4 & 4.7 $\pm$ 3.2 &27.7 $\pm$ 24.5 & 3.2 $\pm$ 1.6 & 6.4 $\pm$ 10.4 & 2.2 $\pm$ 3.8 & Arm & 0 \\
 126 &$ 3^{\rm h}19^{\rm m}36.86^{\rm s}$ & $-19^\circ23^{\prime}43.9^{\prime\prime}$ &1634.5 &2.8 &5.3 & 2.4 $\pm$ 2.0 &$<$29.6 & 4.6 $\pm$ 3.3 & - & - & Arm & 1 \\
 127 &$ 3^{\rm h}19^{\rm m}36.75^{\rm s}$ & $-19^\circ23^{\prime}43.2^{\prime\prime}$ &1627.6 &3.1 &6.6 & 3.9 $\pm$ 3.8 &21.6 $\pm$ 16.2 & 5.0 $\pm$ 3.1 & 3.4 $\pm$ 6.6 & 0.8 $\pm$ 1.4 & Arm & 0 \\
 128 &$ 3^{\rm h}19^{\rm m}36.82^{\rm s}$ & $-19^\circ23^{\prime}42.9^{\prime\prime}$ &1633.0 &3.0 &5.5 & 4.9 $\pm$ 2.3 &44.0 $\pm$ 15.5 & 7.2 $\pm$ 1.9 & 11.1 $\pm$ 11.2 & 1.7 $\pm$ 1.7 & Arm & 0 \\
 129 &$ 3^{\rm h}19^{\rm m}37.04^{\rm s}$ & $-19^\circ23^{\prime}41.5^{\prime\prime}$ &1629.7 &4.0 &6.6 & 4.7 $\pm$ 1.7 &34.6 $\pm$ 8.2 & 8.4 $\pm$ 1.4 & 7.9 $\pm$ 6.1 & 1.0 $\pm$ 0.8 & Arm & 0 \\
 130 &$ 3^{\rm h}19^{\rm m}37.32^{\rm s}$ & $-19^\circ24^{\prime}43.5^{\prime\prime}$ &1620.4 &2.2 &5.8 & 5.5 $\pm$ 2.0 &50.8 $\pm$ 9.0 & 8.7 $\pm$ 1.5 & 16.2 $\pm$ 11.7 & 2.1 $\pm$ 1.4 & other & 0 \\
 131 &$ 3^{\rm h}19^{\rm m}36.70^{\rm s}$ & $-19^\circ24^{\prime}37.2^{\prime\prime}$ &1630.0 &2.2 &6.2 & 6.7 $\pm$ 3.4 &$<$29.6 & 3.2 $\pm$ 4.1 & - & - & other & 1 \\
 132 &$ 3^{\rm h}19^{\rm m}36.65^{\rm s}$ & $-19^\circ24^{\prime}36.7^{\prime\prime}$ &1621.8 &4.2 &11.4 & 5.4 $\pm$ 1.6 &48.2 $\pm$ 15.4 & 11.2 $\pm$ 3.3 & 14.7 $\pm$ 10.0 & 1.5 $\pm$ 0.8 & other & 0 \\
 133 &$ 3^{\rm h}19^{\rm m}36.48^{\rm s}$ & $-19^\circ24^{\prime}35.4^{\prime\prime}$ &1631.9 &1.7 &4.2 & 5.2 $\pm$ 4.2 &$<$29.6 & 4.1 $\pm$ 3.3 & - & - & other & 1 \\
 134 &$ 3^{\rm h}19^{\rm m}36.50^{\rm s}$ & $-19^\circ24^{\prime}34.7^{\prime\prime}$ &1625.8 &2.3 &6.9 & 8.8 $\pm$ 4.0 &$<$29.6 & 3.8 $\pm$ 4.9 & - & - & Bar & 1 \\
 135 &$ 3^{\rm h}19^{\rm m}36.51^{\rm s}$ & $-19^\circ24^{\prime}33.9^{\prime\prime}$ &1627.7 &3.4 &8.6 & 6.8 $\pm$ 3.3 &17.4 $\pm$ 10.7 & 6.5 $\pm$ 3.9 & 8.5 $\pm$ 10.8 & 1.5 $\pm$ 1.8 & Bar & 0 \\
 136 &$ 3^{\rm h}19^{\rm m}36.15^{\rm s}$ & $-19^\circ24^{\prime}33.3^{\prime\prime}$ &1636.0 &2.0 &4.7 & 4.8 $\pm$ 2.2 &8.7 $\pm$ 8.5 & 2.0 $\pm$ 1.0 & 2.1 $\pm$ 3.0 & 1.2 $\pm$ 1.5 & BE & 0 \\
 137 &$ 3^{\rm h}19^{\rm m}36.00^{\rm s}$ & $-19^\circ24^{\prime}03.8^{\prime\prime}$ &1635.7 &1.9 &5.6 & 4.4 $\pm$ 2.6 &$<$29.6 & 1.4 $\pm$ 0.6 & - & - & Arm & 1 \\
 138 &$ 3^{\rm h}19^{\rm m}36.21^{\rm s}$ & $-19^\circ24^{\prime}00.4^{\prime\prime}$ &1637.4 &1.5 &4.3 & 4.8 $\pm$ 2.4 &$<$29.6 & 2.2 $\pm$ 1.0 & - & - & Arm & 1 \\
 139 &$ 3^{\rm h}19^{\rm m}35.98^{\rm s}$ & $-19^\circ23^{\prime}59.5^{\prime\prime}$ &1630.1 &2.1 &6.3 & 3.9 $\pm$ 2.7 &28.0 $\pm$ 21.9 & 3.4 $\pm$ 2.1 & 4.5 $\pm$ 6.8 & 1.5 $\pm$ 2.3 & Arm & 0 \\
 140 &$ 3^{\rm h}19^{\rm m}35.96^{\rm s}$ & $-19^\circ23^{\prime}59.0^{\prime\prime}$ &1633.6 &2.0 &5.4 & 6.7 $\pm$ 5.2 &27.5 $\pm$ 24.8 & 2.3 $\pm$ 2.9 & 13.0 $\pm$ 25.3 & 6.2 $\pm$ 11.4 & Arm & 0 \\
 141 &$ 3^{\rm h}19^{\rm m}36.68^{\rm s}$ & $-19^\circ23^{\prime}58.0^{\prime\prime}$ &1619.1 &2.3 &6.8 & 4.3 $\pm$ 3.0 &52.5 $\pm$ 23.9 & 5.5 $\pm$ 2.3 & 9.9 $\pm$ 14.1 & 2.0 $\pm$ 2.9 & Arm & 0 \\
 142 &$ 3^{\rm h}19^{\rm m}36.71^{\rm s}$ & $-19^\circ23^{\prime}57.6^{\prime\prime}$ &1616.4 &2.2 &5.9 & 5.6 $\pm$ 5.5 &27.8 $\pm$ 22.0 & 3.4 $\pm$ 2.6 & 9.2 $\pm$ 21.1 & 3.0 $\pm$ 5.8 & Arm & 0 \\
 143 &$ 3^{\rm h}19^{\rm m}36.34^{\rm s}$ & $-19^\circ23^{\prime}56.8^{\prime\prime}$ &1632.7 &2.0 &5.4 & 7.2 $\pm$ 3.5 &$<$29.6 & 4.7 $\pm$ 2.8 & - & - & Arm & 1 \\
 144 &$ 3^{\rm h}19^{\rm m}36.35^{\rm s}$ & $-19^\circ23^{\prime}55.5^{\prime\prime}$ &1631.3 &2.8 &7.8 & 6.4 $\pm$ 2.1 &44.1 $\pm$ 20.9 & 10.9 $\pm$ 3.7 & 19.1 $\pm$ 13.2 & 2.0 $\pm$ 1.2 & Arm & 0 \\
 145 &$ 3^{\rm h}19^{\rm m}36.50^{\rm s}$ & $-19^\circ23^{\prime}52.7^{\prime\prime}$ &1632.9 &2.1 &5.5 & 5.9 $\pm$ 3.3 &$<$29.6 & 2.2 $\pm$ 0.8 & - & - & Arm & 1 \\
 146 &$ 3^{\rm h}19^{\rm m}36.33^{\rm s}$ & $-19^\circ23^{\prime}52.2^{\prime\prime}$ &1634.4 &2.1 &5.3 & 6.5 $\pm$ 2.6 &32.5 $\pm$ 12.6 & 4.3 $\pm$ 1.0 & 14.2 $\pm$ 12.9 & 3.7 $\pm$ 3.2 & Arm & 0 \\
 147 &$ 3^{\rm h}19^{\rm m}36.60^{\rm s}$ & $-19^\circ23^{\prime}47.3^{\prime\prime}$ &1633.0 &3.0 &7.2 & 4.5 $\pm$ 4.4 &24.3 $\pm$ 20.5 & 3.7 $\pm$ 5.4 & 5.2 $\pm$ 11.5 & 1.6 $\pm$ 3.5 & Arm & 0 \\
 148 &$ 3^{\rm h}19^{\rm m}36.67^{\rm s}$ & $-19^\circ23^{\prime}47.3^{\prime\prime}$ &1622.2 &4.9 &11.4 & 4.9 $\pm$ 1.3 &51.9 $\pm$ 15.0 & 13.4 $\pm$ 4.3 & 13.1 $\pm$ 8.5 & 1.1 $\pm$ 0.7 & Arm & 0 \\
 149 &$ 3^{\rm h}19^{\rm m}36.63^{\rm s}$ & $-19^\circ23^{\prime}43.4^{\prime\prime}$ &1632.6 &2.4 &5.0 & 4.5 $\pm$ 2.0 &$<$29.6 & 3.3 $\pm$ 1.3 & - & - & Arm & 1 \\
 150 &$ 3^{\rm h}19^{\rm m}36.30^{\rm s}$ & $-19^\circ23^{\prime}40.2^{\prime\prime}$ &1634.5 &2.7 &4.8 & 4.6 $\pm$ 2.2 &27.0 $\pm$ 11.5 & 4.8 $\pm$ 1.5 & 6.0 $\pm$ 6.6 & 1.4 $\pm$ 1.4 & Arm & 0 \\
 151 &$ 3^{\rm h}19^{\rm m}37.36^{\rm s}$ & $-19^\circ24^{\prime}50.5^{\prime\prime}$ &1627.9 &2.0 &4.4 & 5.2 $\pm$ 2.9 &$<$29.6 & 2.7 $\pm$ 1.3 & - & - & other & 1 \\
 152 &$ 3^{\rm h}19^{\rm m}37.09^{\rm s}$ & $-19^\circ24^{\prime}32.0^{\prime\prime}$ &1602.6 &3.3 &10.7 & 9.9 $\pm$ 5.0 &$<$29.6 & 10.4 $\pm$ 13.4 & - & - & Bar & 1 \\
 153 &$ 3^{\rm h}19^{\rm m}37.01^{\rm s}$ & $-19^\circ24^{\prime}31.5^{\prime\prime}$ &1607.2 &4.4 &13.0 & 9.0 $\pm$ 4.3 &$<$29.6 & 23.7 $\pm$ 15.8 & - & - & Bar & 1 \\
 154 &$ 3^{\rm h}19^{\rm m}37.14^{\rm s}$ & $-19^\circ24^{\prime}33.2^{\prime\prime}$ &1596.8 &1.9 &5.3 & 4.6 $\pm$ 3.0 &31.7 $\pm$ 16.3 & 3.7 $\pm$ 2.4 & 7.0 $\pm$ 9.7 & 2.1 $\pm$ 2.7 & Bar & 0 \\
 155 &$ 3^{\rm h}19^{\rm m}37.13^{\rm s}$ & $-19^\circ24^{\prime}32.6^{\prime\prime}$ &1584.1 &1.4 &4.0 & 4.3 $\pm$ 3.3 &$<$29.6 & 1.1 $\pm$ 0.9 & - & - & Bar & 1 \\
 156 &$ 3^{\rm h}19^{\rm m}36.21^{\rm s}$ & $-19^\circ24^{\prime}31.7^{\prime\prime}$ &1634.4 &2.0 &4.7 & 2.4 $\pm$ 2.0 &$<$29.6 & 1.1 $\pm$ 0.6 & - & - & BE & 1 \\
 157 &$ 3^{\rm h}19^{\rm m}36.55^{\rm s}$ & $-19^\circ24^{\prime}30.4^{\prime\prime}$ &1620.2 &2.6 &8.0 & 5.2 $\pm$ 3.5 &25.8 $\pm$ 25.4 & 3.1 $\pm$ 2.3 & 7.2 $\pm$ 11.6 & 2.6 $\pm$ 3.7 & Bar & 0 \\
 158 &$ 3^{\rm h}19^{\rm m}36.54^{\rm s}$ & $-19^\circ24^{\prime}29.5^{\prime\prime}$ &1618.0 &1.6 &4.6 & 3.9 $\pm$ 3.7 &$<$29.6 & 0.9 $\pm$ 0.7 & - & - & Bar & 1 \\
 159 &$ 3^{\rm h}19^{\rm m}36.59^{\rm s}$ & $-19^\circ24^{\prime}32.1^{\prime\prime}$ &1609.7 &2.1 &5.8 & 2.7 $\pm$ 2.2 &32.1 $\pm$ 24.9 & 3.0 $\pm$ 1.3 & 2.4 $\pm$ 5.0 & 0.9 $\pm$ 1.9 & Bar & 0 \\
 160 &$ 3^{\rm h}19^{\rm m}36.53^{\rm s}$ & $-19^\circ24^{\prime}30.7^{\prime\prime}$ &1608.2 &2.2 &6.3 & 4.8 $\pm$ 5.1 &44.4 $\pm$ 38.7 & 2.5 $\pm$ 2.7 & 10.6 $\pm$ 28.5 & 4.7 $\pm$ 13.1 & Bar & 0 \\
 161 &$ 3^{\rm h}19^{\rm m}36.65^{\rm s}$ & $-19^\circ24^{\prime}30.7^{\prime\prime}$ &1604.2 &3.9 &11.8 & 4.9 $\pm$ 6.7 &17.4 $\pm$ 21.5 & 8.1 $\pm$ 11.6 & 4.4 $\pm$ 14.7 & 0.6 $\pm$ 2.1 & Bar & 0 \\
 162 &$ 3^{\rm h}19^{\rm m}36.72^{\rm s}$ & $-19^\circ24^{\prime}30.7^{\prime\prime}$ &1598.5 &4.6 &13.4 & 5.0 $\pm$ 2.4 &81.9 $\pm$ 23.0 & 40.2 $\pm$ 24.4 & 21.5 $\pm$ 19.7 & 0.6 $\pm$ 0.6 & Bar & 0 \\
 163 &$ 3^{\rm h}19^{\rm m}36.81^{\rm s}$ & $-19^\circ24^{\prime}31.2^{\prime\prime}$ &1599.3 &2.2 &7.0 & 6.1 $\pm$ 5.4 &35.7 $\pm$ 21.4 & 5.7 $\pm$ 5.4 & 14.0 $\pm$ 25.2 & 2.7 $\pm$ 5.2 & Bar & 0 \\
 164 &$ 3^{\rm h}19^{\rm m}36.60^{\rm s}$ & $-19^\circ24^{\prime}19.1^{\prime\prime}$ &1624.9 &1.2 &4.2 & 7.5 $\pm$ 6.1 &$<$29.6 & 1.0 $\pm$ 0.6 & - & - & other & 1 \\
 165 &$ 3^{\rm h}19^{\rm m}36.76^{\rm s}$ & $-19^\circ23^{\prime}57.2^{\prime\prime}$ &1614.1 &2.9 &7.8 & 6.8 $\pm$ 3.0 &34.6 $\pm$ 15.2 & 5.1 $\pm$ 2.1 & 16.4 $\pm$ 17.2 & 3.6 $\pm$ 3.7 & Arm & 0 \\
 166 &$ 3^{\rm h}19^{\rm m}36.83^{\rm s}$ & $-19^\circ23^{\prime}56.1^{\prime\prime}$ &1614.9 &2.2 &6.2 & 3.5 $\pm$ 1.9 &96.7 $\pm$ 38.0 & 14.8 $\pm$ 7.4 & 12.6 $\pm$ 14.3 & 1.0 $\pm$ 1.1 & Arm & 0 \\
 167 &$ 3^{\rm h}19^{\rm m}36.21^{\rm s}$ & $-19^\circ23^{\prime}55.2^{\prime\prime}$ &1630.1 &2.0 &5.6 & 8.9 $\pm$ 3.4 &20.1 $\pm$ 11.9 & 3.1 $\pm$ 0.9 & 16.5 $\pm$ 16.6 & 5.9 $\pm$ 5.3 & Arm & 0 \\
 168 &$ 3^{\rm h}19^{\rm m}36.15^{\rm s}$ & $-19^\circ23^{\prime}55.0^{\prime\prime}$ &1634.7 &2.2 &6.3 & 2.4 $\pm$ 2.0 &$<$29.6 & 1.4 $\pm$ 0.4 & - & - & Arm & 1 \\
 169 &$ 3^{\rm h}19^{\rm m}36.23^{\rm s}$ & $-19^\circ23^{\prime}50.5^{\prime\prime}$ &1635.8 &1.7 &4.4 & 4.7 $\pm$ 3.1 &8.1 $\pm$ 12.4 & 0.9 $\pm$ 0.7 & 1.9 $\pm$ 4.6 & 2.4 $\pm$ 5.5 & Arm & 0 \\
 170 &$ 3^{\rm h}19^{\rm m}36.27^{\rm s}$ & $-19^\circ23^{\prime}50.0^{\prime\prime}$ &1625.7 &2.0 &5.1 & 2.7 $\pm$ 2.2 &60.5 $\pm$ 18.1 & 5.5 $\pm$ 2.7 & 4.4 $\pm$ 7.5 & 0.9 $\pm$ 1.6 & Arm & 0 \\
 171 &$ 3^{\rm h}19^{\rm m}36.95^{\rm s}$ & $-19^\circ23^{\prime}44.7^{\prime\prime}$ &1619.7 &3.3 &6.6 & 3.7 $\pm$ 4.2 &38.3 $\pm$ 26.4 & 4.8 $\pm$ 5.3 & 5.4 $\pm$ 13.3 & 1.2 $\pm$ 3.2 & Arm & 0 \\
 172 &$ 3^{\rm h}19^{\rm m}37.08^{\rm s}$ & $-19^\circ23^{\prime}46.9^{\prime\prime}$ &1619.8 &2.2 &4.3 & 3.2 $\pm$ 4.0 &30.0 $\pm$ 23.1 & 3.3 $\pm$ 3.9 & 3.1 $\pm$ 7.8 & 1.1 $\pm$ 2.7 & Arm & 0 \\
 173 &$ 3^{\rm h}19^{\rm m}37.02^{\rm s}$ & $-19^\circ23^{\prime}45.5^{\prime\prime}$ &1617.7 &4.8 &10.2 & 3.6 $\pm$ 1.8 &97.7 $\pm$ 22.5 & 28.9 $\pm$ 9.1 & 13.5 $\pm$ 13.4 & 0.5 $\pm$ 0.5 & Arm & 0 \\
 174 &$ 3^{\rm h}19^{\rm m}37.13^{\rm s}$ & $-19^\circ23^{\prime}47.3^{\prime\prime}$ &1614.6 &3.0 &6.3 & 5.3 $\pm$ 1.9 &42.1 $\pm$ 12.8 & 7.3 $\pm$ 1.7 & 12.2 $\pm$ 10.1 & 1.9 $\pm$ 1.3 & Arm & 0 \\
 175 &$ 3^{\rm h}19^{\rm m}36.99^{\rm s}$ & $-19^\circ23^{\prime}46.6^{\prime\prime}$ &1614.4 &2.9 &5.9 & 5.1 $\pm$ 2.7 &27.0 $\pm$ 18.5 & 7.1 $\pm$ 4.3 & 7.3 $\pm$ 8.0 & 1.1 $\pm$ 1.3 & Arm & 0 \\
 176 &$ 3^{\rm h}19^{\rm m}37.04^{\rm s}$ & $-19^\circ23^{\prime}43.6^{\prime\prime}$ &1624.4 &2.9 &5.3 & 5.0 $\pm$ 1.8 &48.9 $\pm$ 12.6 & 8.5 $\pm$ 1.5 & 12.8 $\pm$ 10.1 & 1.7 $\pm$ 1.3 & Arm & 0 \\
 177 &$ 3^{\rm h}19^{\rm m}36.25^{\rm s}$ & $-19^\circ23^{\prime}40.5^{\prime\prime}$ &1627.2 &2.3 &4.0 & 4.1 $\pm$ 4.7 &$<$29.6 & 3.1 $\pm$ 1.2 & - & - & Arm & 1 \\
 178 &$ 3^{\rm h}19^{\rm m}37.09^{\rm s}$ & $-19^\circ23^{\prime}39.1^{\prime\prime}$ &1628.7 &3.0 &4.4 & 4.5 $\pm$ 3.1 &$<$29.6 & 2.6 $\pm$ 1.3 & - & - & other & 1 \\
 179 &$ 3^{\rm h}19^{\rm m}37.78^{\rm s}$ & $-19^\circ24^{\prime}45.8^{\prime\prime}$ &1620.1 &1.9 &4.8 & 3.4 $\pm$ 2.0 &20.9 $\pm$ 11.8 & 3.2 $\pm$ 1.4 & 2.5 $\pm$ 3.5 & 0.9 $\pm$ 1.2 & other & 0 \\
 180 &$ 3^{\rm h}19^{\rm m}37.28^{\rm s}$ & $-19^\circ24^{\prime}42.5^{\prime\prime}$ &1622.4 &2.2 &6.0 & 3.9 $\pm$ 1.2 &48.0 $\pm$ 14.6 & 4.9 $\pm$ 1.3 & 7.6 $\pm$ 5.5 & 1.7 $\pm$ 1.1 & other & 0 \\
 181 &$ 3^{\rm h}19^{\rm m}36.97^{\rm s}$ & $-19^\circ24^{\prime}39.3^{\prime\prime}$ &1627.7 &1.9 &5.5 & 4.2 $\pm$ 4.6 &25.3 $\pm$ 30.2 & 1.5 $\pm$ 1.5 & 4.6 $\pm$ 12.1 & 3.5 $\pm$ 7.9 & other & 0 \\
 182 &$ 3^{\rm h}19^{\rm m}36.95^{\rm s}$ & $-19^\circ24^{\prime}39.0^{\prime\prime}$ &1624.3 &1.7 &5.1 & 5.3 $\pm$ 3.8 &44.2 $\pm$ 29.0 & 3.8 $\pm$ 2.6 & 13.1 $\pm$ 22.2 & 3.8 $\pm$ 5.6 & other & 0 \\
 183 &$ 3^{\rm h}19^{\rm m}36.88^{\rm s}$ & $-19^\circ24^{\prime}38.6^{\prime\prime}$ &1624.2 &2.6 &7.1 & 5.0 $\pm$ 1.1 &56.0 $\pm$ 9.5 & 9.3 $\pm$ 1.1 & 14.4 $\pm$ 7.4 & 1.7 $\pm$ 0.8 & other & 0 \\
 184 &$ 3^{\rm h}19^{\rm m}36.75^{\rm s}$ & $-19^\circ24^{\prime}35.5^{\prime\prime}$ &1612.5 &2.5 &7.1 & 5.4 $\pm$ 1.4 &51.1 $\pm$ 10.4 & 8.3 $\pm$ 1.1 & 15.4 $\pm$ 9.0 & 2.1 $\pm$ 1.1 & Bar & 0 \\
 185 &$ 3^{\rm h}19^{\rm m}35.97^{\rm s}$ & $-19^\circ24^{\prime}23.1^{\prime\prime}$ &1627.7 &2.1 &5.4 & 4.6 $\pm$ 2.5 &$<$29.6 & 1.9 $\pm$ 0.6 & - & - & BE & 1 \\
 186 &$ 3^{\rm h}19^{\rm m}36.19^{\rm s}$ & $-19^\circ24^{\prime}19.7^{\prime\prime}$ &1624.6 &1.8 &5.1 & 4.0 $\pm$ 1.4 &56.9 $\pm$ 12.0 & 6.6 $\pm$ 1.6 & 9.7 $\pm$ 7.1 & 1.6 $\pm$ 1.1 & BE & 0 \\
 \hline
\end{tabular}
\end{table*}

\begin{table*}
    \contcaption{GMC catalog}
    \label{tab:catalog}
    \begin{tabular}{lcccccccccccc}
    \hline
   ID & RA  & Dec & $v_{\rm LSR}$ & $T_{\rm peak}$ & S/N & $\sigma_v$ & $R$ & $M_{\rm mol}$ & $M_{\rm vir}$ & $\alpha_{\rm vir}$ & Reg & Flag \\
      & (J2000) & (J2000) & ($\rm km~s^{-1}$) & (K) & & ($\rm km~s^{-1}$) & (pc) & ($10^5~M_\odot$) & ($10^5~M_\odot$) &  & &\\
   \hline
 187 &$ 3^{\rm h}19^{\rm m}36.55^{\rm s}$ & $-19^\circ24^{\prime}02.2^{\prime\prime}$ &1616.3 &1.4 &4.9 & 5.3 $\pm$ 2.4 &43.4 $\pm$ 14.4 & 3.3 $\pm$ 1.0 & 12.9 $\pm$ 11.9 & 4.4 $\pm$ 3.6 & Arm & 0 \\
 188 &$ 3^{\rm h}19^{\rm m}36.88^{\rm s}$ & $-19^\circ23^{\prime}46.3^{\prime\prime}$ &1624.8 &2.7 &5.7 & 6.3 $\pm$ 2.3 &50.0 $\pm$ 14.5 & 5.5 $\pm$ 1.8 & 20.8 $\pm$ 17.6 & 4.2 $\pm$ 3.5 & Arm & 0 \\
 189 &$ 3^{\rm h}19^{\rm m}36.57^{\rm s}$ & $-19^\circ24^{\prime}33.2^{\prime\prime}$ &1617.9 &2.3 &6.1 & 6.2 $\pm$ 2.2 &81.9 $\pm$ 10.4 & 11.5 $\pm$ 2.7 & 32.5 $\pm$ 25.7 & 3.1 $\pm$ 2.5 & Bar & 0 \\
 190 &$ 3^{\rm h}19^{\rm m}36.89^{\rm s}$ & $-19^\circ24^{\prime}29.0^{\prime\prime}$ &1622.6 &1.7 &4.8 & 6.3 $\pm$ 4.1 &$<$29.6 & 1.7 $\pm$ 0.6 & - & - & Bar & 1 \\
 191 &$ 3^{\rm h}19^{\rm m}35.75^{\rm s}$ & $-19^\circ24^{\prime}15.0^{\prime\prime}$ &1621.2 &1.6 &4.3 & 5.7 $\pm$ 2.6 &$<$29.6 & 2.1 $\pm$ 1.0 & - & - & Arm & 1 \\
 192 &$ 3^{\rm h}19^{\rm m}37.31^{\rm s}$ & $-19^\circ24^{\prime}13.3^{\prime\prime}$ &1617.7 &1.6 &4.9 & 4.5 $\pm$ 1.9 &$<$29.6 & 1.3 $\pm$ 0.5 & - & - & other & 1 \\
 193 &$ 3^{\rm h}19^{\rm m}35.99^{\rm s}$ & $-19^\circ24^{\prime}08.3^{\prime\prime}$ &1620.1 &1.6 &4.8 & 4.5 $\pm$ 2.8 &$<$29.6 & 1.4 $\pm$ 0.7 & - & - & Arm & 1 \\
 194 &$ 3^{\rm h}19^{\rm m}36.05^{\rm s}$ & $-19^\circ24^{\prime}07.3^{\prime\prime}$ &1619.4 &1.7 &5.1 & 3.8 $\pm$ 1.7 &$<$29.6 & 2.9 $\pm$ 1.4 & - & - & Arm & 1 \\
 195 &$ 3^{\rm h}19^{\rm m}36.71^{\rm s}$ & $-19^\circ24^{\prime}06.3^{\prime\prime}$ &1614.7 &1.6 &4.8 & 2.6 $\pm$ 1.4 &52.4 $\pm$ 16.3 & 3.8 $\pm$ 1.9 & 3.6 $\pm$ 4.3 & 1.0 $\pm$ 1.2 & other & 0 \\
 196 &$ 3^{\rm h}19^{\rm m}36.73^{\rm s}$ & $-19^\circ24^{\prime}05.4^{\prime\prime}$ &1614.5 &1.6 &4.9 & 2.0 $\pm$ 1.6 &$<$29.6 & 0.6 $\pm$ 0.2 & - & - & other & 1 \\
 197 &$ 3^{\rm h}19^{\rm m}36.97^{\rm s}$ & $-19^\circ23^{\prime}54.2^{\prime\prime}$ &1614.1 &2.6 &6.4 & 3.4 $\pm$ 1.5 &36.0 $\pm$ 14.1 & 5.1 $\pm$ 1.7 & 4.4 $\pm$ 4.6 & 1.0 $\pm$ 0.9 & Arm & 0 \\
 198 &$ 3^{\rm h}19^{\rm m}37.04^{\rm s}$ & $-19^\circ23^{\prime}54.0^{\prime\prime}$ &1612.9 &2.4 &6.2 & 4.5 $\pm$ 3.0 &35.7 $\pm$ 16.5 & 4.4 $\pm$ 1.6 & 7.6 $\pm$ 9.7 & 2.0 $\pm$ 2.5 & Arm & 0 \\
 199 &$ 3^{\rm h}19^{\rm m}37.14^{\rm s}$ & $-19^\circ23^{\prime}43.9^{\prime\prime}$ &1613.9 &3.2 &5.3 & 8.4 $\pm$ 2.5 &27.7 $\pm$ 12.8 & 5.6 $\pm$ 1.5 & 20.3 $\pm$ 16.5 & 4.0 $\pm$ 3.1 & Arm & 0 \\
 200 &$ 3^{\rm h}19^{\rm m}38.30^{\rm s}$ & $-19^\circ24^{\prime}38.1^{\prime\prime}$ &1589.1 &2.9 &7.8 & 5.9 $\pm$ 5.3 &$<$29.6 & 10.4 $\pm$ 12.6 & - & - & Bar & 1 \\
 201 &$ 3^{\rm h}19^{\rm m}38.44^{\rm s}$ & $-19^\circ24^{\prime}37.0^{\prime\prime}$ &1567.9 &2.4 &5.4 & 3.3 $\pm$ 4.7 &32.3 $\pm$ 25.9 & 6.9 $\pm$ 6.1 & 3.7 $\pm$ 10.9 & 0.6 $\pm$ 1.7 & Bar & 0 \\
 202 &$ 3^{\rm h}19^{\rm m}38.41^{\rm s}$ & $-19^\circ24^{\prime}37.6^{\prime\prime}$ &1563.1 &2.4 &5.9 & 6.6 $\pm$ 3.6 &48.5 $\pm$ 28.2 & 6.9 $\pm$ 4.9 & 21.7 $\pm$ 31.6 & 3.5 $\pm$ 4.5 & Bar & 0 \\
 203 &$ 3^{\rm h}19^{\rm m}37.13^{\rm s}$ & $-19^\circ24^{\prime}37.9^{\prime\prime}$ &1620.0 &2.2 &6.6 & 3.1 $\pm$ 0.9 &$<$29.6 & 2.0 $\pm$ 0.5 & - & - & Bar & 1 \\
 204 &$ 3^{\rm h}19^{\rm m}37.12^{\rm s}$ & $-19^\circ24^{\prime}31.0^{\prime\prime}$ &1611.4 &1.6 &5.0 & 5.9 $\pm$ 1.9 &$<$29.6 & 3.1 $\pm$ 1.1 & - & - & Bar & 1 \\
 205 &$ 3^{\rm h}19^{\rm m}37.17^{\rm s}$ & $-19^\circ24^{\prime}07.3^{\prime\prime}$ &1608.3 &2.1 &6.6 & 5.0 $\pm$ 2.0 &$<$29.6 & 3.6 $\pm$ 0.7 & - & - & other & 1 \\
 206 &$ 3^{\rm h}19^{\rm m}37.21^{\rm s}$ & $-19^\circ23^{\prime}49.7^{\prime\prime}$ &1615.0 &3.5 &7.4 & 2.8 $\pm$ 0.8 &61.0 $\pm$ 8.9 & 9.7 $\pm$ 1.7 & 5.0 $\pm$ 3.1 & 0.6 $\pm$ 0.4 & Arm & 0 \\
 207 &$ 3^{\rm h}19^{\rm m}37.23^{\rm s}$ & $-19^\circ23^{\prime}44.5^{\prime\prime}$ &1607.0 &3.4 &6.1 & 7.5 $\pm$ 2.3 &67.0 $\pm$ 12.0 & 16.7 $\pm$ 2.3 & 38.9 $\pm$ 25.4 & 2.6 $\pm$ 1.6 & Arm & 0 \\
 208 &$ 3^{\rm h}19^{\rm m}36.74^{\rm s}$ & $-19^\circ24^{\prime}32.4^{\prime\prime}$ &1606.9 &1.8 &5.0 & 6.3 $\pm$ 3.1 &28.1 $\pm$ 13.4 & 5.6 $\pm$ 2.2 & 11.5 $\pm$ 13.8 & 2.3 $\pm$ 2.5 & Bar & 0 \\
 209 &$ 3^{\rm h}19^{\rm m}36.83^{\rm s}$ & $-19^\circ24^{\prime}32.3^{\prime\prime}$ &1606.1 &2.1 &6.4 & 3.6 $\pm$ 1.8 &30.1 $\pm$ 10.7 & 4.6 $\pm$ 1.5 & 4.1 $\pm$ 4.9 & 1.0 $\pm$ 1.1 & Bar & 0 \\
 210 &$ 3^{\rm h}19^{\rm m}36.91^{\rm s}$ & $-19^\circ24^{\prime}05.4^{\prime\prime}$ &1614.8 &1.3 &4.2 & 3.0 $\pm$ 1.8 &$<$29.6 & 0.9 $\pm$ 0.5 & - & - & other & 1 \\
 211 &$ 3^{\rm h}19^{\rm m}36.97^{\rm s}$ & $-19^\circ24^{\prime}05.3^{\prime\prime}$ &1610.0 &2.0 &5.9 & 5.4 $\pm$ 2.5 &$<$29.6 & 2.6 $\pm$ 0.6 & - & - & other & 1 \\
 212 &$ 3^{\rm h}19^{\rm m}36.58^{\rm s}$ & $-19^\circ23^{\prime}50.2^{\prime\prime}$ &1613.6 &1.7 &4.3 & 4.6 $\pm$ 2.3 &$<$29.6 & 1.6 $\pm$ 0.6 & - & - & Arm & 1 \\
 213 &$ 3^{\rm h}19^{\rm m}37.39^{\rm s}$ & $-19^\circ23^{\prime}44.7^{\prime\prime}$ &1607.9 &3.7 &6.3 & 3.6 $\pm$ 1.1 &54.2 $\pm$ 11.2 & 10.1 $\pm$ 1.7 & 7.1 $\pm$ 4.5 & 0.8 $\pm$ 0.5 & Arm & 0 \\
 214 &$ 3^{\rm h}19^{\rm m}36.95^{\rm s}$ & $-19^\circ24^{\prime}34.1^{\prime\prime}$ &1603.4 &1.7 &5.2 & 6.1 $\pm$ 2.4 &16.5 $\pm$ 10.5 & 2.6 $\pm$ 0.9 & 6.4 $\pm$ 6.7 & 2.8 $\pm$ 2.5 & Bar & 0 \\
 215 &$ 3^{\rm h}19^{\rm m}37.60^{\rm s}$ & $-19^\circ24^{\prime}30.8^{\prime\prime}$ &1601.1 &2.3 &6.5 & 6.5 $\pm$ 1.3 &63.4 $\pm$ 8.1 & 13.7 $\pm$ 1.6 & 27.7 $\pm$ 11.7 & 2.3 $\pm$ 0.8 & Bar & 0 \\
 216 &$ 3^{\rm h}19^{\rm m}38.09^{\rm s}$ & $-19^\circ24^{\prime}27.0^{\prime\prime}$ &1599.6 &1.6 &5.0 & 7.2 $\pm$ 4.2 &$<$29.6 & 1.5 $\pm$ 0.5 & - & - & other & 1 \\
 217 &$ 3^{\rm h}19^{\rm m}38.20^{\rm s}$ & $-19^\circ24^{\prime}25.9^{\prime\prime}$ &1604.8 &1.9 &5.1 & 3.4 $\pm$ 2.1 &$<$29.6 & 2.1 $\pm$ 0.5 & - & - & other & 1 \\
 218 &$ 3^{\rm h}19^{\rm m}37.03^{\rm s}$ & $-19^\circ24^{\prime}05.0^{\prime\prime}$ &1600.6 &2.1 &5.7 & 5.8 $\pm$ 2.5 &21.4 $\pm$ 7.8 & 3.2 $\pm$ 0.9 & 7.4 $\pm$ 7.4 & 2.6 $\pm$ 2.4 & other & 0 \\
 219 &$ 3^{\rm h}19^{\rm m}37.06^{\rm s}$ & $-19^\circ24^{\prime}01.6^{\prime\prime}$ &1609.1 &2.4 &6.5 & 2.3 $\pm$ 1.2 &35.9 $\pm$ 11.2 & 3.5 $\pm$ 0.7 & 2.0 $\pm$ 2.3 & 0.6 $\pm$ 0.7 & other & 0 \\
 220 &$ 3^{\rm h}19^{\rm m}37.18^{\rm s}$ & $-19^\circ23^{\prime}52.2^{\prime\prime}$ &1602.6 &2.0 &4.6 & 4.4 $\pm$ 1.7 &24.5 $\pm$ 10.9 & 4.1 $\pm$ 1.3 & 4.9 $\pm$ 4.5 & 1.3 $\pm$ 1.3 & Arm & 0 \\
 221 &$ 3^{\rm h}19^{\rm m}38.06^{\rm s}$ & $-19^\circ24^{\prime}31.5^{\prime\prime}$ &1592.3 &1.5 &4.1 & 4.7 $\pm$ 1.6 &67.8 $\pm$ 14.3 & 6.9 $\pm$ 2.7 & 15.4 $\pm$ 11.4 & 2.5 $\pm$ 1.8 & Bar & 0 \\
 222 &$ 3^{\rm h}19^{\rm m}37.79^{\rm s}$ & $-19^\circ24^{\prime}27.5^{\prime\prime}$ &1599.8 &1.5 &4.4 & 1.9 $\pm$ 0.9 &29.2 $\pm$ 15.0 & 1.7 $\pm$ 0.8 & 1.0 $\pm$ 1.2 & 0.7 $\pm$ 0.7 & Bar & 0 \\
 223 &$ 3^{\rm h}19^{\rm m}37.26^{\rm s}$ & $-19^\circ23^{\prime}45.7^{\prime\prime}$ &1602.8 &2.8 &5.1 & 4.5 $\pm$ 2.8 &23.2 $\pm$ 11.8 & 3.9 $\pm$ 1.3 & 5.0 $\pm$ 7.0 & 1.4 $\pm$ 1.8 & Arm & 0 \\
 224 &$ 3^{\rm h}19^{\rm m}37.03^{\rm s}$ & $-19^\circ24^{\prime}32.9^{\prime\prime}$ &1594.2 &2.3 &7.2 & 4.1 $\pm$ 1.5 &24.5 $\pm$ 11.5 & 3.5 $\pm$ 0.7 & 4.2 $\pm$ 3.9 & 1.3 $\pm$ 1.1 & Bar & 0 \\
 225 &$ 3^{\rm h}19^{\rm m}37.79^{\rm s}$ & $-19^\circ23^{\prime}50.9^{\prime\prime}$ &1585.5 &2.7 &4.6 & 8.8 $\pm$ 3.9 &35.5 $\pm$ 17.9 & 5.8 $\pm$ 2.7 & 28.8 $\pm$ 33.1 & 5.5 $\pm$ 6.2 & other & 0 \\
 226 &$ 3^{\rm h}19^{\rm m}38.28^{\rm s}$ & $-19^\circ24^{\prime}36.6^{\prime\prime}$ &1580.6 &2.1 &5.6 & 5.0 $\pm$ 5.0 &28.0 $\pm$ 20.7 & 3.9 $\pm$ 3.4 & 7.1 $\pm$ 15.9 & 2.1 $\pm$ 4.2 & Bar & 0 \\
 227 &$ 3^{\rm h}19^{\rm m}38.29^{\rm s}$ & $-19^\circ24^{\prime}36.9^{\prime\prime}$ &1567.0 &1.6 &4.4 & 8.1 $\pm$ 10.5 &48.0 $\pm$ 33.8 & 5.3 $\pm$ 9.8 & 32.5 $\pm$ 88.4 & 6.8 $\pm$ 19.7 & Bar & 0 \\
 228 &$ 3^{\rm h}19^{\rm m}38.39^{\rm s}$ & $-19^\circ24^{\prime}35.7^{\prime\prime}$ &1584.1 &2.2 &5.6 & 5.1 $\pm$ 2.6 &25.3 $\pm$ 8.2 & 3.7 $\pm$ 0.8 & 6.8 $\pm$ 7.7 & 2.1 $\pm$ 2.2 & Bar & 0 \\
 229 &$ 3^{\rm h}19^{\rm m}37.94^{\rm s}$ & $-19^\circ24^{\prime}35.4^{\prime\prime}$ &1580.6 &1.8 &4.7 & 8.7 $\pm$ 3.9 &9.8 $\pm$ 7.8 & 4.0 $\pm$ 1.2 & 7.7 $\pm$ 9.6 & 2.2 $\pm$ 2.4 & Bar & 0 \\
 230 &$ 3^{\rm h}19^{\rm m}38.17^{\rm s}$ & $-19^\circ24^{\prime}33.4^{\prime\prime}$ &1587.7 &2.0 &5.7 & 4.7 $\pm$ 3.4 &$<$29.6 & 1.4 $\pm$ 0.5 & - & - & Bar & 1 \\
 231 &$ 3^{\rm h}19^{\rm m}37.84^{\rm s}$ & $-19^\circ24^{\prime}32.7^{\prime\prime}$ &1585.2 &2.3 &6.5 & 4.8 $\pm$ 2.7 &$<$29.6 & 1.4 $\pm$ 0.3 & - & - & Bar & 1 \\
 232 &$ 3^{\rm h}19^{\rm m}37.01^{\rm s}$ & $-19^\circ24^{\prime}14.3^{\prime\prime}$ &1581.9 &1.4 &4.7 & 8.8 $\pm$ 4.8 &$<$29.6 & 1.3 $\pm$ 0.7 & - & - & other & 1 \\
 233 &$ 3^{\rm h}19^{\rm m}36.59^{\rm s}$ & $-19^\circ23^{\prime}37.3^{\prime\prime}$ &1578.5 &3.1 &4.9 & 5.6 $\pm$ 2.6 &$<$29.6 & 2.9 $\pm$ 1.8 & - & - & Arm & 1 \\
 \hline
 \end{tabular}
\end{table*}












\bsp	
\label{lastpage}
\end{document}